\documentclass[aps,prd,groupedaddress,preprintnumbers,
              floatfix, nofootinbib,longbibliography]{revtex4} 
\usepackage{textcomp}
\usepackage[latin1]{inputenc}                    
\usepackage{graphicx}                            
\usepackage{epsfig}
\usepackage[export]{adjustbox}
\usepackage{mathrsfs}

\usepackage{epsf}
\usepackage{mathalfa}
\usepackage{latexsym}                            
\usepackage{amsfonts}                            
\usepackage{amssymb}                             
\usepackage{amsmath}                             
\usepackage{slashed}
\usepackage[mathscr]{eucal}                      
\usepackage{dcolumn}                             
\usepackage{multirow}                               
\usepackage{bm}                                  
\usepackage{hyperref}                            

\graphicspath{{.}{Graphics/}}                    
\newcommand{\msbar}{\overline{\mbox{\rm MS}}}  
\newcommand{\mmsbar}{{\rm M}\overline{\mbox{\rm MS}}}  
\newcommand{\RI}{{\rm RI}'}  

\newcommand{\MSb}{\overline{\mathrm{MS}}}
\newcommand{\MMSb}{{\rm M}\overline{\mathrm{MS}}}

\newcommand{\be}{\begin{equation}}
\newcommand{\ee}{\end{equation}}
\newcommand{\bea}{\begin{eqnarray}}
\newcommand{\eea}{\end{eqnarray}}

\newcommand{\myeq}{\overset{0\ll \tau \ll t_s}{=}}
  
\newcommand{\hsn}{\hspace{-0.25cm}}

\def \3{\ss }

\newcommand*{\backin}{\scriptsize{\rotatebox[origin=c]{-180}{$\perp$}}}
\newcommand{\beqn}{\begin{eqnarray}}
\newcommand{\eeqn}{\end{eqnarray}}

\renewcommand{\arraystretch}{1.8}

\def\cyp{a}
\def\cyi{b}
\def\poz{c}
\def\teml{d}
\def\desy{e}
\def\wup{f}
\def\bon{g}

\begin{document}

\begin{titlepage}
  \begin{flushright}
    {\footnotesize DESY 19-001}
  \end{flushright}
  {\vspace{-0.5cm} \normalsize
  \hfill \parbox{60mm}{
}}\\[10mm]
  \begin{center}
    \begin{LARGE}
            \textbf{Systematic uncertainties in parton distribution functions from lattice QCD simulations at the physical point}
    \end{LARGE}
  \end{center}

\vspace*{1cm}

 \vspace{-0.8cm}
  \baselineskip 20pt plus 2pt minus 2pt
  \begin{center}
    \textbf{
      Constantia Alexandrou$^{(\cyp, \cyi)}$,
      Krzysztof Cichy$^{(\poz)}$,
      Martha Constantinou$^{(\teml)}$,\\
      Kyriakos Hadjiyiannakou$^{(\cyi)}$,
      Karl Jansen$^{(\desy)}$,
      Aurora Scapellato$^{(\cyp,\wup)}$,
      Fernanda Steffens$^{(\bon)}$ 
      }
\end{center}

  \begin{center}
    \begin{footnotesize}
      \noindent 	
 	$^{(\cyp)}$ Department of Physics, University of Cyprus, P.O. Box 20537, 1678 Nicosia, Cyprus\\	
 	$^{(\cyi)}$ Computation-based Science and Technology Research Center, The Cyprus Institute, 20 Kavafi Str., Nicosia 2121, Cyprus \\
 	$^{(\poz)}$  Faculty of Physics, Adam Mickiewicz University, Umultowska 85, 61-614 Pozna\'{n}, Poland\\
 	$^{(\teml)}$ Temple University,1925 N. 12th Street, Philadelphia, PA 19122-1801, USA \\
 	$^{(\desy)}$ NIC, DESY, Platanenallee 6, D-15738 Zeuthen, Germany \\
 	$^{(\wup)}$ University of Wuppertal, Gau\ss str. 20, 42119 Wuppertal, Germany\\
 	$^{(\bon)}$ Institut f\"{u}r Strahlen- und Kernphysik, Rheinische Friedrich-Wilhelms-Universit\"{a}t Bonn,  Nussallee 14-16, 53115 Bonn\\
     \vspace{0.2cm}
    \end{footnotesize}
  \end{center}

\centerline{\today}

\begin{abstract}

\vspace*{0.2cm}
We present a detailed study of the helicity-dependent and helicity-independent collinear parton distribution functions (PDFs) of the nucleon, using the quasi-PDF approach.
The lattice QCD computation is performed employing twisted mass fermions with a physical value of the light quark mass.
We give a systematic and in-depth account of the salient features entering in the 
evaluation of quasi-PDFs and their relation to the light-cone PDFs.
In particular, we give details for the computation of the matrix elements, including 
the study of the various sources of systematic uncertainties, such as excited states contamination. 
In addition, we discuss the non-perturbative renormalization scheme used here and its systematics, 
effects of truncating the Fourier transform and different matching prescriptions. 
Finally, we show improved results for the PDFs and discuss future directions, challenges  and prospects for evaluating precisely PDFs from lattice QCD with fully quantified uncertainties.

\end{abstract}
\maketitle 
\end{titlepage}

\section{Introduction}

Quantum Chromodynamics (QCD), being the fundamental theory of the strong interactions,  describes the interactions among quarks and gluons both in the perturbative regime, as well as in the non-perturbative regime with the emergence of complex structures like hadrons. 
As the latter requires a non-perturbative approach,  the use of a discretized Euclidean spacetime to numerically solve QCD has been proven to be extremely successful in the computation of physical  properties of hadrons, such as their masses and decay constants. 
The rich and diverse structure of hadrons is described in terms of observables like various kinds of form factors, revealing different structural aspects, such as the average distribution of electric charge or parton (quark and gluon) momentum and spin inside the hadron.
Even more information is contained in distribution functions, such as parton distribution functions (PDFs), 
generalized parton distributions (GPDs) and transverse momentum dependents PDFs (TMDs), 
which decompose the aggregate observables into probability densities for 
partons with specific longitudinal momentum, transverse momentum and  momentum transfer within a scattering event.
The simplest of these are the PDFs, which have been studied intensively and continuously in experimental facilities over the last few decades, most notably for the proton, and also provide input in collider experiments.
It is well-established that the main source of information on PDFs are global QCD analyses, providing accurate results due to theoretical advances and new data emerging from accelerators, covering different kinematical regions (see e.g. Refs.~\cite{Perez:2012um,DeRoeck:2011na,Alekhin:2011sk,Ball:2012wy,Aidala:2012mv,Forte:2013wc,Jimenez-Delgado:2013sma,Rojo:2015acz,Butterworth:2015oua,Accardi:2016ndt,Gao:2017yyd,Lin:2017snn}). 
Such analyses rely on the factorization framework, in which scattering cross sections, generally obtaining contributions from all energy scales, are written as convolutions of perturbatively computable coefficient functions describing the high-energy scales, and the low-energy PDFs.
Despite the tremendous progress in phenomenological parametrizations of PDFs, the procedures are not without ambiguities~\cite{Jimenez-Delgado:2013sma}, as there are still kinematical regions that are not easily accessible experimentally, e.g.\ the large Bjorken-$x$ region. 
Uncertainties on the PDFs in the large-$x$ range  propagate  to smaller-$x$ regions when QCD evolution is applied. 

A calculation of PDFs from first principles is a valuable addition to the global fitting analyses and of crucial importance for the deeper understanding of the inner structure of hadrons.
Apart from providing  a fundamental framework for the study of these quantities, it can also serve as input for experimental analyses in collision experiments. 
The non-perturbative nature of PDFs makes lattice QCD an ideal \textit{ab initio} formulation to determine them, utilizing large-scale simulations.
A novel method to extract parton distribution functions from lattice QCD was proposed five years ago by Ji~\cite{Ji:2013dva}.
It is based on considering  matrix elements probing purely spatial correlations, 
making them accessible in Euclidean lattice QCD.

The first studies on quasi-PDFs within lattice 
QCD~\cite{Lin:2014zya,Alexandrou:2014pna,Alexandrou:2015rja} have appeared soon after their proposal by Ji. 
These papers and the follow-up ones~\cite{Chen:2016utp,Alexandrou:2016jqi} focused mainly on the 
feasibility of the approach only calculating bare nucleon matrix elements of boosted nucleons 
for nonsinglet operators. 
The first computations applying a proper procedure for renormalization and matching to $\MSb$ 
light-cone PDFs and using simulations with physical pion mass were published in 
Ref.~\cite{Alexandrou:2018pbm} for the unpolarized isovector PDFs, $u(x)-d(x)$, and helicity, $\Delta u(x) - \Delta d(x)$ and in Ref.~\cite{Alexandrou:2018eet} for the isovector transversity PDF, $h_1^u(x) - h_1^d(x)$.
These works works were published in shorter letters, where many of the technical and methodological
details of the calculations 
needed to be left out. 

It is the purpose of this paper to fill this gap and provide the before mentioned details in a 
comprehensive way, give the
present understanding and control of systematics effects appearing in the computations
and to present the status of the calculations of PDFs for our twisted mass
setup. 
In particular, 
we extend the work of Refs.~\cite{Alexandrou:2018pbm,Alexandrou:2018eet} and give a 
systematic and in-depth account of the salient features entering in the evaluation of 
quasi-PDFs and their relation to the light-cone PDFs.  
We explain in detail the computation of the matrix elements including the study of 
the various sources of systematic uncertainties, for example excited states contamination. 
An improved renormalization procedure is presented using three ensembles 
for the non-perturbative extraction of the renormalization functions, allowing to take the chiral limit, 
as well as an investigation of other sources of systematic uncertainties related to the 
renormalization, such as the finite volume and scale dependence.
In addition, different prescriptions for applying the Fourier transform and the matching 
prescription are employed and critically compared.
 
On the physics side, 
in order to compare with existing phenomenological estimates, we convert our results in the RI$^{\prime}$ scheme into the $\MSb$ scheme following the procedure given in Ref.~\cite{Constantinou:2017sej}. We then apply appropriate matching to relate quasi-PDFs in the $\MSb$ scheme to light-cone PDFs extracted from global fits.
With the detailed analysis of systematic effects, as decribed in this work, having a physical value of the pion mass and the improved renormalization and 
matching prescriptions, we are finally in position to make indeed contact to the phenomenological analyses of PDFs and we present final 
results of our lattice PDF calculations in Sec.~\ref{sec:results} below. 
In addition to our work, results on the quasi-distribution approach were reported in Refs.~\cite{Zhang:2017bzy,Chen:2017mzz,Lin:2017ani,Chen:2017gck,Chen:2018xof,Chen:2018fwa,Lin:2018qky,Liu:2018uuj,Fan:2018dxu,Liu:2018hxv,Petreczky:2018jqc} that include, besides the nucleon PDFs, exploratory studies  of pion distribution amplitudes and PDFs as well as of gluon PDFs.
For complete references and an overview of the theoretical and numerical developments, see Ref.~\cite{Cichy:2018mum}.

On more general grounds, it is very important to realize that quasi-PDFs and light-cone PDFs have been shown to share the same infrared physics~\cite{Xiong:2013bka,Briceno:2017cpo}, which is the fundamental observation that allows one to relate both quantities using perturbation theory, provided that the hadron is moving with a large, although necessarily finite, momentum in a chosen spatial direction.
It has also been proven that quasi-PDFs can be extracted from lattice QCD in Euclidean spacetime \cite{Briceno:2017cpo} and that they do not suffer from power-divergent mixings with lower-dimensional operators \cite{Ji:2017rah,Radyushkin:2018nbf,Karpie:2018zaz}.
A factorization formula makes it possible to extract the PDFs from the quasi-PDFs, an operation called matching~\cite{Xiong:2013bka,Ji:2015qla,Xiong:2015nua,Wang:2017qyg,Stewart:2017tvs,Izubuchi:2018srq,Alexandrou:2018pbm,Alexandrou:2018eet,Liu:2018uuj,Liu:2018hxv}. 
In general, this procedure is based on a newly developed large-momentum effective theory (LaMET)~\cite{Ji:2014gla}, and it is renormalizable to all 
orders in perturbation theory~\cite{Ishikawa:2017faj,Ji:2017oey,Zhang:2018diq,Li:2018tpe}. 
Other approaches for a direct computation of the $x$-dependence of PDFs include the hadronic tensor~\cite{Liu:1993cv,Liu:1998um,Liu:1999ak}, fictitious heavy quark~\cite{Detmold:2005gg}, auxiliary light quark~\cite{Braun:2007wv}, good lattice cross sections~\cite{Ma:2014jla,Ma:2017pxb} (closely related to the auxiliary light quark method), ``OPE without OPE''~\cite{Chambers:2017dov} and pseudo-PDFs~\cite{Radyushkin:2016hsy,Radyushkin:2017ffo,Radyushkin:2017cyf,Radyushkin:2017lvu}, where the latter can be seen as a generalization of PDFs off the light-cone. 
These alternative approaches have been explored in lattice QCD, and recent results can be found in Refs.~\cite{Liang:2017mye,Detmold:2018kwu,Bali:2017gfr,Bali:2018spj,Chambers:2017dov,Orginos:2017kos,Karpie:2017bzm,Radyushkin:2018cvn,Karpie:2018zaz}.
The new formulation and its successful implementation within  lattice QCD  has also led to a wider interest on phenomenological  studies using models and toy theories of QCD~\cite{Gamberg:2014zwa,Gamberg:2015opc,Jia:2015pxx,Jia:2018qee,Bacchetta:2016zjm,Radyushkin:2016hsy,Radyushkin:2017ffo,Broniowski:2017wbr,Broniowski:2017gfp,Bhattacharya:2018zxi}. 
A detailed overview of the current status of lattice QCD calculation of PDFs and other partonic distributions can be found in the recent reviews of Refs.~\cite{Monahan:2018euv,Cichy:2018mum}.

The remainder of the paper is organized as follows: 
In Sec.~\ref{sec:methodology}, we provide the general theoretical aspects, lattice QCD action and parameters.
We discuss the computation of the bare matrix elements, along with the lattice techniques necessary for attaining high statistical accuracy. 
In Sec.~\ref{sec:bare_ME}, we present our results on bare matrix elements, together with various analyses methods that are required to control excited states contamination. Sec.~\ref{sec:renorm} describes in detail the non-perturbative renormalization program, which investigates pion mass dependence, volume effects and renormalization scale dependence. 
The matching procedure is presented in Sec.~\ref{sec:matching} and we demonstrate that a modified $\MSb$ scheme is preferable.
Different prescriptions are presented and we compare the effect on the final PDFs. 
Sec.~\ref{sec:matching} also includes an investigation of systematic uncertainties resulting from the Fourier transform. 
In Sec.~\ref{sec:results}, we present our final results with discussion on the systematic uncertainties, while in Sec.~\ref{sec:conclusions} we conclude.

\section{Methodology}
\label{sec:methodology}

\subsection{From PDF to quasi-PDF}
\label{sec:pdfs_quasi}
The original definition of quark momentum distribution within a hadron can be derived from the
operator product expansion (OPE) of hadronic deep-inelastic scattering and is given by
\begin{equation}
\label{def_pdf}
q(x)=\int_{-\infty}^{+\infty}\,d\xi^{-}\,e^{-ixP^{+}\xi^{-}}\,\langle N\vert \, \overline{\psi}(\xi^{-})\,\Gamma W(\xi^{-},0)\,\psi(0)\,\vert N\rangle \, ,
\end{equation}
where $x$ is the momentum fraction of the quark, $\vert N\rangle$ the hadron state at rest assumed to be the nucleon, the  momentum $P^{+} {=} (P^{0} + P^{3} )/\sqrt{2} {=} M/\sqrt{2}$, $\xi^{-}{=} (\xi^{0} - \xi^{3})/\sqrt{2}$,
$W(\xi^{-},0)$ is the Wilson line connecting  $\xi^{-}$ and $0$, and $\Gamma$ is a Dirac matrix whose structure defines the momentum distributions of quarks having spin parallel or perpendicular to the nucleon momentum.
The factorization scale is implicit.
Eq.~(\ref{def_pdf}) is light-cone dominated, as it receives  contributions only in the region where $\xi^2 {=} t^2 - \vec{r}^{\,2} {=} 0$. Consequently, a direct implementation of this equation is impossible on the lattice, where only a single point can satisfy the Euclidean light-cone condition $\xi^2 {=} t^2 + \vec{r}^{\,2}{=} 0$. The alternative approach proposed by Ji overcomes this difficulty by reconstructing light-cone distributions from correlation functions where the bilinear fermionic fields, $\psi$ and $\bar{\psi}$, are separated by a purely spatial distance $z$ and the nucleon is boosted with a finite momentum. Quasi-PDFs are defined as
\begin{equation}
\label{eq:quasi_pdf}
\tilde{q}(x,P_3)=\,\int_{-z_{\rm max}}^{+z_{\rm max}}\,\frac{dz}{4\pi}\,e^{-ixP_3z}\,\langle N\vert \, \overline{\psi}(0,z)\,\Gamma W(z,0)\,\psi(0,0)\,\vert N\rangle\, ,
\end{equation}
where $\vert N\rangle$ represents a nucleon, which is boosted in the $z$-direction with momentum $P {=} (P_0, 0, 0, P_3)$. On the lattice, quasi-PDFs are thus computed using matrix elements of non-local operators containing a straight Wilson line of finite length $z$, that varies from 0 to some maximum value, $z_{\rm max}$. This aspect will be further discussed in the following sections. The standard light-cone PDFs can be recovered from quasi-PDFs using the Large Momentum Effective Theory~\cite{Ji:2014gla}, developed specifically for these operators. Eventually, quasi-PDFs are matched to the physical PDFs through the factorization 
\begin{equation}
q(x,\mu)=\int_{-\infty}^\infty\,\frac{d\xi}{\vert \xi\vert}\,C\left( \xi,\frac{\mu}{P_3}\right) \tilde{q}\left( \frac{x}{\xi},\mu,P_3\right) + O\left( \frac{m_N^2}{P_3^2},\frac{\Lambda_{QCD}^2}{P_3^2}\right) \,,
\end{equation}
where $q(x,\mu)$ is the light-cone PDF at the scale $\mu$ and we follow the usual choice of common factorization and renormalization scales, $\mu_F=\mu_R\equiv\mu$. $C$ is the matching kernel, which can be computed perturbatively and  has so far  been evaluated to one-loop level. For sufficiently large momenta, ideally much larger than the nucleon mass, $m_N$, and $\Lambda_{QCD}$, higher order corrections in $\alpha_s$ are very small and the extraction of quark distributions is fully robust. Presently, such a situation cannot be accomplished and it is part of this paper to investigate whether practically reachable momenta are sufficient.

\subsection{Lattice gauge ensemble}
The analysis is performed using  a gauge ensemble of two dynamical degenerate light quarks  ($N_f{=}2$), generated by the ETM~\footnote{Effective from this year, the European Twisted Mass Collaboration has officially changed its name to Extended Twisted Mass Collaboration, as it comprises now members also from non-European institutions. Along with the name change, there is also a new logo.} Collaboration~\cite{Abdel-Rehim:2015pwa}. The light quark mass has been tuned to reproduce the physical value of the pion mass, henceforth referred to as the \textit{physical point}. The lattice volume is $48^3 \times 96$, the lattice spacing $a=0.0938(2)(3)$ fm~\cite{Alexandrou:2017xwd}, the spatial lattice extent $L\approx4.5$ fm and $\,\,m_\pi L = 2.98$. Based on a model prediction of excited states effects in certain hadron structure quantities~\cite{Hansen:2016qoz}, the dependence on $m_\pi L$ is mild (see, e.g., Fig. 3 of Ref.~\cite{Green:2018vxw}).
The complete list of parameters of this ensemble is reported in Table~\ref{Table:params}. 

\begin{table}[h]
\begin{center}
\renewcommand{\arraystretch}{1.5}
\renewcommand{\tabcolsep}{5.5pt}
\begin{tabular}{c|lc}
\hline\hline
\multicolumn{3}{c}{ 
$\beta=2.10$, $c_{\rm SW} = 1.57751$, $a=0.0938(3)(2)$~fm, ${r_0/a}=5.32(5)$}\\
\hline
\multirow{4}{*}{$48^3\times 96$, $L\approx4.5$~fm\,}  & $\,\,a\mu = 0.0009$   \\
                                              & $\,\,m_\pi = 0.1304(4)$~GeV     \\
                              			      & $\,\,m_\pi L = 2.98$    \\
                        					  & $\,\,m_N = 0.932(4)$~GeV     \\
\hline\hline
\end{tabular}
\begin{center}
\begin{minipage}{15cm}
\vspace*{-0.45cm}
\hspace*{3cm}
\caption{\small{Parameters of the ensemble used in this work. The nucleon mass $(m_N)$, the pion mass $(m_\pi)$ and the lattice spacing $(a)$ have been determined in Ref.~\cite{Alexandrou:2017xwd}.}}
\label{Table:params}
\end{minipage}
\end{center}
\end{center}
\end{table} 

\noindent
The gauge configurations were generated with the Iwasaki improved gauge action~\cite{Iwasaki:2011np}. In the fermionic sector, the twisted mass fermion action~\cite{Frezzotti:2000nk,Frezzotti:2003ni} with a clover term~\cite{Sheikholeslami:1985ij} was employed. The fermion action is given by:
\be\label{eq:S_tml}
S_F\left[\chi,\overline{\chi},U \right]= a^4\sum_x
\overline{\chi}(x)\left(D_W[U] + i \mu_l \gamma^5\tau^3 - \frac{1}{4}
c_{\rm SW}\sigma^{\mu\nu}\mathcal{F}^{\mu\nu}[U] \right) \chi(x)\;,
\ee
where $D_W$ is the Wilson-Dirac operator, $\mu_l$ is the bare twisted mass for the light quarks, $\tau^3={\rm diag}(1,-1)$ is the third Pauli matrix in flavor space and the last term includes the field strength tensor $\mathcal{F}^{\mu\nu}[U]$ weighted by $c_{\rm SW}$, known as the Sheikoleslami-Wohlert (clover) coefficient. In the gauge ensemble used in this work, the clover parameter is taken to be $c_{\rm SW}{=}1.57551$~\cite{Aoki:2005et}. In Eq.~(\ref{eq:S_tml}) $\chi(x){=}(u,d)^T$ denotes the light quark doublet in the ``twisted basis'' at maximal twist. The fermion fields in the ``physical basis'', denoted by $\psi(x)$, are recovered by the following chiral rotation 
\be
\psi(x)\,\equiv \, e^{i\frac{\alpha}{2}\gamma^5\tau^3}\chi(x),\qquad \qquad \overline{\psi}(x)\, \equiv \, \overline{\chi}(x)\, e^{i\frac{\alpha}{2}\gamma^5\tau^3}\,,
\ee
where $\alpha{=}\pi/2$ at maximal twist. In the next sections of this paper, the interpolating fields and the nucleon matrix elements have to be understood with quark fields in the physical basis, unless otherwise specified. 

The introduction of the twisted mass term in the lattice action has a series of advantages in hadron structure calculations, such as  excluding zero eigenvalues from the spectrum of the Wilson Dirac operator allowing a speed-up in the numerical simulations. Moreover, it also simplifies renormalization properties of operators~\cite{Frezzotti:2003ni, Jansen:2005cg, Farchioni:2004us, Farchioni:2005bh} and at maximal twist it provides an automatic $O(a)$ improvement. However, the isospin symmetry breaking of the twisted mass formulation could lead to instabilities when simulations are carried out at light quark masses close to the physical value. The isospin breaking, which manifests itself in the neutral pion being lighter than the charged one, is reduced  with the addition of the clover term in the fermion action~\cite{Abdel-Rehim:2015pwa}.

\subsection{Nucleon bare matrix elements}
\label{sec:matrix_elements}
\begin{figure}[h!]
\includegraphics[scale=0.95]{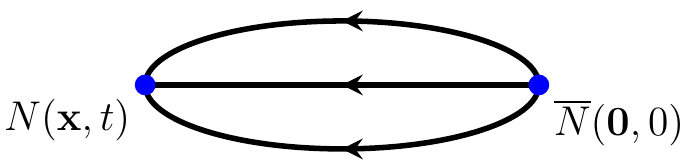}
\hspace*{1cm}
\includegraphics[scale=0.95]{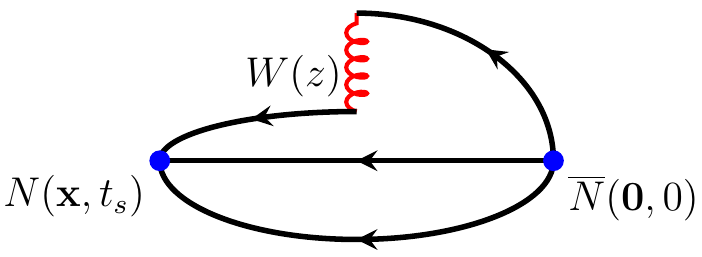}
\caption{\small{Schematic representation of the nucleon two- (left) and three-point (right) functions. The creation point of the nucleon is at $(\mathbf{0},0)$, while the annihilation at $(\mathbf{x},t_s)$ and $(\mathbf{x},t_s)$ for the two- and three-point functions, respectively. The solid lines represent the quark propagators and the curly line denotes the Wilson line of length $z$.}}
\label{fig:diagram}
\end{figure}

In this work, we evaluate the unpolarized, helicity and transversity quasi-PDFs using three values for the nucleon momentum, that is, $P_3{=}6\pi/L$, $P_3{=}8\pi/L$ and $P_3{=}10\pi/L$, corresponding in physical units to 0.83 GeV, 1.11 GeV and 1.38 GeV, respectively.  The matrix element of interest is given by
\begin{equation}
\label{eq:matrix_elements}
h_\Gamma(P_3,z)\,=\,\langle N\vert \, \overline{\psi}(0,z)\,\Gamma W(z,0)\,\psi(0,0)\,\vert N\rangle\, ,
\end{equation}
for a straight Wilson line, $W$, with varying length from $z{=}0$, corresponding to the standard ultra-local operators, up to half of the spatial extension, $L/2$. $\Gamma$ is the Dirac structure leading to different types of PDFs, as discussed below. The matrix elements of Eq.~(\ref{eq:matrix_elements}) are extracted from a ratio of two- and three-point functions, averaged over the gauge field ensemble.  The two-point and three-point functions are given by
\bea
C^{\rm 2pt}(\mathbf{P},t,0) &=& {\cal P}_{\alpha\beta}\,\sum_\mathbf{x}\,e^{-i\mathbf{P}\cdot \mathbf{x}}\langle 0\vert N_\alpha(\mathbf{x},t) \overline{N}_\beta(\mathbf{0},0)\vert 0\rangle \,, 
\label{eq:2pt}
\\[2ex]
C^{\rm 3pt}(\mathbf{P};t_s,\tau,0) &=&  \tilde{{\cal P}}_{\alpha\beta}\,\sum_{\mathbf{x},\mathbf{y}}\,e^{-i\mathbf{P}\cdot \mathbf{x}}\,\langle 0\vert N_{\alpha}(\mathbf{x},t_s) \mathcal{O}(\mathbf{y},\tau;z)\overline{N}_{\beta}(\mathbf{0},0)\vert 0\rangle\,,
\label{eq:3pt}
\eea
where $N_{\alpha}(x)$ is the proton interpolating field, $t$ ($t_s$) is the time separation of the sink relative to the source in the two-point (three-point) function\footnote{We use different symbols to emphasize that the three-point function is computed only for selected values of the source-sink separation, $t_s/a{=}8,9,10,12$ in this work, while the two-point function is evaluated for the full range of time separations to constrain more precisely the gap between the ground state and the first excited state in two-state fits, see Sec.~\ref{sub:excited_states}.}, $\tau$ is the insertion time of the $\mathcal{O}$ operator and ${\cal P}_{\alpha\beta}$, $\tilde{\cal P}_{\alpha\beta}$ are the parity projectors for the two- and three-point functions, respectively. For the two-point functions, we use the plus and minus parity projectors ${\cal P}^\pm=\frac{1\pm\gamma^0}{2}$ and average over the forward and backward correlators. The projector $\tilde{\cal P}_{\alpha\beta}$ depends on the operator under study. The operator $\mathcal{O}(\mathbf{y},\tau;z)$ has the following form
\be
\label{Op}
\mathcal{O}(\mathbf{y,\tau;z})\,=\,\overline{\psi}\,(\mathbf{y+z,\tau})\,\Gamma W(\mathbf{y}+z,\mathbf{y})\,\psi(\mathbf{y},\tau)\, ,
\ee
where the Wilson line is aligned to the direction of the nucleon momentum and $\psi$ is the fermion doublet of flavors $(u,d)$. We compute the isovector combination $u{-}d$, inserting a $\tau_3$ in flavor space. With this choice, the disconnected diagrams cancel and the total contribution is obtained from the connected diagram shown in Fig.~\ref{fig:diagram}\,.

As mentioned in Section~\ref{sec:pdfs_quasi}, the choice for the Dirac structure in the operator of Eq.~(\ref{def_pdf}) gives access to different quark distribution functions. In fact, to extract a given PDF, different $\Gamma$ matrices may be used, for example $\gamma^3$ (parallel to the Wilson line direction) or $\gamma^0$ (temporal direction) for the unpolarized PDF. However, some choices are preferable (e.g., $\gamma^0$) as they avoid finite mixing under renormalization, that was found to be allowed  in lattice regularization~\cite{Constantinou:2017sej}. In the absence of mixing, the operators renormalize multiplicatively, and one avoids systematic uncertainties related to the elimination of mixing. In this work, we compute matrix elements of the operators with
\begin{itemize}
\item ${\Gamma}=\gamma^3\,,\gamma^0$ for the unpolarized distribution, $\tilde{q}(x)=\tilde{q}(x)_{\uparrow} + \tilde{q}_{\downarrow}(x)$\,,
\item ${\Gamma}=\gamma^5\gamma^3$ for the helicity distribution, $\Delta \tilde{q}(x)=\tilde{q}_{\uparrow}(x)-\tilde{q}_{\downarrow}(x)$\,,
\item ${\Gamma}=\sigma^{3j}$ for the transversity distribution, $\delta\tilde{q}(x)=\tilde{q}_{\perp}(x)+\tilde{q}_{\backin}(x)$\,.
\end{itemize}
$q_{\uparrow}$ ($q_{\downarrow}$) and ${q}_{\perp}$ (${q}_{\backin}$) indicate quarks with helicity aligned (anti-aligned) with that of a longitudinally and transversely polarized proton, respectively. The index $j$ entering the matrix element of the transversity distribution is purely spatial and denotes the direction of the quark spin, orthogonal to the proton momentum. Each choice of an operator requires an appropriate parity projector $\tilde{\cal P}_{\alpha\beta}$, that is, $\frac{1+\gamma^0}{2}$ for the unpolarized, $i\gamma^3\gamma^5\frac{1+\gamma^0}{2}$ for the helicity, and $i\gamma^k\frac{1+\gamma^0}{2}$ for the transversity PDF.
The desired matrix elements of Eq.~(\ref{eq:matrix_elements}) are finally extracted from fitting  the ratio of three- over two-point functions
to isolate the ground state. One choice is a constant fit as a function of the time insertion of the operator over which the ratio is time independent  (plateau region). Other choices are two-state fits that take into account the first excited state contributions, as discussed in  Sec.\ \ref{sec:bare_ME} with a comparison among them.  Thus, the desired matrix element in the plateau fit is given by
\begin{equation}
\label{eq:ratio}
h_\Gamma(p,z) \,\myeq\,\mathcal{K}\,\frac{\langle C^{\rm 3pt}({p;t_s,\tau,0})\rangle_G}{\langle C^{\rm 2pt}(p,t_s,0)\rangle_G}\,,
\end{equation}
where $\mathcal{K}$ is a kinematic factor equal to $\mathcal{K}\,{=}\,iE/p$ (with $E{=}\sqrt{m_N^2+p^2}$) for the vector current $\gamma^3$, whereas $\mathcal{K}{=}1$ for all other choices of the Dirac structure  used. A gauge average ($\langle\cdots\rangle_G$) is performed on the two- and three-point functions prior to taking the ratio.

\subsection{Lattice techniques}
\label{sec:lattice_techniques}

To evaluate the correlation functions of Eqs.~(\ref{eq:2pt}),(\ref{eq:3pt}), a state with the same quantum numbers as the nucleon is created  at the source and annihilated at the sink. In our calculation, we employ point sources generated with the proton interpolating field $N_{\alpha}(x){=}\epsilon ^{abc}u^a _\alpha(x)\left( d^{b^{T}}(x)\mathcal{C}\gamma^5u^c(x)\right)$, with $\mathcal{C}{=}\gamma^0\gamma^2$ the charge conjugation matrix. In general, the overlap of the states generated with such interpolating fields with the desired ground state is improved by employing smearing. 
Gaussian smearing~\cite{Gusken:1989qx,Alexandrou:1992ti} on quark fields, used in combination with APE smearing on the gauge field, is a very effective approach
to improve ground state dominance. For nucleon states at rest, previous studies \cite{Abdel-Rehim:2015owa} performed on the gauge ensemble employed in this work, have extracted the optimal parameters for both Gaussian and APE smearing, tuned to approximately reproduce a nucleon state with root mean square (rms) radius of about 0.5 fm and
maximize the overlap with the ground state. These parameters are $(N_G,\alpha_G)=(50,4)$ for Gaussian smearing and $(N_{APE},\alpha_{APE}=0.5)$ for APE smearing. 

\begin{figure}[!h]
\begin{center}
  \begin{minipage}[t]{0.476\linewidth}
    \includegraphics[width=\textwidth]{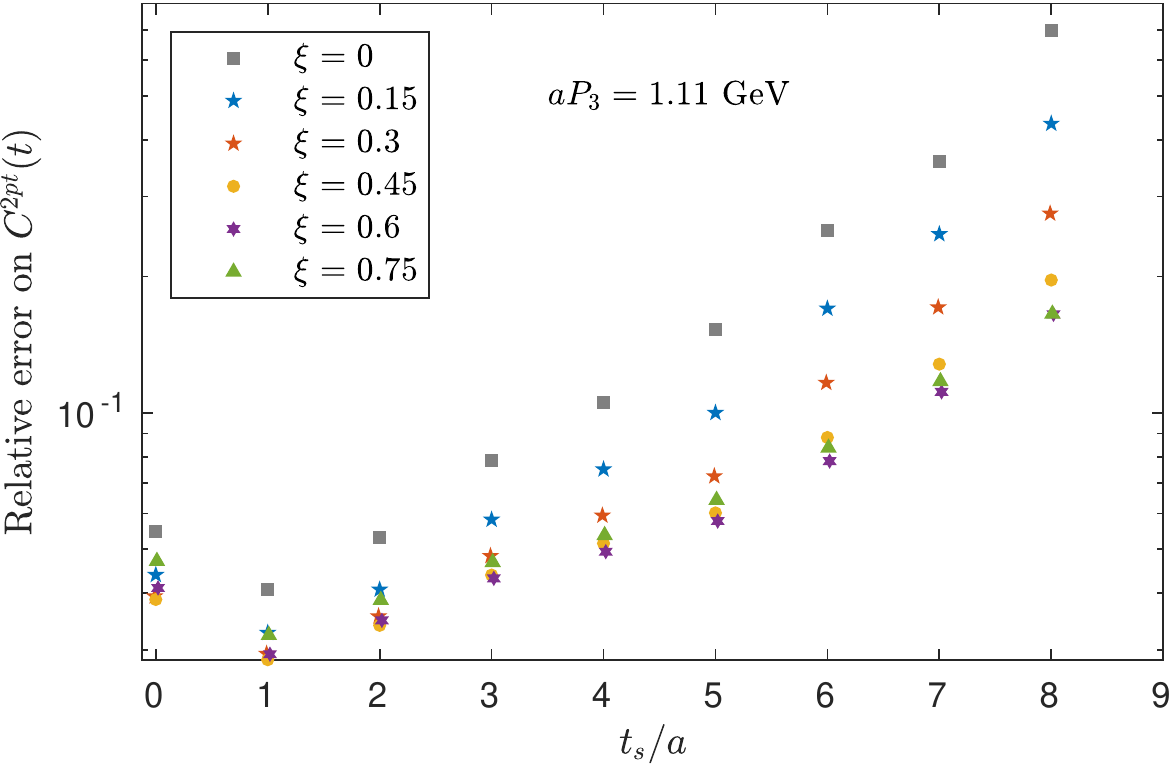}
  \end{minipage}
  \hfill
  \begin{minipage}[t]{0.475\linewidth}
    \includegraphics[width=\textwidth]{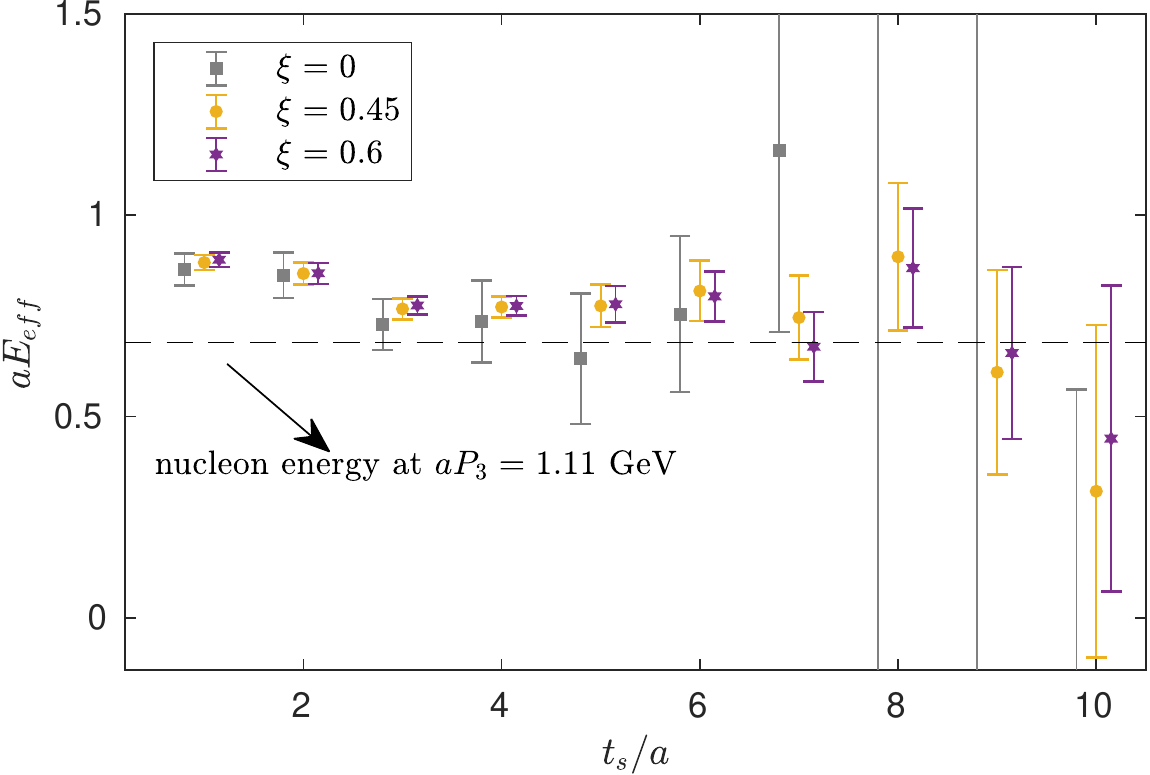}
  \end{minipage}
\hfill
\end{center}  
\vspace*{-.5cm}
\caption{Left: Relative error of the nucleon two-point correlator as a function of the Euclidean time $t/a$, for $P_3{=}8\pi/L$ and different values of the momentum smearing parameter $\xi$. The value $\xi{=}0$ (black points) corresponds to the standard Gaussian smearing. Right: Effective energy of the nucleon with boost $P_3{=}8\pi/L$ for different values of $\xi$.}
\label{fig:mom_smearing}
\end{figure}

However, in the computation of PDFs, we need to optimize the overlap to a boosted nucleon state keeping the statistical noise minimal.
We, thus, modify the standard Gaussian smearing by  using  momentum-smeared interpolating fields~\cite{Bali:2016lva}. This has proven  extremely advantageous when the nucleon is boosted with a large momentum, as is the case for quasi-PDFs. 
The momentum smearing  modifies the standard Gaussian smearing function by including a complex phase factor that affects only the gauge links along the direction of the boost, namely
\begin{equation}
\label{eq:mom_smearing}
\mathcal{S}_{mom}\,=\,\frac{1}{1+6\alpha_G}\left( \psi(x)+\alpha_G\sum_j\,U_j(x)\,e^{i\xi\mathbf{P}\cdot \mathbf{j}}\,\psi(x+\hat{j})\right) \, ,
\end{equation}
where $U_j$ denotes a gauge link in the $j$-direction. The value of the free momentum smearing parameter $\xi$ depends, in general, on the parameters of the gauge ensemble and on the nucleon momentum. Tuning of $\xi$ is necessary in order to improve the overlap of our interpolator with the boosted proton. We, thus, have optimized the parameter $\xi$ for each value of the momentum employed by minimizing the statistical errors of  the nucleon two-point functions. 
In Fig.~\ref{fig:mom_smearing}, we demonstrate the effect of the momentum smearing, by plotting the scaling of the error of the two-point correlator and the effective energy. The results shown have been extracted using $100$ measurements for the nucleon boost $P_3{=}8\pi/L$. 
For comparison, we include also the results using the standard Gausian smearing ($\xi{=}0$).
As can be seen, the errors in the correlation functions reduce dramatically as the value of $\xi$ increases and convergence is observed in the range $\xi\in[0.6{-}0.75]$. Thus, any value of $\xi$ in this window of values leads to a similar signal-to-noise ratio. The tuning procedure for the other two momenta, $P_3{=}6\pi/L$ and $P_3{=}10\pi/L$, leads to the same conclusions and we fix $\xi{=}0.6$ through out this work.

The three-point functions are computed using the sequential method~\cite{Martinelli:1988rr}, which  has the advantage of summing automatically the sink spatial volume to produce the sequential propagator for all insertion points.  Compared to the stochastic method~\cite{Alexandrou:2013xon}, the sequential method  does not introduce stochastic noise, but has the drawback that new inversions of the Dirac matrix have to be carried out for different source-sink separations and nucleon momenta. For details on the comparison between the stochastic method (with pure Gaussian smearing) and the sequential method (with and without momentum smearing), we refer to Refs.~\cite{Alexandrou:2014pna,Alexandrou:2016jqi}. 
In addition, the sequential method is preferable when the momentum smearing technique on the quark fields is employed. 

\begin{figure}[!ht]
\begin{center}
  \begin{minipage}[t]{0.475\linewidth}
    \includegraphics[width=\textwidth]{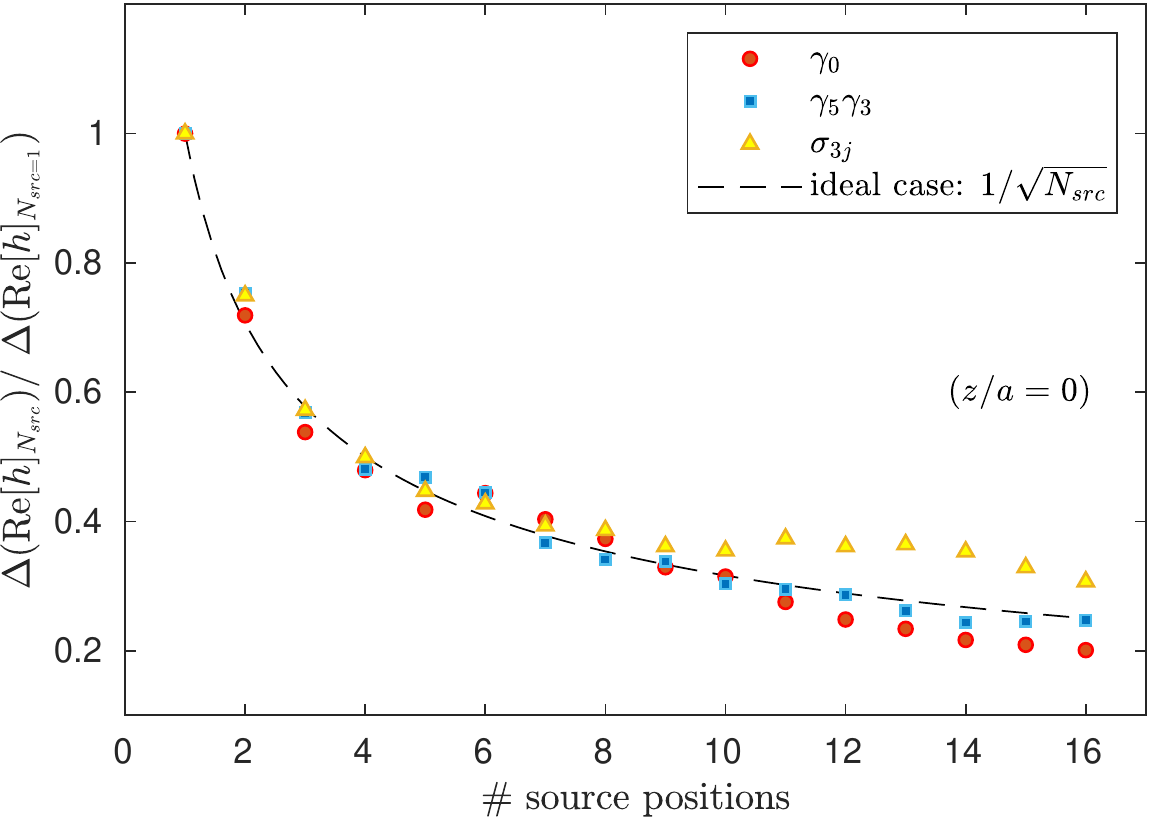}
  \end{minipage}
  \hfill
  \begin{minipage}[t]{0.475\linewidth}
    \includegraphics[width=\textwidth]{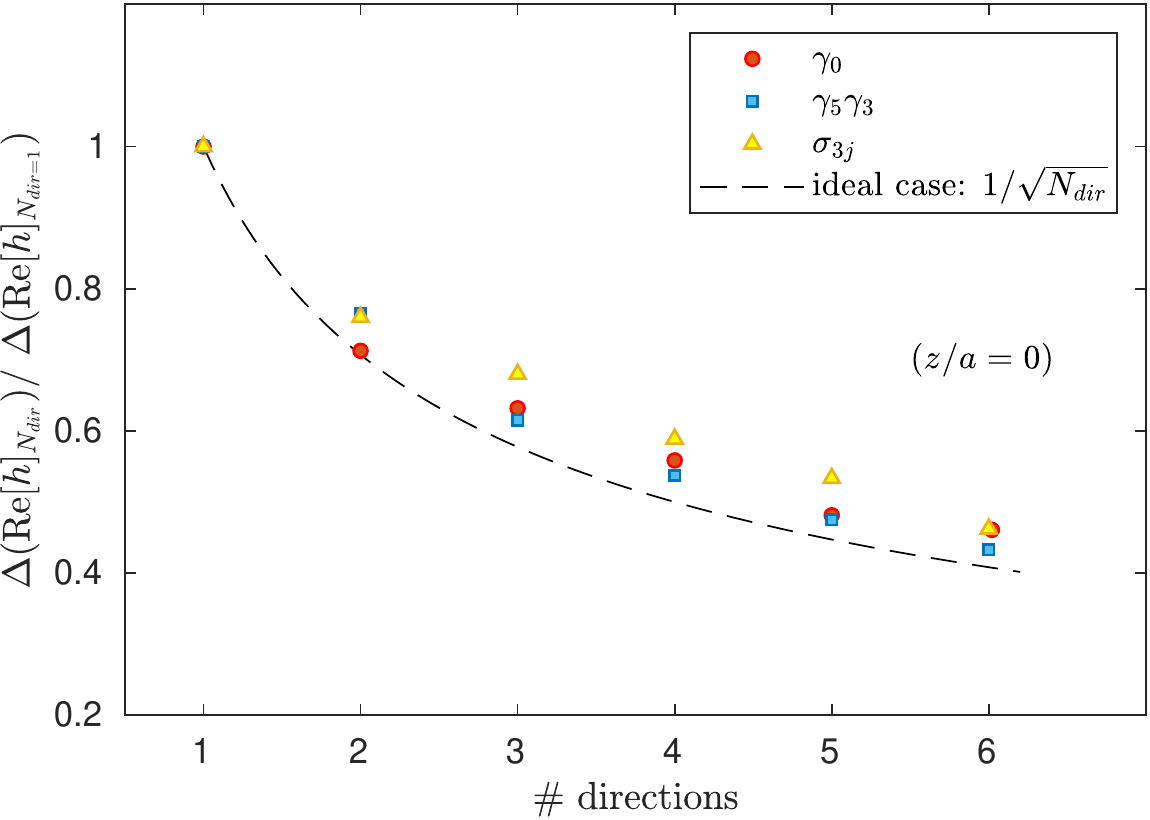}
  \end{minipage}
\hfill
\end{center}  
\vspace*{-.5cm}
\caption{Scaling of the statistical errors of the matrix elements, for the operator with $z{=}0$, varying the number of source positions on each configuration (left) and averaging over the directions of the nucleon boost (right). The absolute errors for the unpolarized, helicity and transversity matrix elements (red circles, blue squares, yellow triangles, respectively) are normalized to the one obtained from only one point source or one boost direction. The results are compared with the ideal scaling (dashed line in both figures) expected for uncorrelated measurements.}
\label{fig:scaling_src}
\end{figure}

To increase the number of measurements of the three-point functions, we use multiple source positions ($N_{src}$) for each configuration. To confirm that the data extracted from different source positions on the same configuration are statistically independent, we  study the scaling of the statistical error with the increase of $N_{src}$. In  Fig.~\ref{fig:scaling_src} (left), we show  the error on the matrix elements for $z{=}0$, obtained from the analysis of a sample of gauge configurations for nucleon boost $6\pi/L$. For comparison purposes, the absolute error is normalized to the one obtained using only one source position per configuration. As can be seen, the errors for the three choices of $\Gamma$ structure  decrease approximately as $1/\sqrt{N_{src}}$, which is the scaling expected if the measurements are not correlated.

To further decrease the statistical uncertainties, we boost the nucleon along different spatial directions and orientations, that is, $\pm x$, $\pm y$ and $\pm z$, with the Wilson line taken always along the axis in which the spatial component of the momentum is nonzero. The correlation functions obtained from these $N_{dir}{=}6$ possible directions lead to the same physical results due to the spatial rotational symmetry on the lattice, and therefore can be averaged within each configuration. The statistical error reduction fluctuates is close to the ideal behavior $1/\sqrt{N_{dir}}$, as shown in the right panel of Fig.~\ref{fig:scaling_src}.

Despite the use of the momentum and APE smearing, the exponential decrease of the signal-to-noise ratio persists as the nucleon momentum and source-sink separation increase. To increase statistics at a reduced cost, we employ, together with the momentum and APE smearing,  the Covariant Approximation Averaging (CAA)~\cite{Blum:2012uh} that belongs to the class of truncated solver methods with a bias correction. For each configuration, $N_{LP}$ low-precision (LP) inversions of the Dirac matrix are carried out from a set of random source positions $\left\lbrace X_{src}\right\rbrace $ and the bias from the measurements is removed using a small number $N_{HP}$ of high-precision (HP) inversions. Denoting with $C_{LP}$ and $C_{HP}$ the correlation functions produced with low and high precision of the solver, respectively, the improved correlation functions $C^{imp}$ for each configuration are defined by
\be
C^{imp}=\frac{1}{N_{LP}}\,\sum_{i=1}^{N_{LP}}\,C_{i,LP}\,+\,\frac{1}{N_{HP}}\,\sum_{i=1}^{N_{HP}}\,(C_{i,HP}\,-\,C_{i,LP})\,\ ,
\ee
where, in the second sum, $C_{LP}$ and $C_{HP}$ are computed on the same source position, otherwise the bias cannot be corrected. 
The error for a given observable scales  with the ratio $N_{HP}/N_{LP}$ as
\be 
{\rm error}\propto \sqrt{\,2(1-r_c)\,+\,\frac{N_{HP}}{N_{LP}}}\, ,
\ee
where $r_c$ is the correlation coefficient among nucleon correlators computed at high and low precision. A compromise is needed to have $r_c\simeq 1$, while keeping the inversions as fast as possible and $N_{LP}\gg N_{HP}$. Thus, a tuning of the precision of the solver has to be  carried out. To invert the Dirac operator, we use the adaptive multigrid solver with twisted mass fermion support~\cite{Alexandrou:2016izb} and require the residual to be $r_{HP}{=} 10^{-10}$  for HP inversions. After testing different values of the residual for LP inversions, we find that the stopping criterion
\be
 r_{ LP}\equiv \frac{|{\rm residue}|_{LP} }{|{\rm source}|} = 2\cdot 10^{-3}
 \ee
 guarantees a correlation coefficient $r_c \geq 0.999$ with a considerable speed-up in the inversion time. Moreover, taking 15 HP inversions as the reference setup, a comparison of the HP and CAA estimates for the two- and three-point correlators verified that the bias introduced from LP inversions is negligible compared to the gauge noise of our measurements when one HP inversion for each configuration is performed. Thus, to extract the nucleon matrix elements for quasi-PDFs at momenta $8\pi/L$ and $10\pi/L$, we use the CAA setup with $(N_{HP},N_{LP}) {=} (1,16)$. 

\begin{table}[ht!]
\begin{center}
\renewcommand{\arraystretch}{1.4}
\renewcommand{\tabcolsep}{6pt}
\begin{tabular}{cccc|ccccc|ccccc}
 & $P_3=\frac{6\pi}{L}$ &  &  & & & \hsn\hsn$P_3=\frac{8\pi}{L}$\hsn\hsn &  & & & & \hsn\hsn$P_3=\frac{10\pi}{L}$\hsn\hsn & \\
\hline
Ins. & \hsn$N_{\rm conf}$ & \hsn$N_{\rm HP}$  & \hsn$N_{\rm meas}$ & Ins. & \hsn$N_{\rm conf}$ & \hsn$N_{\rm HP}$ & \hsn$N_{\rm LP}$ & \hsn$N_{\rm meas}$ & Ins. & \hsn$N_{\rm conf}$  &  \hsn$N_{\rm HP}$ & \hsn$N_{\rm LP}$ & \hsn$N_{\rm meas}$ \\
\hline
$\gamma^3$\hsn                 & \hsn100\hsn   & \hsn16\hsn & \hsn9600\hsn    & $\gamma^3$\hsn                & \hsn425\hsn  & \hsn1\hsn & \hsn16\hsn & \hsn38250\hsn & $\gamma^3$\hsn                & \hsn811\hsn & \hsn1\hsn & \hsn16\hsn & \hsn72990\hsn \\
$\gamma^0$\hsn                & \hsn50\hsn   &\hsn16\hsn & \hsn4800\hsn    & $\gamma^0$\hsn               & \hsn425\hsn & \hsn1\hsn & \hsn16\hsn  & \hsn38250\hsn & $\gamma^0$\hsn               & \hsn811\hsn & \hsn1\hsn & \hsn16\hsn & \hsn72990\hsn \\
$\gamma^5\gamma^3$\hsn & \hsn65\hsn   & \hsn16\hsn &\hsn6240\hsn   & $\gamma^5\gamma^3$\hsn & \hsn425\hsn & \hsn1\hsn & \hsn16\hsn  & \hsn38250\hsn & $\gamma^5\gamma^3$\hsn & \hsn811\hsn  & \hsn1\hsn & \hsn16\hsn & \hsn72990\hsn \\
$\sigma^{3j}$\hsn & \hsn50\hsn   &\hsn16\hsn & \hsn9600\hsn   & $\sigma^{3j}$\hsn & \hsn425\hsn & \hsn1\hsn & \hsn16\hsn  & \hsn38250\hsn & $\sigma^{3j}$\hsn & \hsn811\hsn & \hsn1\hsn & \hsn16\hsn & \hsn72990\hsn\\
\hline
\end{tabular}
\caption{\small{The statistics of our calculation at source-sink separation $t_s{=}12a$, for each current insertion and each momentum.
$N_{\rm conf}$ is the number of gauge configurations, $N_{\rm HP}$ ($N_{\rm LP}$) the number of high (low) precision measurements, and $N_{\rm meas}$ the total number of measurements. A factor of 6 is included in the total measurements due to the averaging of data with the Wilson line and momentum boost in the $\pm x$, $\pm y$, $\pm z$-directions. An additional factor of 2 compared to all other cases is included for the transversity at $P_3=\frac{6\pi}{L}$ to take into account the averages over the two possible Dirac structures, as explained in text.}}
\label{tab:statistics}
\end{center}
\end{table}

The quasi-PDFs are computed at different source-sink time separations $t_s$ in order to investigate excited states effects. In particular, we use $t_s/a{=}8,9,10,12$ for the unpolarized and $t_s/a{=}8,10,12$ for the helicity and transversity cases. For a detailed discussion on the excited states effects, we refer to Sec.~\ref{sub:excited_states}. In the remaining part of this section and unless otherwise stated, we focus on the results extracted from $t_s{=}12a$, which is the one where excited states are found to be suppressed for all  three PDFs when  the statistical precision is around $10\%$. The number of configurations and the total statistics for each operator and momentum is reported in Table~\ref{tab:statistics}. In the number of total measurements, we include a factor of $6$, coming from the average over correlators computed from the boost aligned along $\pm x$, $\pm y$, $\pm z$-directions, and a factor 16 (15) from the source positions for $P_3=6\pi/L$ ($P_3=[8,10]\pi/L$). This translates into an additional factor of $96$ for $P_3{=}6\pi/L$ and $90$ for $P_3{=}8,10\pi/L$, where the CAA is employed. Moreover, we note that only for the transversity PDF at $P_3{=}6\pi/L$, we have averaged over the matrix elements computed for the two possible choices of Dirac matrices (for example, $\sigma^{31}$ and $\sigma^{32}$ for a boost in the $z$-direction). This contributes with an additional factor of 2 in the total number of measurements at this momentum, as well as in the computational cost (due to different parity projectors needed for $\sigma^{31}$ and $\sigma^{32}$).

\begin{figure}[h!]
\begin{center}
\includegraphics[width=0.52\textwidth]{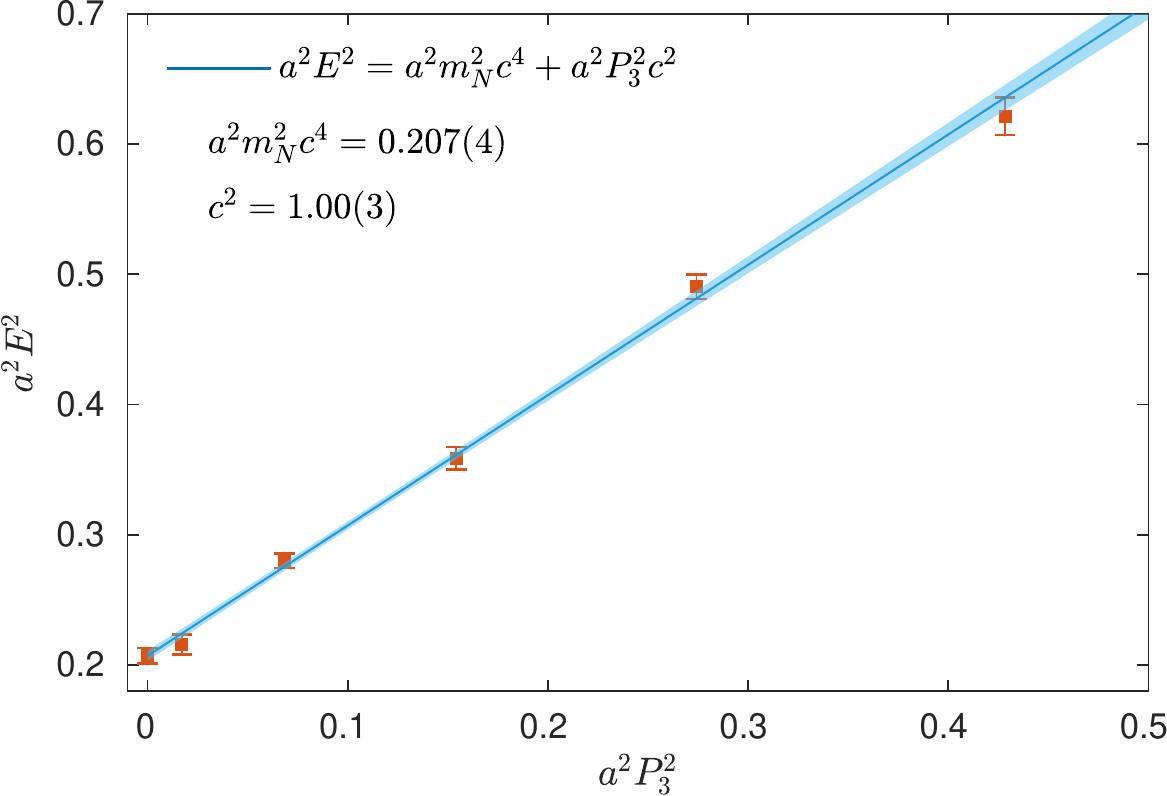}
\end{center}
\vspace*{-0.5cm}
\caption{Nucleon energy as a function of the momentum in lattice units. From the fit of the continuum dispersion relation, $E^2{=}m_N^2c^4+P_3^2c^2$, to the lattice data (red squares), we find a value of the squared speed of light equal to $c^2{=}1.00(3)$, and the nucleon mass in lattice units $a^2m_N^2c^4{=}0.207(4)$, which is also in agreement with the estimate given in Ref.~\cite{Alexandrou:2017xwd}.}
\label{fig:dispersion_relation}
\end{figure}

Although large nucleon momenta are needed to approach light-cone PDFs, high values of the  momentum on a Euclidean lattice may lead to substantial cut-off effects if the condition $P_3\ll 1/a$ is not satisfied. One possible check of cutoff effects is via the computation of the dispersion relation. To this end, we compute the nucleon energies for momenta \{0,2,4,6,8,10\}$\pi/L$ using the momentum smearing method, and check whether the relativistic dispersion relation $E^2{=}m_N^2c^4+P_3^2c^2$ is satisfied for our results. As can be seen in Fig.~\ref{fig:dispersion_relation}, no deviations from the continuum energy-momentum relation are observed for the values employed in this work (up to 1.38 GeV), which is well below the inverse lattice spacing ($10\pi/L\approx 0.65/a$).
Moreover, by performing a two-parameter fit to the lattice data, we obtain $c^2{=}1.00(3)$ for the squared speed of light and $a^2m_N^2c^4{=}0.207(4)$ for the nucleon mass in lattice units, which is compatible with the value $am_N{=}0.4436(11)$ extracted at zero-momentum~\cite{Alexandrou:2017xwd}. Whether this finding of only small cutoff effects in the dispersion relation also holds in case of the PDFs we are interested in, can eventually only be answered when results at several lattice spacing are available.

\section{Techniques for the  evaluation of bare quasi-PDF nucleon matrix elements in lattice QCD}
\label{sec:bare_ME}

In this section, we discuss crucial aspects related to the lattice QCD computation of bare quasi-PDF nucleon matrix elements, such as dependence on the number of stout smearing iterations used in the operator, the choice of the Dirac structure and identification of excited states contamination.

\subsection{Stout smearing}
\begin{figure}[h!]
\begin{center}
\includegraphics[width=0.49\textwidth]{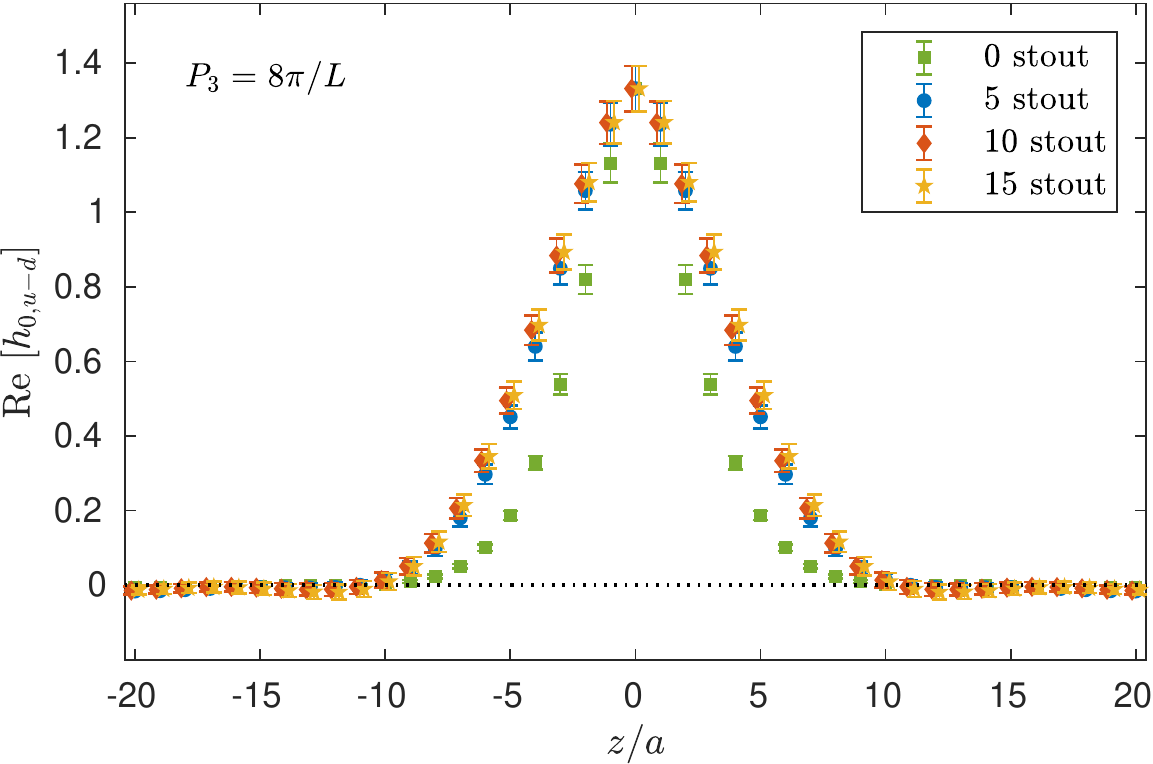}\hspace*{0.2cm}
\includegraphics[width=0.49\textwidth]{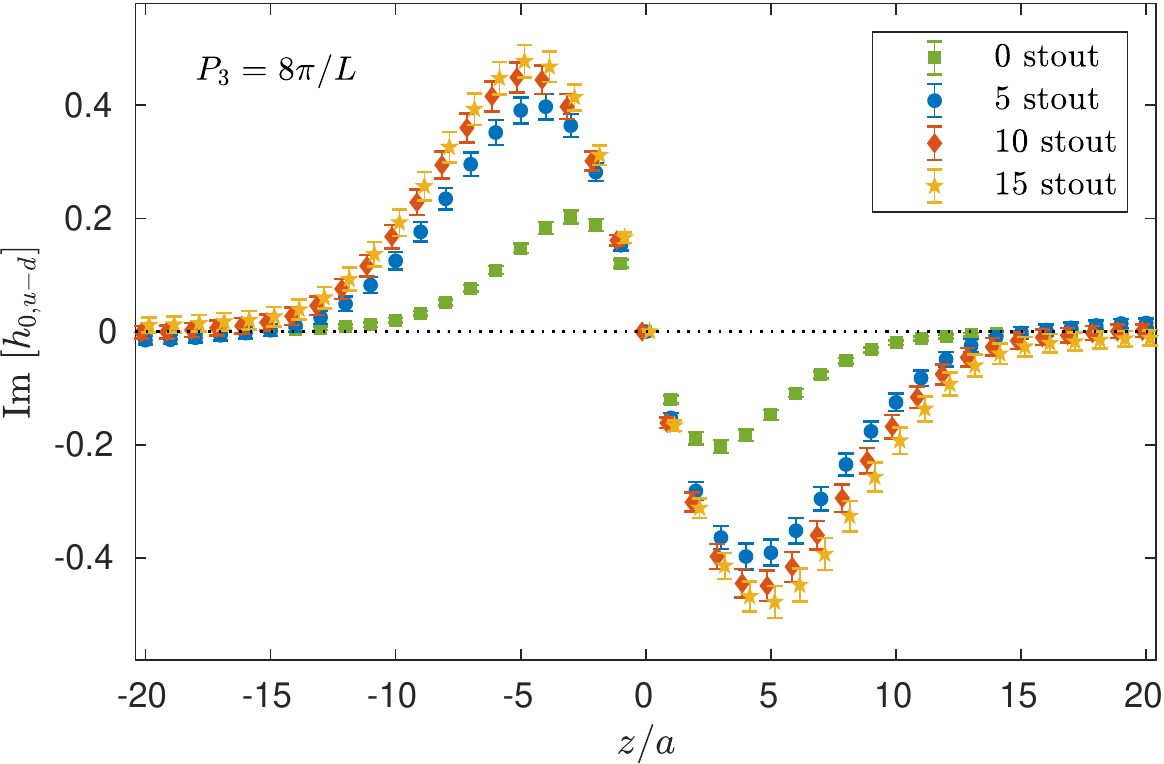}

\vspace*{0.25cm}
\includegraphics[width=0.49\textwidth]{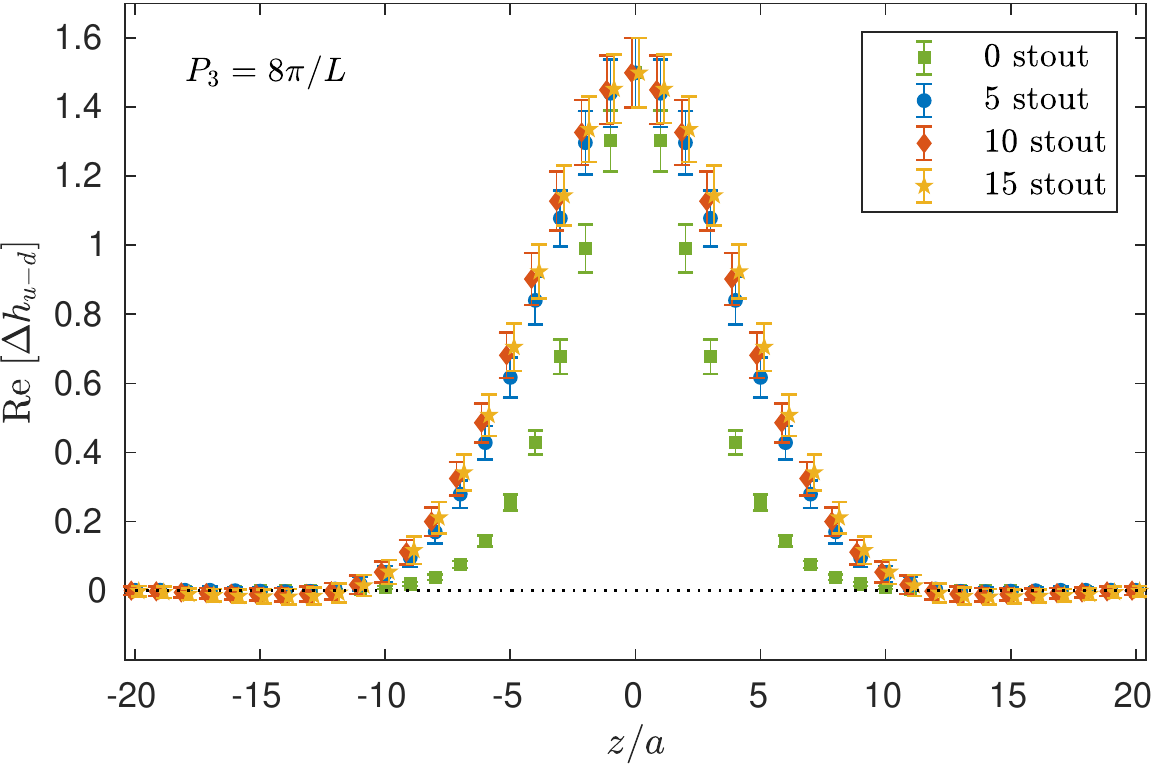}\hspace*{0.2cm}
\includegraphics[width=0.49\textwidth]{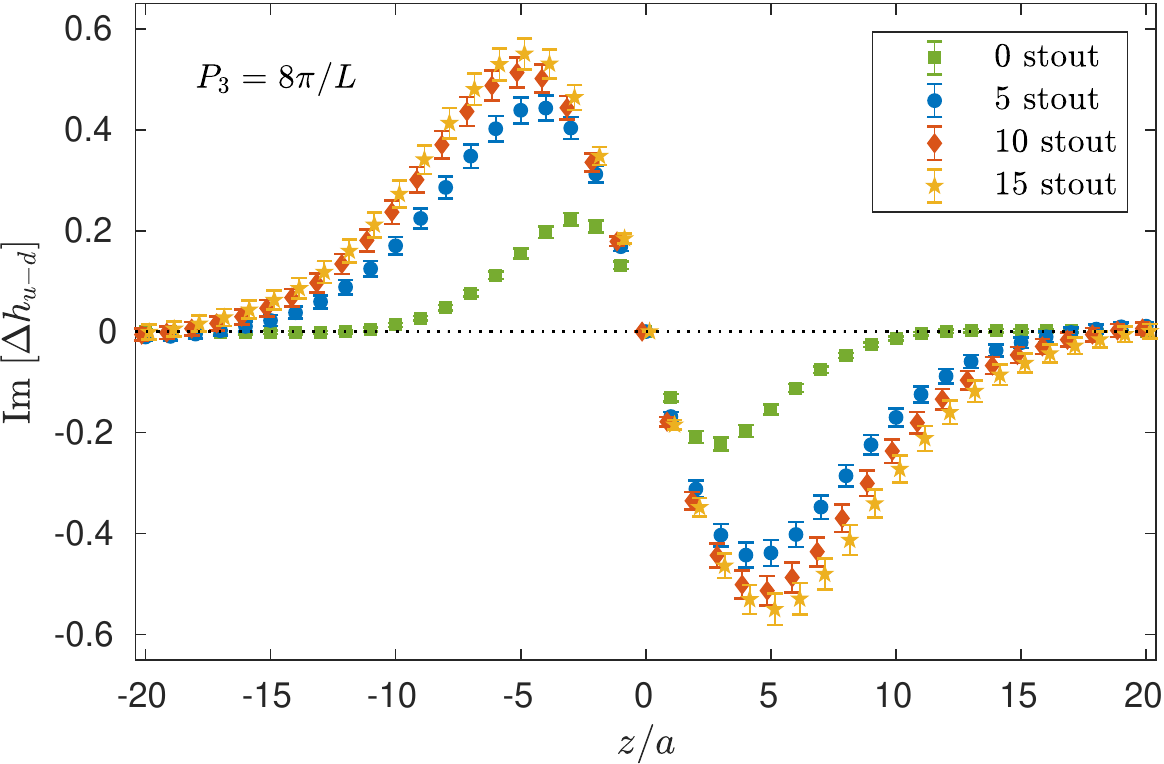}

\vspace*{0.25cm}
\includegraphics[width=0.49\textwidth]{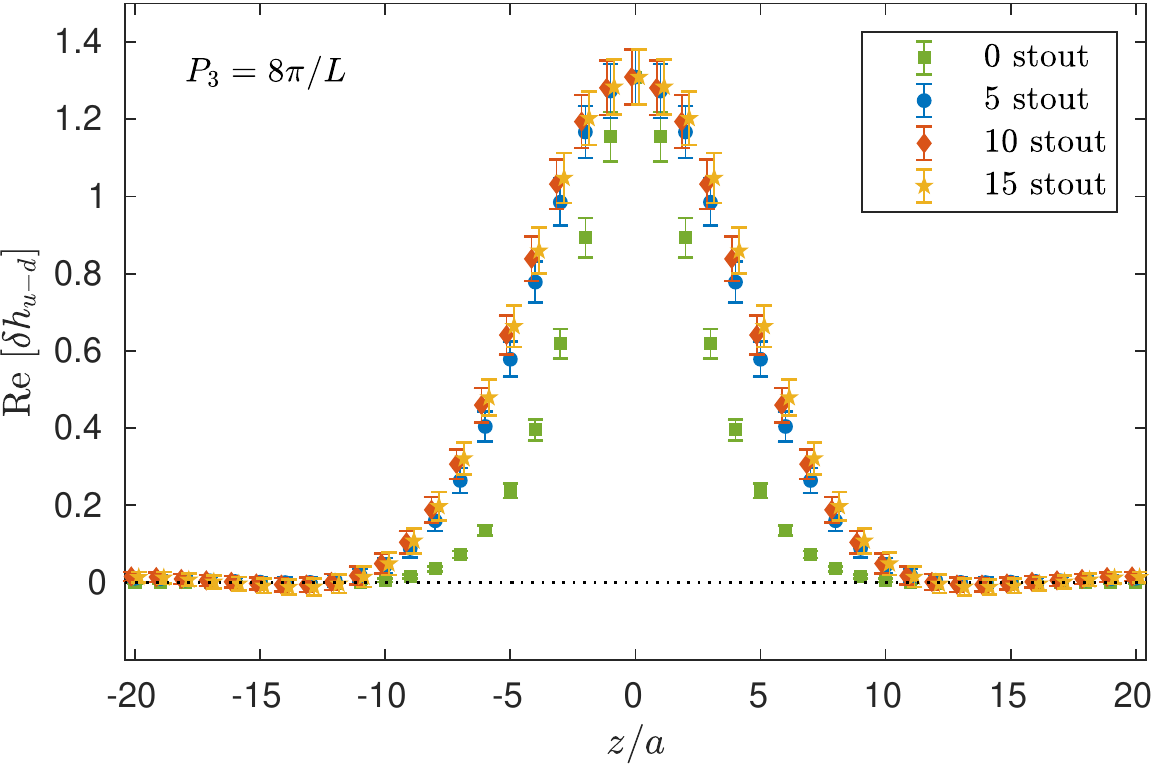}\hspace*{0.2cm}
\includegraphics[width=0.49\textwidth]{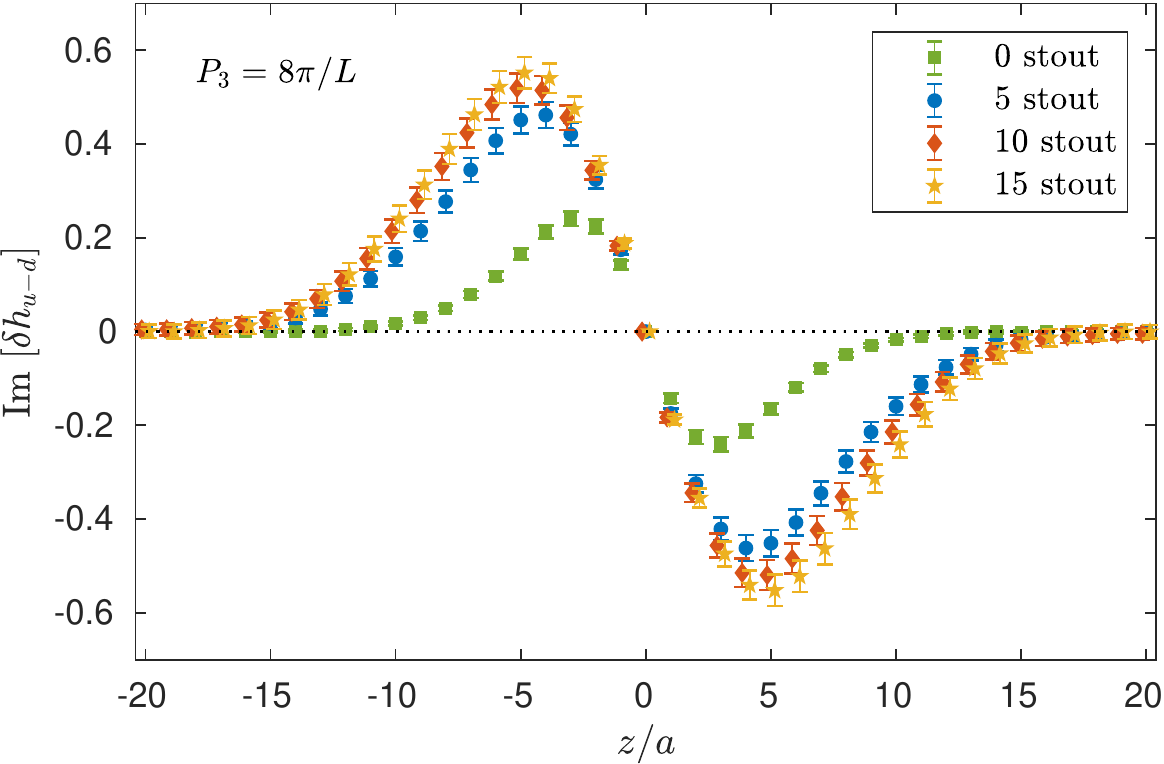}
\vspace*{-0.75cm}
\end{center}
\caption{Real and imaginary part of the bare matrix elements for 0, 5, 10, 15 steps of stout smearing at nucleon momentum $P_3{=}8\pi/L$. From top to bottom:  the bare nucleon matrix element for the  unpolarized, helicity and transversity operators.}
\label{fig:stout}
\end{figure}
We apply three-dimensional stout smearing to the gauge links of the Wilson line  of the operator, following the prescription of 
Ref.~\cite{Morningstar:2003gk}. This reduces the power divergence  that is present in the matrix elements 
of fermion operators with Wilson lines (see, e.g. Refs.~\cite{Dotsenko:1979wR,Brandt:1981kf}, as well as Ref.~\cite{Cichy:2018mum} for a review of recent investigations of the power divergence). 
The application of smearing was especially crucial in the first calculations of quasi-PDFs~\cite{Lin:2014zya,Alexandrou:2015rja,Chen:2016utp,Alexandrou:2016jqi} when  the renormalization procedure was not yet developed. 
Even though the complete renormalization procedure was developed recently~\cite{Alexandrou:2017huk}, a few iterations of stout smearing is still useful for noise reduction in the renormalized matrix elements. In Fig.~\ref{fig:stout}, we show examples of the effect of stout smearing for the case of the bare matrix elements of the unpolarized (insertion $\gamma^0$), helicity and transversity quasi-PDFs for nucleon momentum $P_3{=}8\pi/L$ without stout smearing, using 0, 5, 10 and 15 stout steps. As expected, the stout smearing modifies the matrix elements, increasing the values of  the real and imaginary parts  at each $z$. We find convergence of the matrix elements after a few iterations of stout smearing and, thus, in what follows we discuss in detail results obtained with $5$ steps of stout. The renormalized matrix elements are expected to no longer show any dependence on the smearing. We discuss the details of the renormalization in Sec.~\ref{sec:renorm}.

\subsection{Choice of the Dirac structure}
\begin{figure}[!ht]
\begin{center}
    \includegraphics[width=0.49\textwidth]{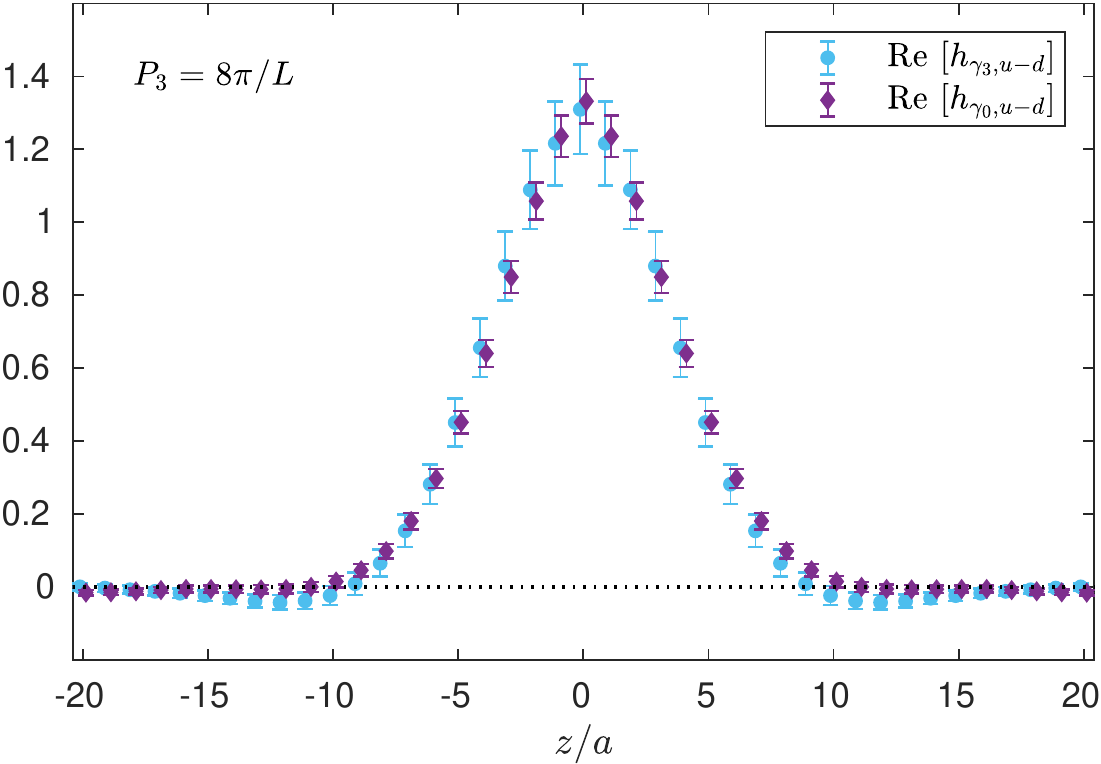}\,\,
    \includegraphics[width=0.495\textwidth]{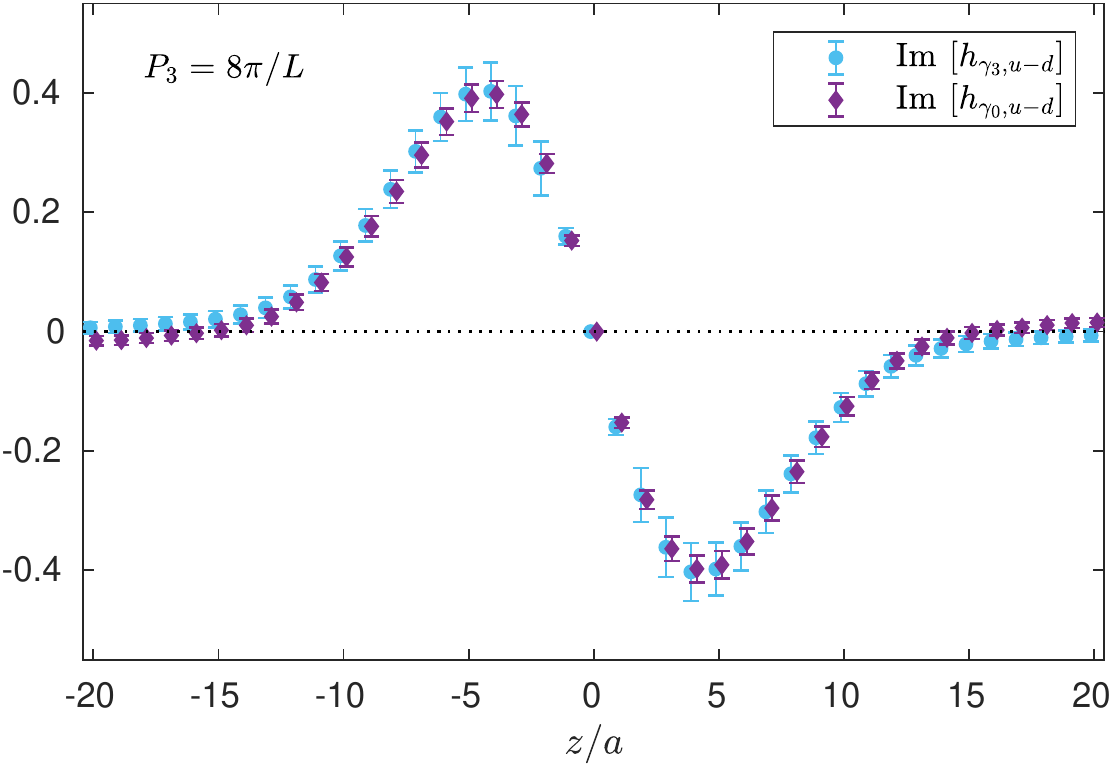}
\end{center}  
\vspace*{-0.4cm}
\caption{Comparison of the spatial $\gamma^3$ (cyan circles) and temporal $\gamma^0$  (purple diamonds) unpolarized bare matrix elements
for 5 stout steps and momentum $8\pi/L$.}
\label{fig:vec_g0}
\end{figure}

Although the natural choice to extract the unpolarized PDF would be $\gamma^{+}{=}(\gamma^0{+}\gamma^3)/\sqrt{2}$, a recent study~\cite{Constantinou:2017sej} has revealed that the matrix element with $\gamma^3$ exhibits mixing with the twist-3 scalar operator (later confirmed by symmetry properties~\cite{Chen:2017mzz}) and the computation of a mixing renormalization matrix is required~\cite{Alexandrou:2017huk}. However, the twisted mass formulation has the advantage that the mixing is between the vector and pseudoscalar operator. The latter vanishes in the continuum limit and only contributes as a discretization effect increasing the gauge noise. Consequently, it may be neglected in the first approximation, since eventually one is interested in the results for $a{\rightarrow}0$. 
In this work, we compute both matrix elements, but use the data for $\gamma^0$ to extract our final results for the unpolarized quasi-PDF. Even though the mixing is a purely lattice artifact for twisted mass fermions, we expect increased noise contamination for the case of $\gamma^3$, which is clearly visible in the lattice data, as shown in Fig.~\ref{fig:vec_g0}. 
This behavior is momentum-independent and we compare the matrix elements for the two $\gamma$-structure at momentum $8\pi/L$. The statistical errors for $\gamma^3$ are twice larger as compared to  the ones for $\gamma^0$ with the same number of measurements. We, thus, conclude that the insertions $\gamma^3$ or $\gamma^{+}$ are not optimal for extracting the unpolarized PDFs within a lattice QCD calculation. 

\begin{figure}[!ht]
\begin{center}
    \includegraphics[width=0.495\textwidth]{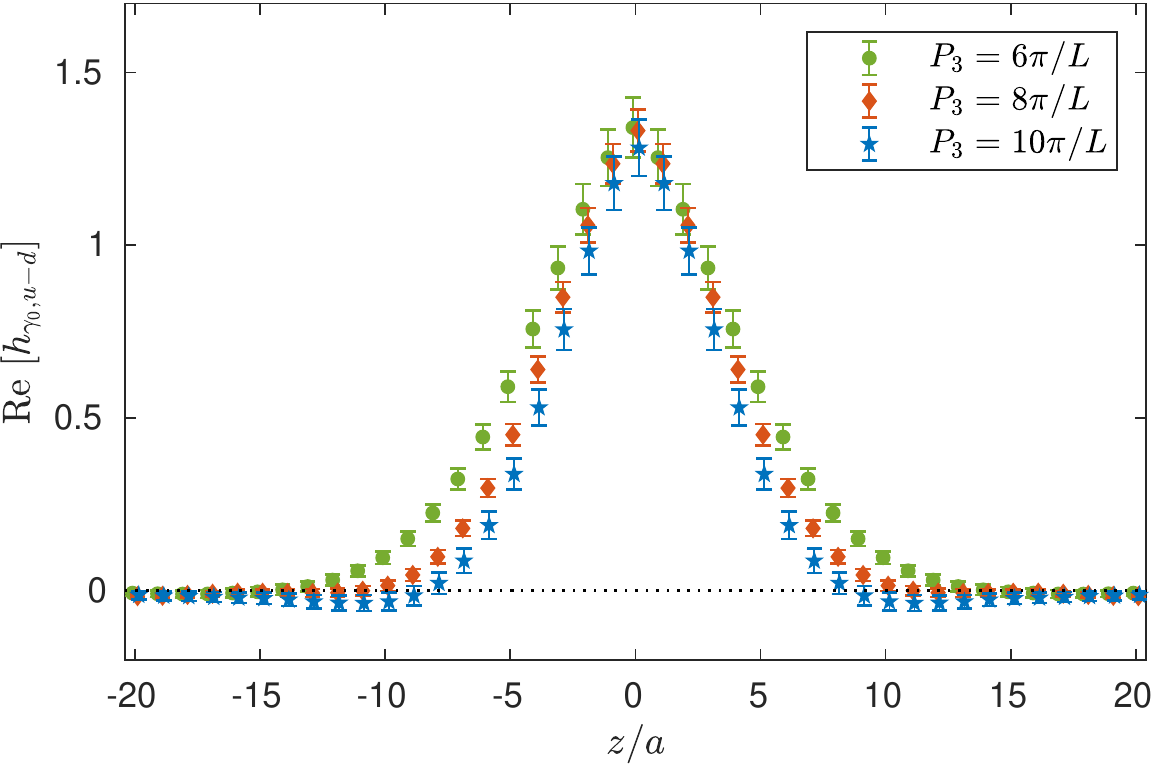}\hspace*{0.2cm}
    \includegraphics[width=0.495\textwidth]{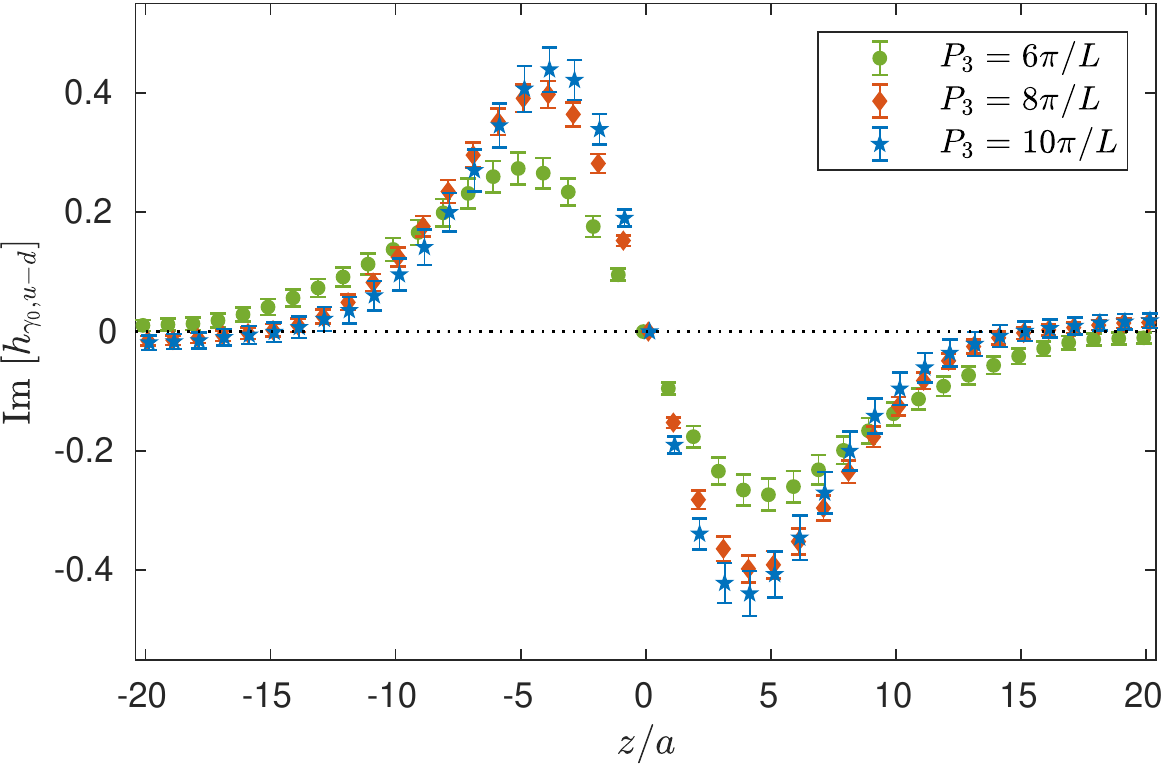}
    
    \vspace*{0.2cm}
    \includegraphics[width=0.495\textwidth]{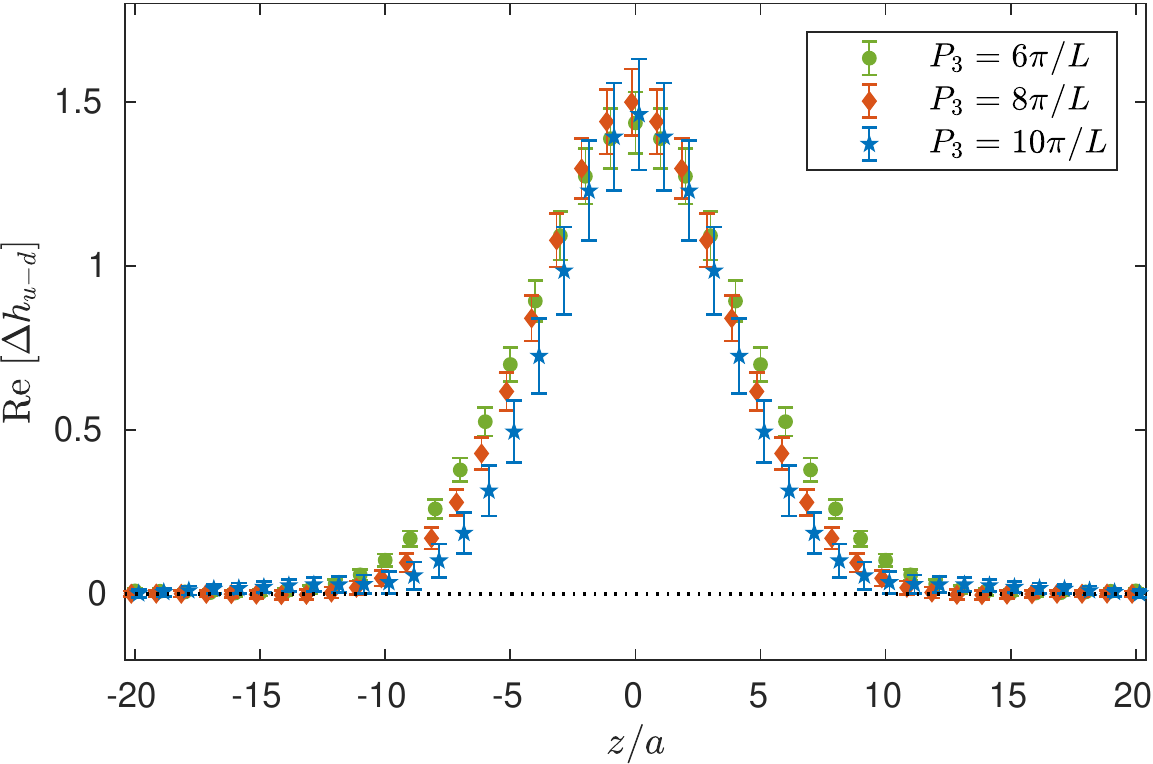}\hspace*{0.2cm}
    \includegraphics[width=0.495\textwidth]{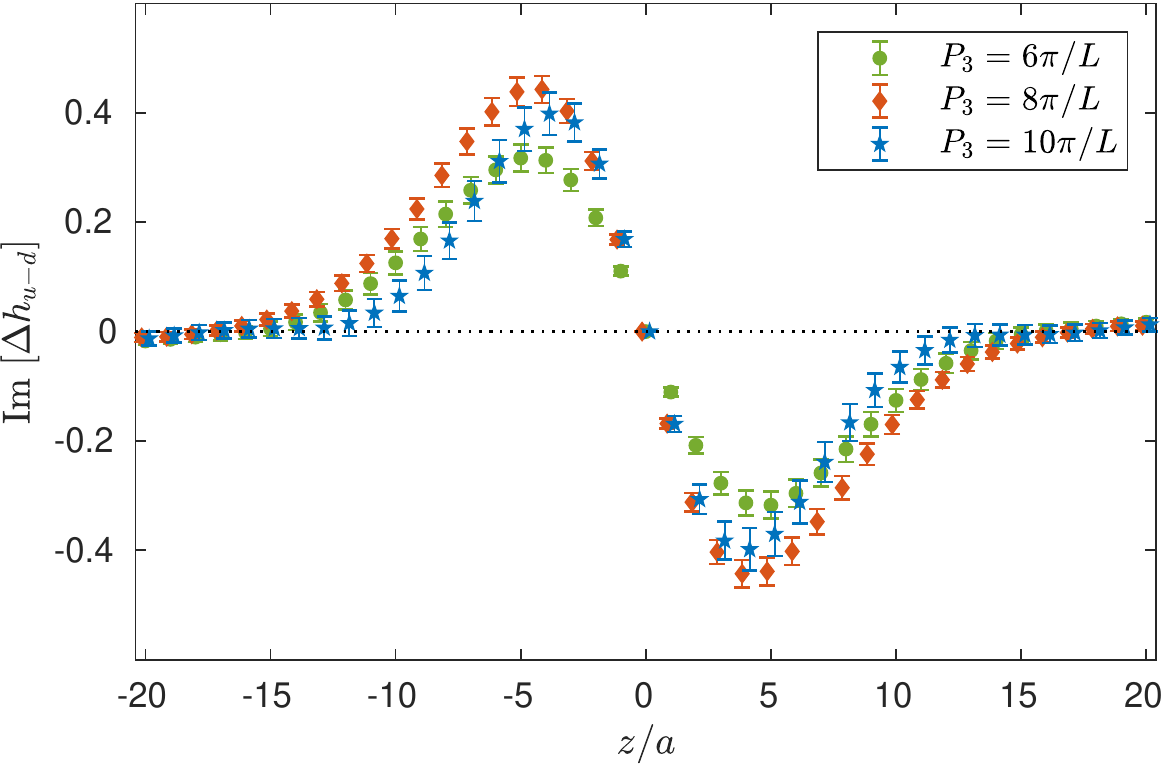}
    
    \vspace*{0.2cm}
   \includegraphics[width=0.495\textwidth]{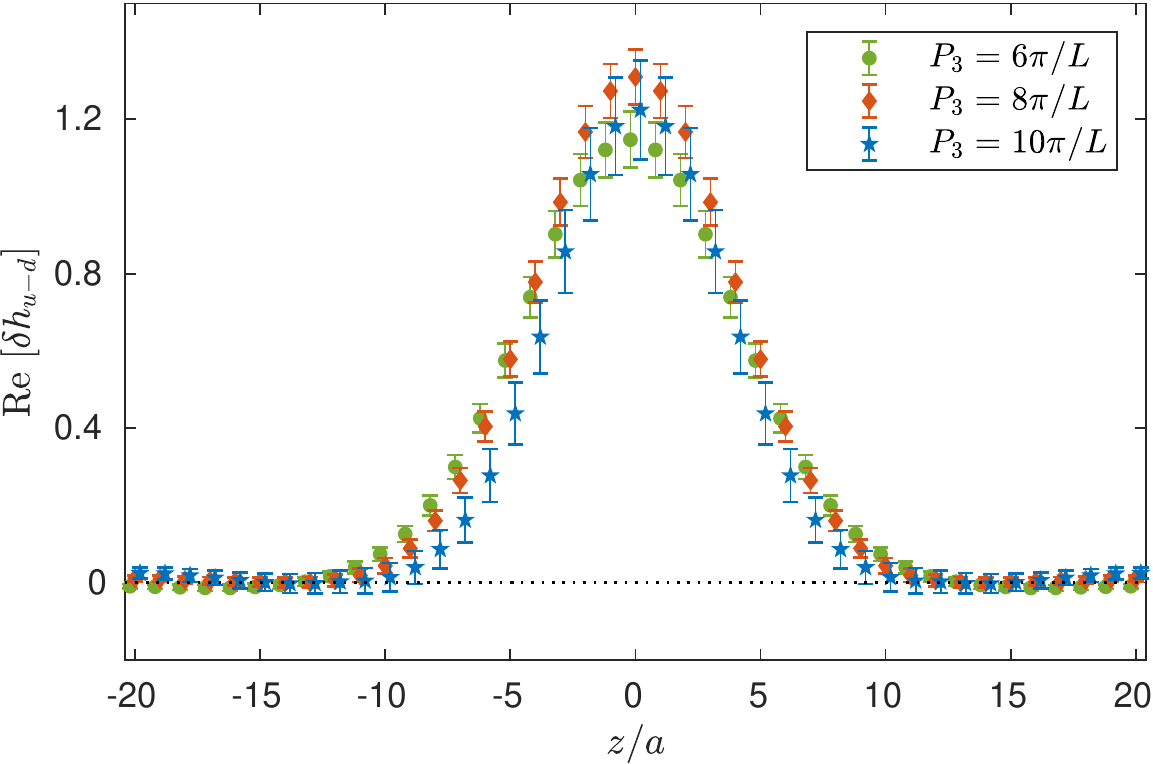}\hspace*{0.2cm}
    \includegraphics[width=0.495\textwidth]{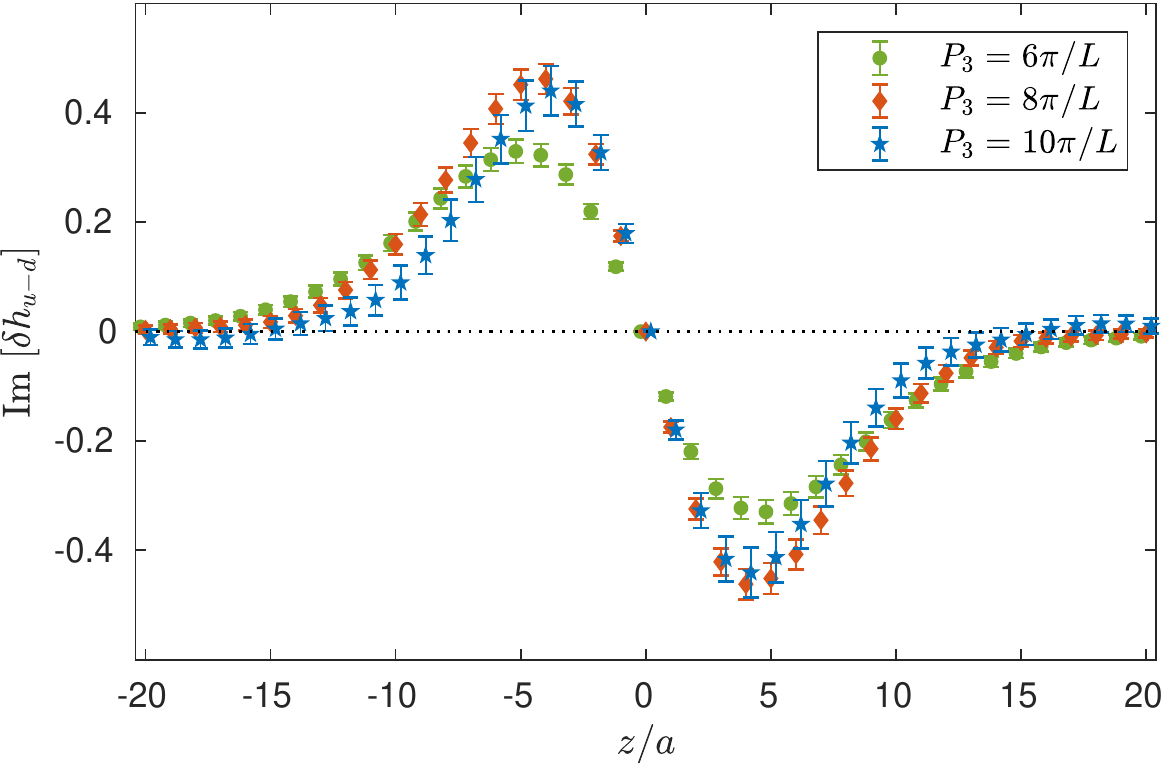} 
\end{center}  
\vspace*{-0.5cm}
\caption{Bare matrix elements for momenta $6\pi/L$ (green circles), $8\pi/L$ (red diamonds), $10\pi/L$ (blue stars), extracted at source-sink separation $t_s{=}12a\simeq 1.13$~fm. From top to bottom: matrix elements of unpolarized, helicity and transversity operators.}
\label{fig:mom_dep}
\end{figure}

We compute the matrix elements for the three values of the nucleon momentum given in Table~\ref{tab:statistics} with the associated number of measurements. The  momentum dependence of the resulting matrix elements is shown in Fig.~\ref{fig:mom_dep} for the three operators considered in this work. We find that with increasing momentum, the matrix elements decay faster to zero and, for the highest momentum employed, both  real and imaginary parts are  compatible with zero for $z\geq 11a\simeq 1.03$~fm.

\subsection{Excited States Contamination}
\label{sub:excited_states}
Identification of the nucleon state is crucial in order to extract the correct nucleon matrix elements from lattice QCD measurements.
This requires a careful analysis of excited states. 
An additional challenge is the need to boost the nucleon with a relatively large momentum, something that it is not needed in e.g.\ studies of nucleon charges and form factors. 
Requiring large nucleon momentum in combination with using simulations at the physical point leads to a more severe contamination due to the excited states, since the spectrum is denser. Therefore, a thorough assessment of the excited states is even more essential for the reliable extraction of PDFs. 
We extend the study of excited states first presented in Ref.~\cite{Alexandrou:2018yuy} in order to eliminate, as much as possible, one of the major systematic uncertainties in the evaluation of nucleon matrix elements from which the PDFs are extracted within lattice QCD. 
In what follows, we review the analysis methods that we employ to isolate the ground state. 
As will be discussed in this subsection, the excited states contributions are milder for the matrix elements of the unpolarized operator as compared to the matrix elements of the helicity and transversity operators.

To extract the ground state contribution in the correlation functions, we employ three analysis methods, briefly described below.

\vspace*{0.2cm}
1. \textit{Plateau method (single-state fit)}. In this method, one seeks for a region where the ratio of Eq.~(\ref{eq:ratio}) becomes independent of the insertion time, $\tau$, which indicates suppression of excited states (possibly partly). The ratio in this region is fitted to a constant value, yielding the matrix element of the ground state. Indeed, inserting in the two- and three-point functions two sets of complete eigenstates of the QCD Hamiltonian ($\vert n(P_3)\rangle$ and $\vert n'(P_3)\rangle$, where we indicated that the states are momentum-dependent, but we keep this dependence implicit below to simplify the notation) with quantum numbers of the nucleon, the ratio of Eq.~(\ref{eq:ratio}) can be written as\footnote{All correlation functions in this section should be understood as averaged over the gauge field configurations ensemble.}
\begin{equation}
\frac{C^{\rm 3pt}(P_3;t_s,\tau)}{C^{\rm 2pt}(P_3;t_s)}=\frac{\sum_{n,n'}\langle N\vert n'\rangle \langle n\vert N\rangle \langle n'\vert \mathcal{O}\vert n\rangle e^{-E_{n'}(t_s-\tau)}e^{-E_{n}\tau}}{\sum_n \vert\langle N\vert n\rangle \vert ^2 e^{-E_n t_s}}\, ,
\label{eq:sum_n}
\end{equation}
where $\vert N\rangle$ is the nucleon state, $E_{n,n'}$ are energies of the eigenstates $\vert n\rangle$, $\vert n'\rangle$ (with $\vert 0\rangle$ -- ground state, $\vert 1\rangle$ -- first excited state, etc.) and the timeslice of the source is set to zero. Isolating in the ratio the contribution of the ground state and expanding the sum up to the first excited state, Eq.(\ref{eq:sum_n}) reads
\begin{equation}
\frac{C^{\rm 3pt}(P_3;t_s,\tau)}{C^{\rm 2pt}(P_3;t_s)}{=}\frac{\langle 0\vert \mathcal{O}\vert 0\rangle +f_{10}\langle 1\vert\mathcal{O}\vert 0\rangle e^{-\Delta E(t_s-\tau)}+f_{10}^{\dagger}\langle 0\vert \mathcal{O}\vert 1\rangle e^{-\Delta E\tau}+ \vert f_{10}\vert ^2 \langle 1\vert \mathcal{O}\vert 1\rangle e^{-\Delta E t_s} +\ldots}{1+\vert f_{10}\vert ^2 e^{-\Delta E t_s}+ \ldots}\, ,
\label{eq:leading_terms}
\end{equation}
where we introduced the constant $f_{10}{=}{\langle N\vert 1\rangle}/{\langle N\vert 0\rangle}$ and $\Delta E=E_1-E_0$. 
As can be seen, the excited states contributions fall off exponentially and  for  $\Delta E \,t_s\gg 0$, $\Delta E\, \tau\gg 0$ and $\Delta E \,(t_s-\tau) \gg 0$, the first time-independent term dominates, yielding the matrix element of the nucleon state, $\langle 0\vert \mathcal{O}\vert 0\rangle $.
Thus, fitting the ratio within the plateau region to a constant value yields the desired nucleon matrix element. 
Since statistical errors grow exponentially with $t_s$, the challenge is to identify the smallest value of $t_s$ that suppresses the contributions of excited states in the ratio to a level negligible as compared to the statistical accuracy.

\vspace*{0.2cm}
2. \textit{Summation method}. This method was introduced in Ref.~\cite{Maiani:1987by} and entails the sum of the ratio of Eq.~(\ref{eq:leading_terms}) over the insertion time $\tau$, excluding the timeslices of the source and the sink. 
The sum is a geometric series in the terms involving the excited states contributions and yields the expression 
\begin{equation}
\label{eq:summation}
\mathcal{R}(P_3;t_s)\equiv\sum_{\tau=a}^{t_s-a}\,\frac{C^{\rm 3pt}(P_3;t_s,\tau)}{C^{\rm 2pt}(P_3;t_s)} = C + \mathcal{M}\,t_s + \mathcal{O}\left(e^{-\Delta E \,t_s}\right)\, ,
\end{equation}
where  $C$ is a constant and the desired matrix element $\mathcal{M}{=}\langle 0\vert \mathcal{O}\vert 0\rangle$ can  be extracted from a linear two-parameter fit. 
As a consequence, this procedure yields, in general, results with larger uncertainties as compared to the plateau method, but has the advantage of suppressing the excited states contamination by a faster decaying factor $e^{-\Delta E \,t_s}$, as compared to the leading exponential factors in Eq.~(\ref{eq:leading_terms}).

\vspace*{0.2cm}
3. \textit{Two-state fits}. In this method, one retains the first excited state contributions in the two-point and three-point correlation functions. 
We use two types of fits: (a) either a fit to the two-point correlator followed by a fit to the three-point function, or (b) a simultaneous fit to both the two- and three-point correlators.

We refer to the (a) type as a sequential fit and  parameterize  the nucleon two-point function with momentum $P_3$ as
\begin{equation}
\label{C2pt}
C^{\rm 2pt}(P_3;t)=\vert A_0\vert^2 e^{-E_0t}\left(1+\vert f_{10}\vert ^2 e^{-\Delta E \,t}\right),
\end{equation}
with fit parameters being the ground state amplitude, $\vert A_0\vert^2=\vert\langle N\vert 0\rangle \vert ^2$, the ground state energy $E_0$, $\vert f_{10}\vert ^2$ and $\Delta E$. 
The three-point correlator can be written as 
\begin{equation}
\label{C3pt}
C^{\rm 3pt}(P_3;t_s,\tau)=\vert A_0\vert^2e^{-E_0 t_s} \Big(\langle 0\vert \mathcal{O}\vert 0\rangle  +f_{10}\langle 1\vert\mathcal{O}\vert 0\rangle e^{-\Delta E\,(t_s- \tau)}+ f_{10}^\dagger\langle 0\vert \mathcal{O}\vert 1\rangle e^{-\Delta E\, \tau}+ \vert f_{10}\vert^2\langle 1\vert \mathcal{O}\vert 1\rangle e^{-\Delta E\, t_s}\Big).
\end{equation}
The amplitudes $\vert A_0\vert^2$ and $\vert f_{10}\vert^2$ and energies $E_0$ and $\Delta E$ are determined from first fitting the two-point function in Eq.~(\ref{C2pt}) and used as inputs into the three-point function, which is fitted separately for the real and imaginary parts. The fit parameters are $\textrm{Re} \langle 0\vert \mathcal{O}\vert 0\rangle$, $\textrm{Re} \widetilde{f}_{10}{\equiv}\textrm{Re} f_{10}\langle 1\vert\mathcal{O}\vert 0\rangle$, $\textrm{Re} \widetilde f_{01}{\equiv}\textrm{Re} f_{10}^{\dagger}\langle 0\vert\mathcal{O}\vert 1\rangle$, $\textrm{Re} \langle 1\vert \mathcal{O}\vert 1\rangle$ and $\textrm{Im} \langle 0\vert \mathcal{O}\vert 0\rangle$, $\textrm{Im} \widetilde f_{10}$, $\textrm{Im} \widetilde f_{01}$, $\textrm{Im} \langle 1\vert \mathcal{O}\vert 1\rangle$, respectively.
Through this fitting procedure that entails 4 fit parameters for the two-point function fit and additional 4 parameters for the three-point function, we extract the desired matrix element, $\langle 0\vert \mathcal{O}\vert 0\rangle$.

The type (b) fit we call simultaneous and it is a combined fit to Eqs.~(\ref{C2pt}) and (\ref{C3pt}), using all 8 parameters.
We expect that both fit types lead to very similar results.
Proper analysis requires the evaluation of two- and three-point correlation functions on exactly the same gauge field configurations and for the same set of source positions. Only in such a case, the correlations among the fit parameters can be probed in a fully consistent manner.
In practice, this is a severe problem for the observables we are interested in, which is due to the exponential increase of the noise with increasing source-sink time separation.
When the time separation  grows by one lattice spacing, the necessary statistics to suppress noise to the same level as before increases by a factor 2-3 in our data.
Thus, small source-sink time separations yield a precise signal already for $\mathcal{O}(10^3)$  measurements, while correlators at $t_s{=}12a$ require $\mathcal{O}(10^5)$ measurements to reach a similar precision.
For a given amount of computational resources, one, therefore, has to make a choice between using the same statistics for all $t_s$ values and thus obtaining very precise data at small values of $t_s$ and considerably less precise ones for larger values of $t_s$, or using different statistics achieving similar precision for all values of $t_s$.
The severe drawback of the former choice is that the two-state fits are dominated by the precise results at small $t_s$ that can lead to biased results for the ground state matrix element. This bias becomes more severe  as the boost increases due to the denser spectrum prohibiting the robust identification of the effects of excited states.

\subsubsection{Two-point correlator}
\label{sec:2pt}

\begin{figure}[h!]
\begin{center}
    \includegraphics[width=0.475\textwidth]{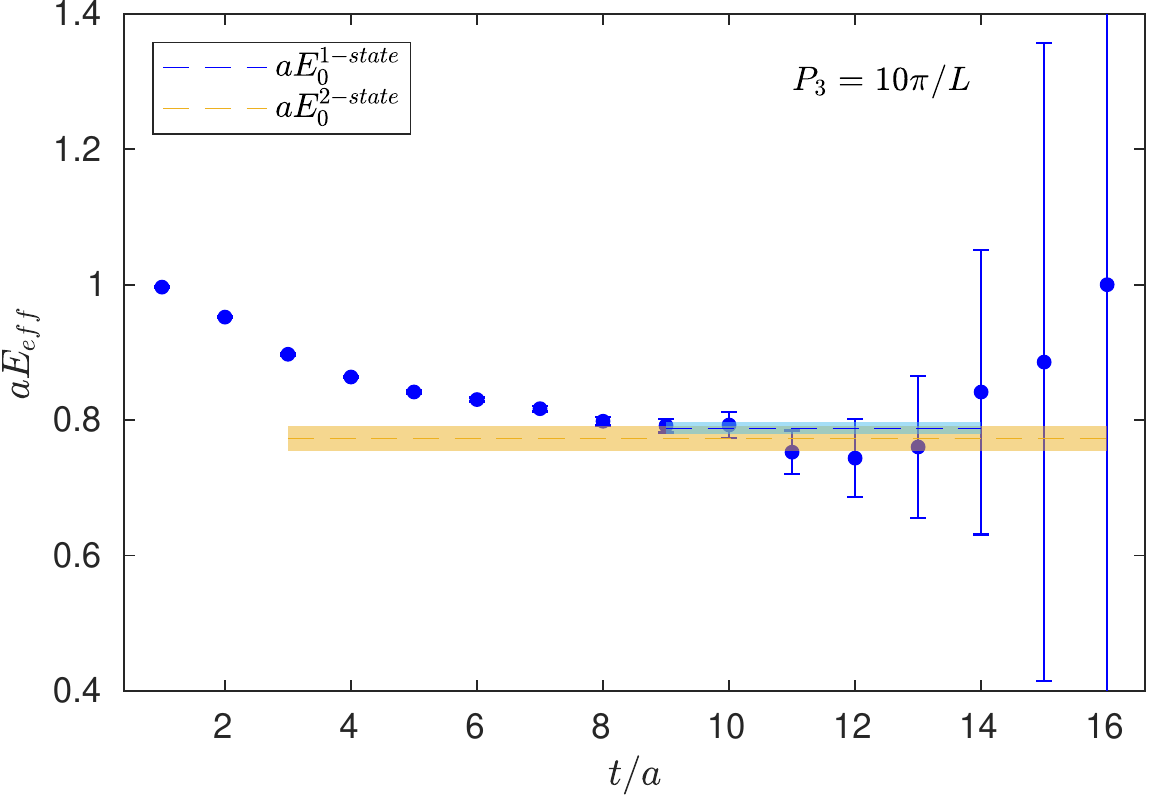}
\end{center}
\vspace*{-0.5cm} 
\caption{Effective mass together with the extracted values of the ground state energy from one-state (blue band) and two-state fits (orange band). The length of the bands indicates the fit range. The nucleon is boosted with $10\pi/L\simeq 1.38$ GeV.}
\label{fig:2pt}
\end{figure}

Before we compare different analysis methods for the extraction of quasi-PDF matrix elements, we discuss the two-state fit of the two-point correlator at our largest momentum, using Eq.\ (\ref{C2pt}).
We obtain the boosted nucleon ground state energy of $aE_0^{\textrm{2-state}}{=}0.773(18)$, which is in line with a single-state fit, leading to a value $aE_0^{\textrm{1-state}}{=}0.788(9)$. These values are illustrated in the left panel of Fig.\ \ref{fig:2pt}.
As we showed in Fig.\ \ref{fig:dispersion_relation}, the obtained ground state energy satisfies the continuum dispersion relation.
The energy difference between the ground state and the first excited state is $a\Delta E{=}0.36(7)$, which in physical units is 0.76(15) GeV.
This can be compared with the expectation from the approximation of noninteracting stable hadrons in a box, which leads, at our volume and the physical pion mass ($m_\pi L\approx3$), to the following splittings (for $N\pi\pi$, $N\pi$, $N\pi\pi\pi\pi$, $N\pi\pi\pi$) with respect to the ground state ($N$): $\approx$ 0.27 GeV, 0.36 GeV, 0.54 GeV, 0.63 GeV, respectively.
The above values hold for a nucleon at rest, whereas the spectrum becomes denser with increasing nucleon momentum.
In the interacting case, the spectrum obviously changes, however its general features are unchanged; for a comprehensive discussion on the spectrum of excitations, we refer to the recent review of Ref.~\cite{Green:2018vxw}.

The energy extracted for the first excited state does not match the energy expected for the two- and three-particle states lying below. Furthermore, it is known that a two-state fit will model, in one exponential, all the states above the ground state that have an overlap with the interpolating field. Thus, one expects a bias in the determination of the first excited state. Consequently, the conclusions from two-state fits need to be checked against the plateau values or three-state fits, if the accuracy of the data allows for this.
If consistency is found, this indicates ground state dominance up to the reached statistical precision.

\subsubsection{Matrix elements of the unpolarized PDF}
\label{subsec:g0_excited_states}
We first present the analysis of excited states in the nucleon matrix elements for the unpolarized operator using the $\gamma^0$ structure and the largest momentum, $P_3{=}10\pi/L$, where excited states effects are expected to be most severe.
We use four values of the source-sink time separation, namely $t_s/a{=}8,9,10,12$ or in physical units $t_s {\simeq} 0.75, 0.84, 0.94, 1.13$~fm.  
We opt to increase the number of measurements as we increase $t_s$, to have  statistical errors that are approximately the same for consequent values of $t_s$. 
This enables us to perform a reliable analysis at each value of $t_s$.
We use the same CAA setup for all separations, as explained above. 
In Table~\ref{tab:g0_statistics}, we collect the statistics for each value of $t_s$.

\begin{table}[ht!]
\begin{center}
\renewcommand{\arraystretch}{1.4}
\renewcommand{\tabcolsep}{6pt}
\begin{tabular}{c | c | c | c | c}
 $P_3=10\pi/L$ & $t_s=8a$  & $t_s=9a$ & $t_s=10a$ & $t_s=12a$ \\
\hline\hline
\hsn$N_{\rm conf}$ & 48  & 98 & 100 & 811\\
\hline
\hsn$N_{\rm meas}$ & 4320  & 8820 & 9000 & 72990\\
\hline
\end{tabular}
\caption{\small{Numbers of measurements of the matrix element for the unpolarized operator ($\gamma^0$) at each source-sink separation and $P_3{=}10\pi/L$.}}
\label{tab:g0_statistics}
\end{center}
\end{table}

\begin{figure}[h!]
\begin{center}
    \includegraphics[width=0.49\textwidth]{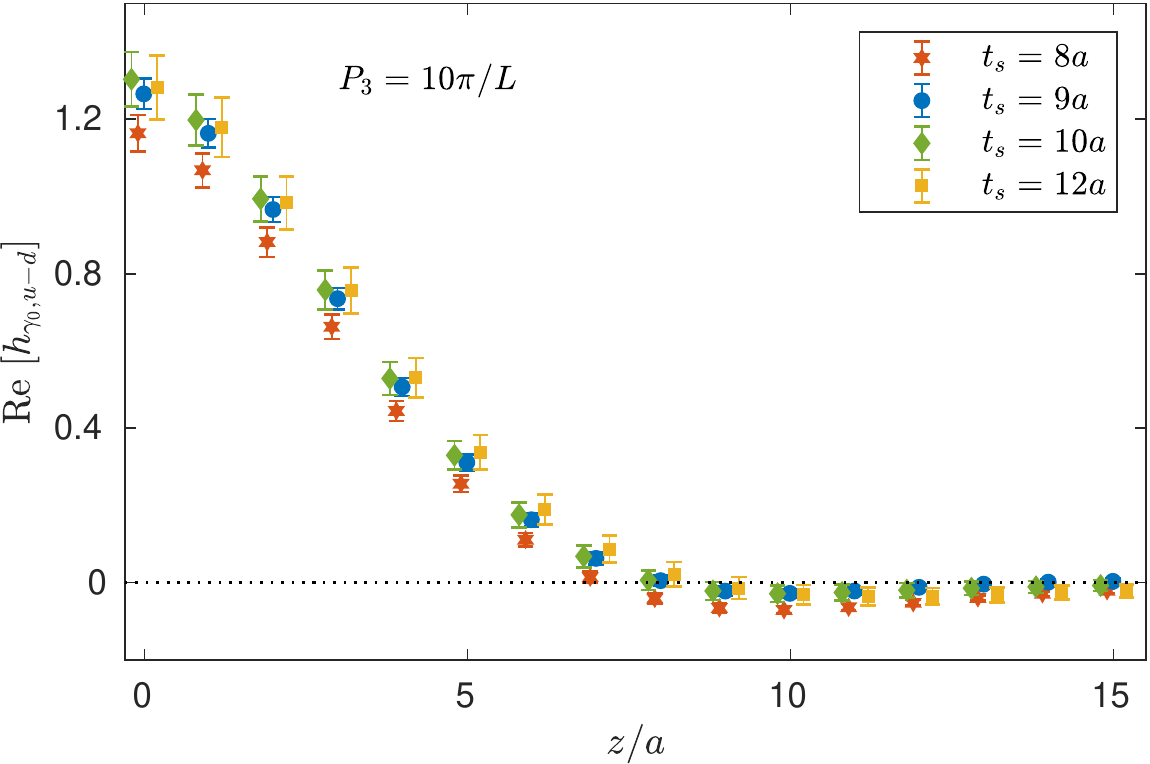}\,\,
    \includegraphics[width=0.49\textwidth]{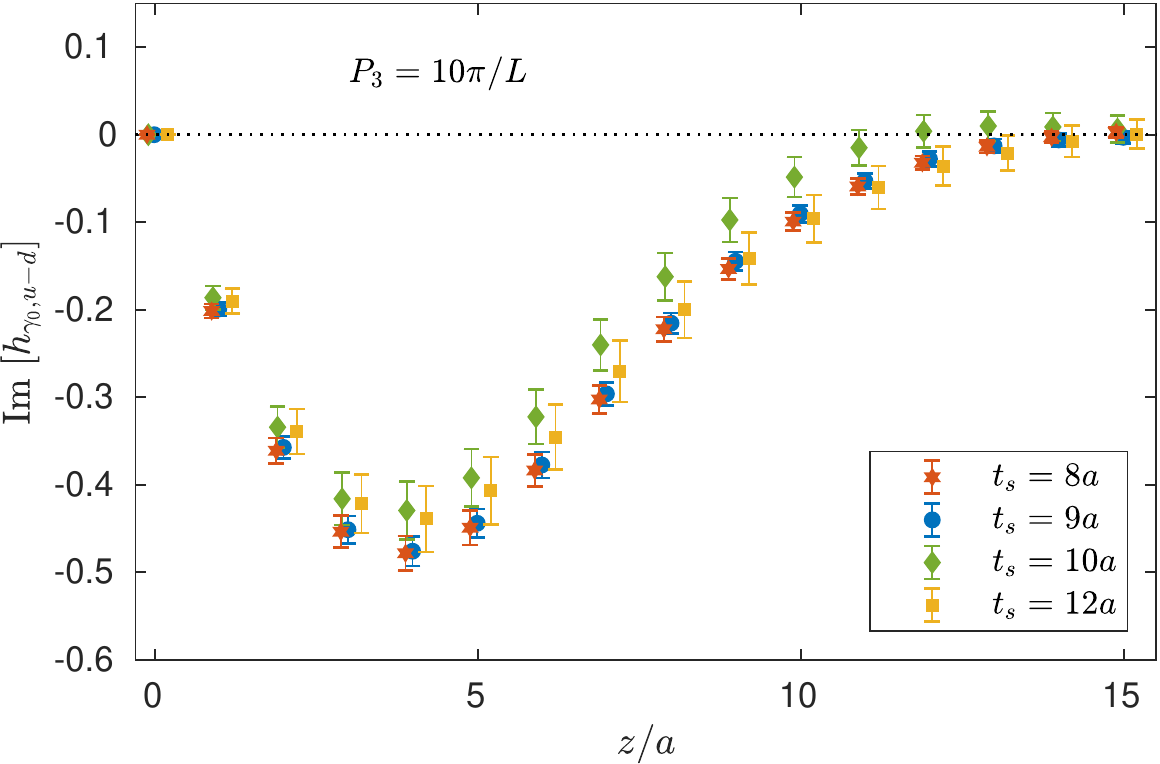}
\end{center}
\vspace*{-0.5cm} 
\caption{Real (left) and imaginary (right) part of the matrix element for the unpolarized PDF, at source-sink separation $8a$ (red stars), $9a$ (blue circles), $10a$ (green diamonds) and $12a$ (yellow squares). The nucleon is boosted with $10\pi/L\simeq 1.38$ GeV.}
\label{fig:tsink_plateau}
\end{figure}

The plateau method is employed to analyze the data from each $t_s$ value  and the results are shown in Fig.~\ref{fig:tsink_plateau}. For $t_s{=}8a$, the real part of the matrix element shifts towards smaller values at each $z/a$ compared to larger $t_s$, while the effect in the imaginary part is less prominent. Within statistical uncertainties, a convergence between the results at $t_s{=}10a$ and $t_s{=}12a$ is observed in both the real and imaginary parts. Comparing the data for the four $t_s$ values, we observe signs of a nonmonotonic behavior that affects the real and imaginary parts differently, depending on the value of $z$. This can introduce a complicated effect in the determination of the PDF. Ideally, one must compute the matrix elements for several large enough values of $t_s$ with equally small statistical errors  and demonstrate convergence to one value. However, the exponential increase of the noise-to-signal ratio seen in Fig.~\ref{fig:mom_smearing} and the need for new sets of sequential inversions for each $t_s$, require computational resources beyond what is currently available, placing limitations on the maximum value of $t_s$.
Nevertheless, if consistency can be found between the converged plateau values of the single-state fits and other methods, ground state dominance can be reliably established.

\begin{figure}[h!]
\begin{center}
    \includegraphics[width=0.49\textwidth]{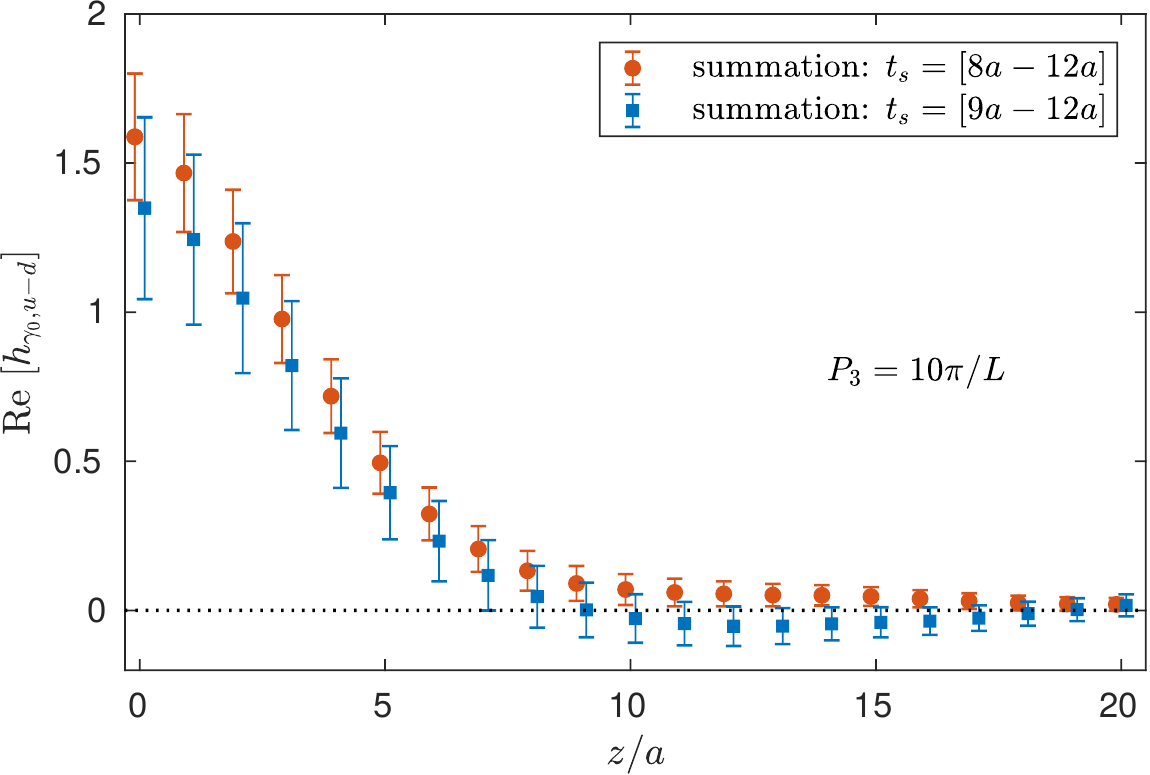}\,\,
    \includegraphics[width=0.49\textwidth]{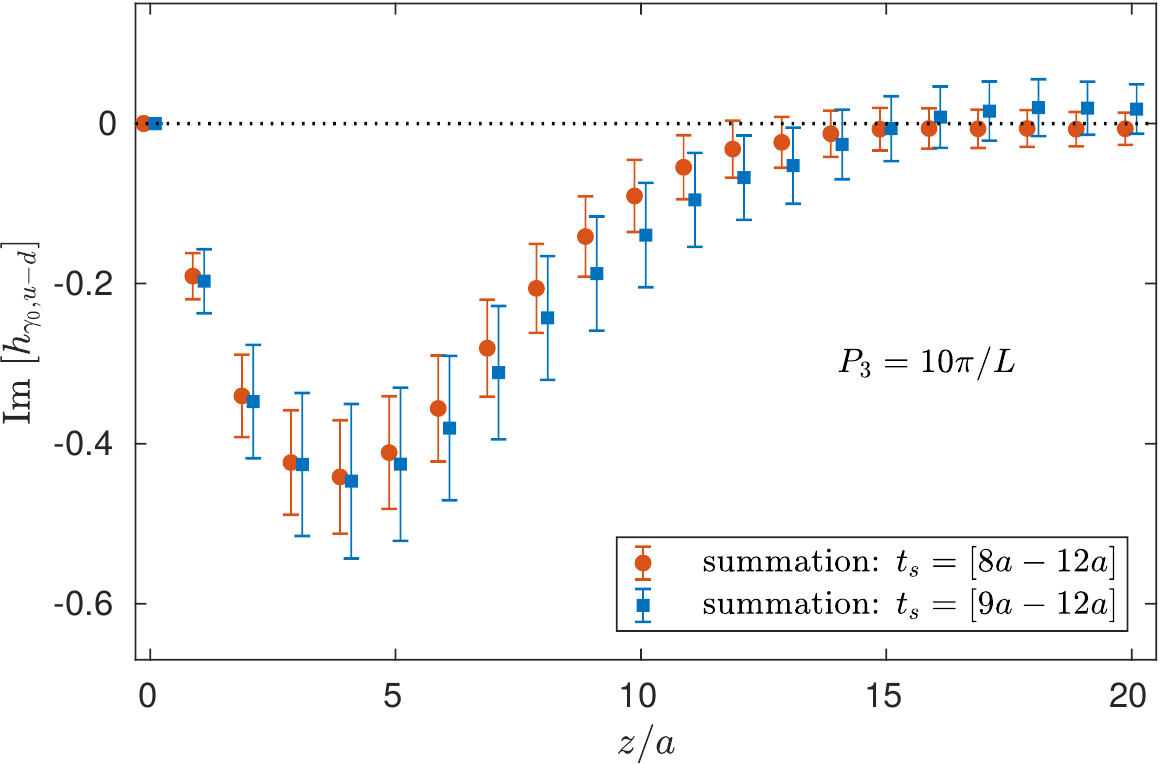}
\end{center}
\vspace*{-0.5cm}
\caption{Real (left) and imaginary (right) part of the matrix element for the unpolarized PDF from the summation method, using source-sink separations $8a-12a$ (red circles) and $9a-12a$ (blue squares). The nucleon is boosted with momentum $10\pi/L{\simeq}1.38$ GeV.}
\label{fig:summation}
\end{figure}

\begin{figure}[h!]
\begin{center}
    \includegraphics[width=0.49\textwidth]{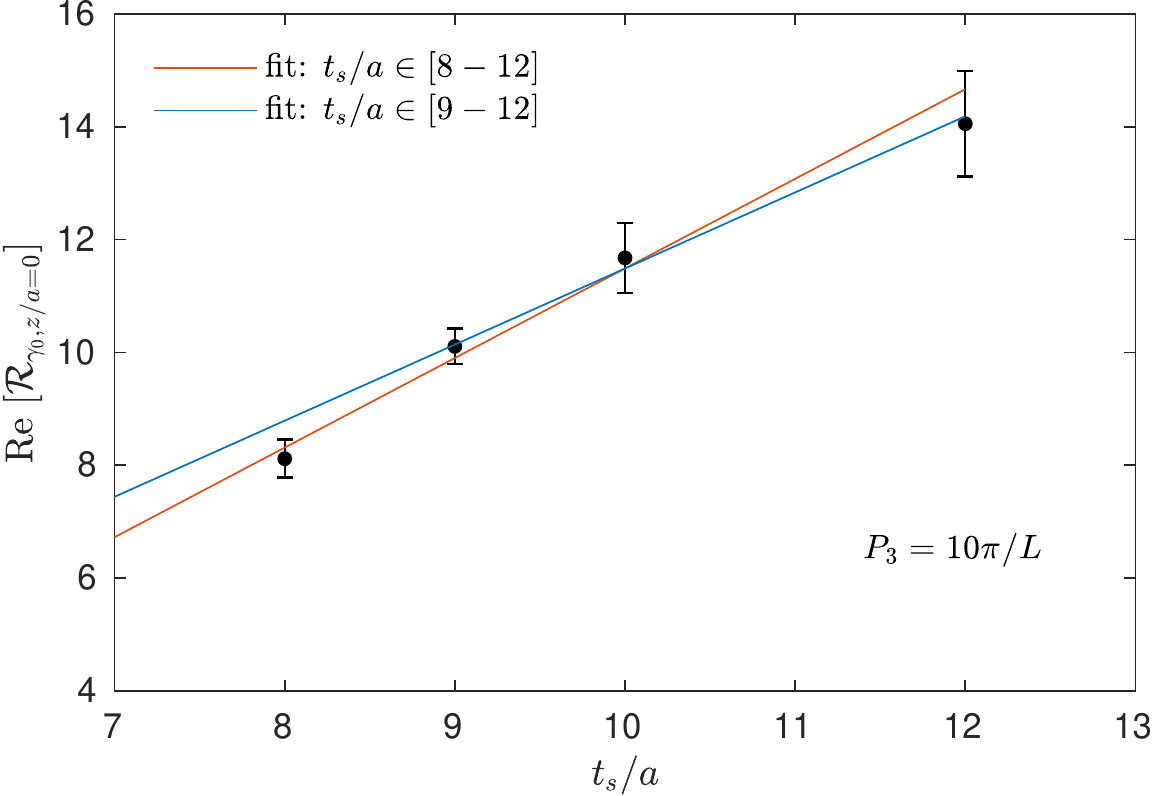}\,\,
    \includegraphics[width=0.49\textwidth]{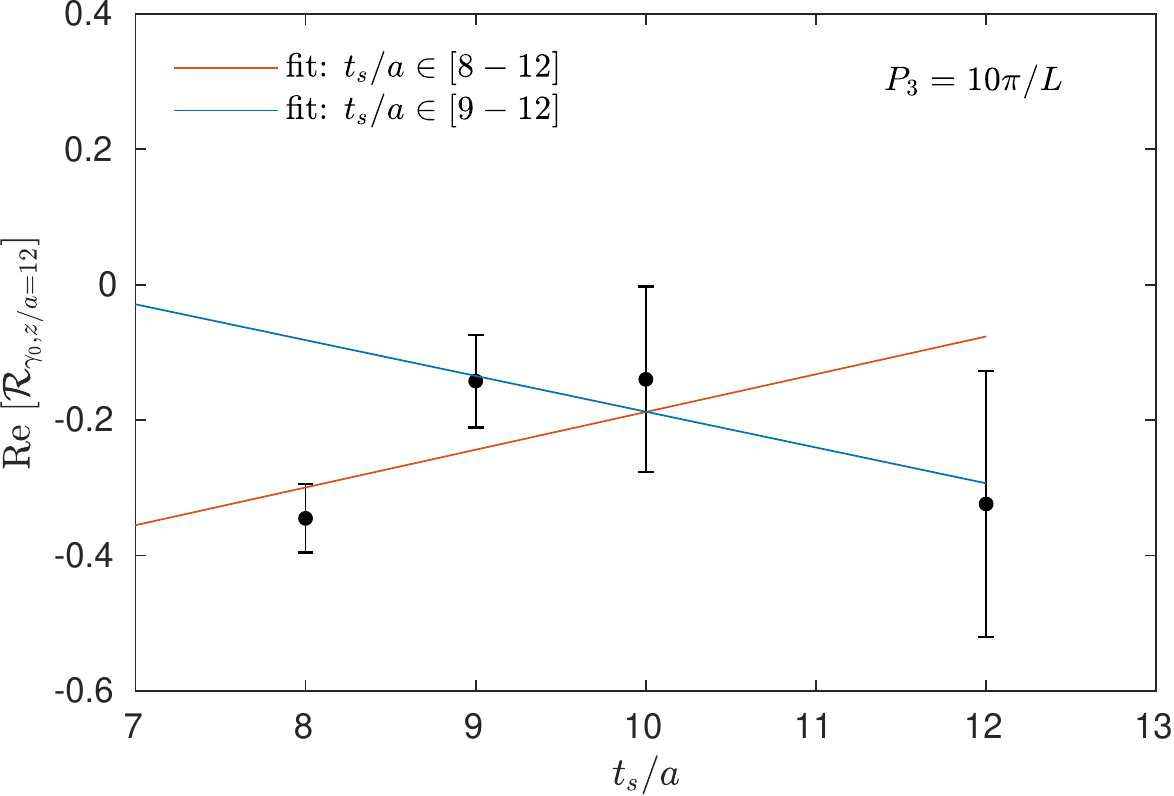}
\end{center}
\vspace*{-0.5cm}
\caption{Summation fits for the real part of the matrix element for the unpolarized PDF, $z/a=0$ (left) and $z/a{=}12$ (right). In both cases, fits (a) to all source-sink separations and fits (b) to the three largest separations are shown. The nucleon boost is $10\pi/L{\simeq}1.38$ GeV.}
\label{fig:summ_z12}
\end{figure}

We apply the summation method for two sets of $t_s$ values, namely: (a) $t_s{=}8a,\,9a,\,10a,\,12a$ and (b) $t_s{=}9a,\,10a,\,12a$, with the results shown in Fig.~\ref{fig:summation}.
The first set yields results that, although consistent with those from the second fit, are systematically higher. 
Furthermore, for large values of $z/a$, the slope obtained using the (b) set is significantly larger than from the (a) set and $\chi^2/{\rm d.o.f.}>2$ for $z/a{\approx}10{-}15$, indicating that the (a) fits do not provide good description of the data. 
This is a consequence of the ratio at $t_s{=}8a$ being considerably below the ones at other source-sink separations.
Fig.\ \ref{fig:summ_z12} shows examples of summation fits for our data.
In the right panel, only the (b) fit yields an acceptable $\chi^2/{\rm d.o.f.}$ of around 0.16, while the full fit to all $t_s$ values has $\chi^2/{\rm d.o.f.}\approx2.4$.
Thus, the fitting ansatz of Eq.\ (\ref{eq:summation}) does not provide a correct description of the data and the large-$z$ values in Fig.~\ref{fig:summation} for $t_s{=}[8a{-}12a]$ are not reliable.
The effect is visible also for small $z/a$ (see the left panel of Fig.\ \ref{fig:summ_z12} for $z/a{=}0$ fits) and may result in overestimating the real part of the matrix elements.
At $z/a{=}0$, the extracted value of the matrix element is 1.21(16) for the (a) set and 1.02(24) for the (b), after applying in both cases the renormalization factor $Z_V{=}0.7565$~\cite{Alexandrou:2015sea}. 
We note that the matrix element should be equal to 1 upon renormalization, which is satisfied only by the set (b), while the value from the summation method that includes the smallest $t_s$  in the fit is considerably larger. 
All the above lead to the conclusion that using too small source-sink separations gives incorrect results. 
We will thus omit $t_s{=}8a$ in our summation method analysis, and take fits (b) as our final estimates from the summation method.
All of these fits, both in the real and imaginary part, have $\chi^2/{\rm d.o.f.}{\lesssim}1$, indicating that the exponential contributions from higher excited states in Eq.~(\ref{eq:summation}) are sufficiently suppressed, yielding a good estimate of the ground state matrix element ${\cal M}$.
However, the precision of the summation method estimates is much worse than the one from the plateau or two-state fits (see below).

\begin{figure}[h!]
\begin{center}
    \includegraphics[width=0.49\textwidth]{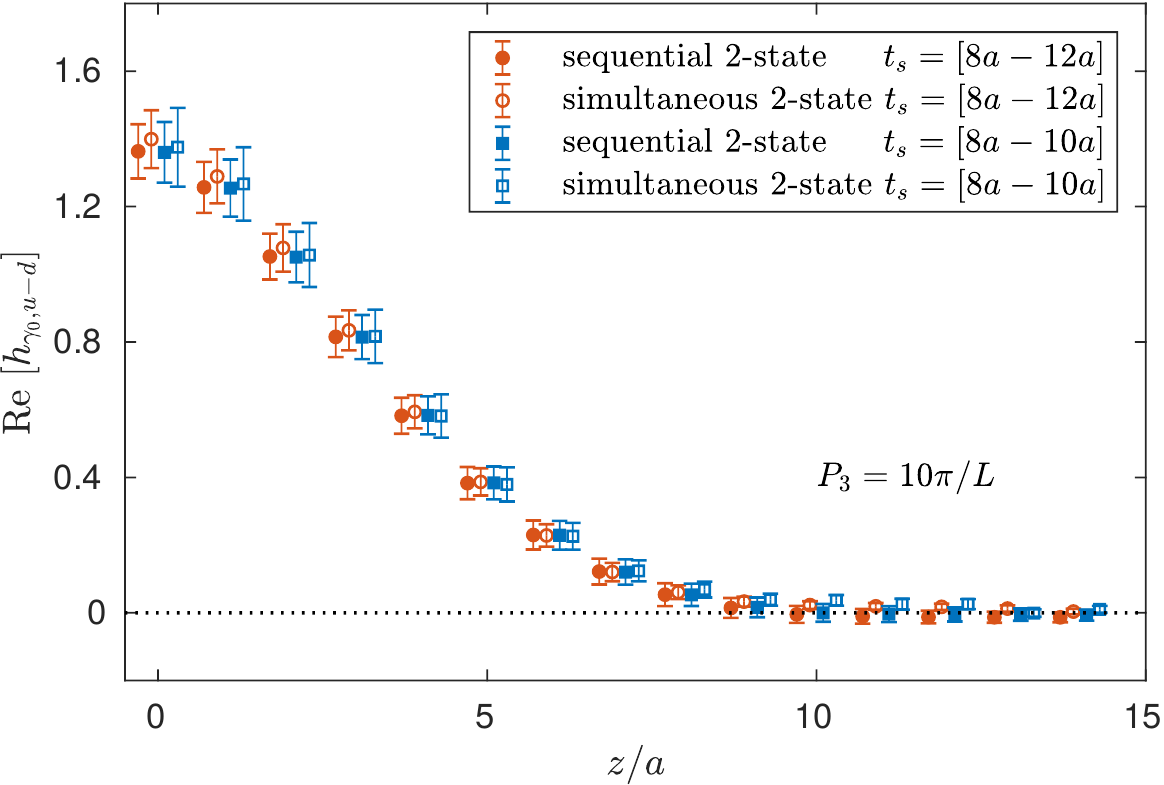}\,\,
    \includegraphics[width=0.495\textwidth]{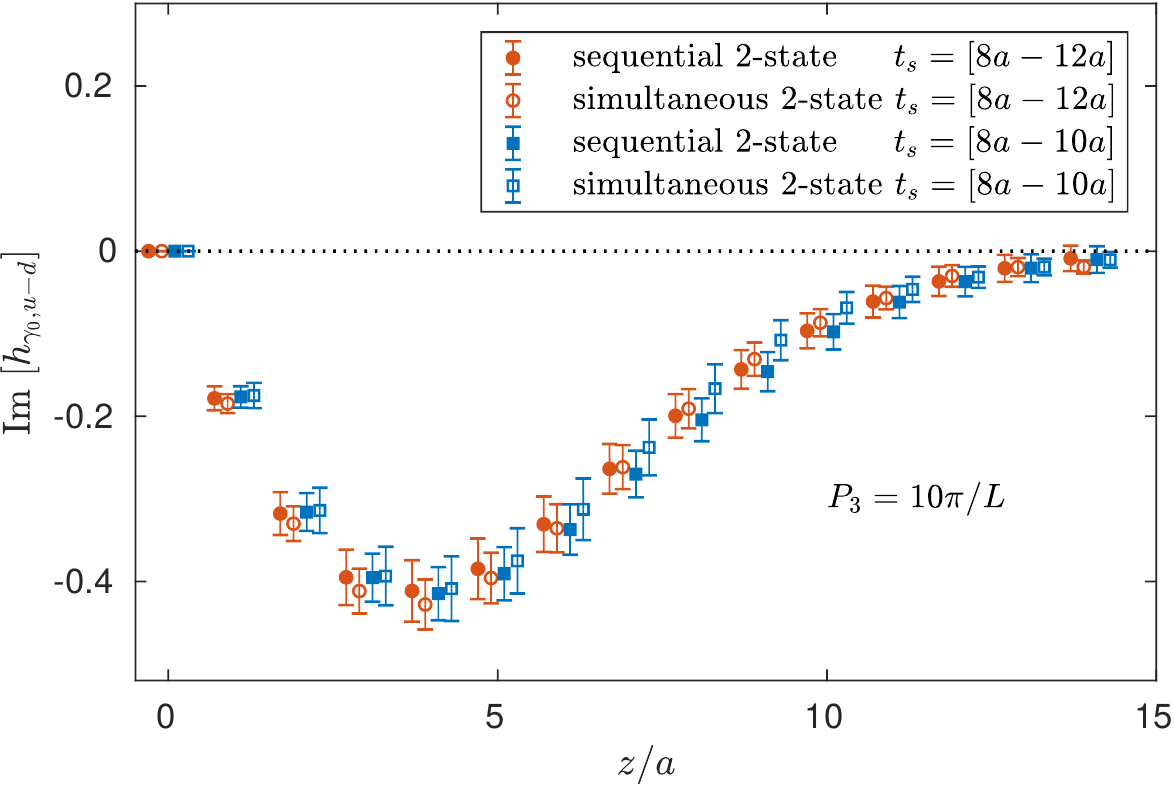}
\end{center}
\vspace*{-0.5cm} 
\caption{Real (left) and imaginary (right) part of the matrix element for the unpolarized PDF from 2-state fits, using source-sink separations $8a-12a$ (circles) and $8a-10a$ (squares). Sequential fit (filled symbols) is also compared to simultaneous fit (open symbols). The nucleon is boosted with $10\pi/L\simeq 1.38$ GeV.}
\label{fig:2state_compare}
\end{figure}

We present now the two-state fits and compare results from the sequential and simultaneous two-state fits. 
In addition, we check the robustness of these fits by using three or four values of $t_s$.
The fits using only two $t_s$ lead to considerably larger errors (with compatible central values) and hence, we do not show them.
The different choices of fits with three $t_s$ values lead to very similar results in terms of the central values and their errors.
Thus, in Fig.~\ref{fig:2state_compare}, we present results from one choice of three source-sink separations, namely $t_s{=}8a,\,9a,\,10a$, and from the fit to all $t_s{=}8a,\,9a,\,10a,\,12a$.
We conclude that the simultaneous and sequential fits lead to statistically  identical results for both sets of $t_s$ values. 
Including the results for the largest $t_s$ yields consistent fits, but with errors that are slightly smaller (up to 10-15\%) for most $z$ values. 
This is visible particularly for the simultaneous fits at all $z$ and for sequential fits at small $z$ values in the real part.
Note that the central values are essentially the same, suggesting that excited states are sufficiently suppressed.

\begin{figure}[h!]
\begin{center}
   \includegraphics[width=0.49\textwidth]{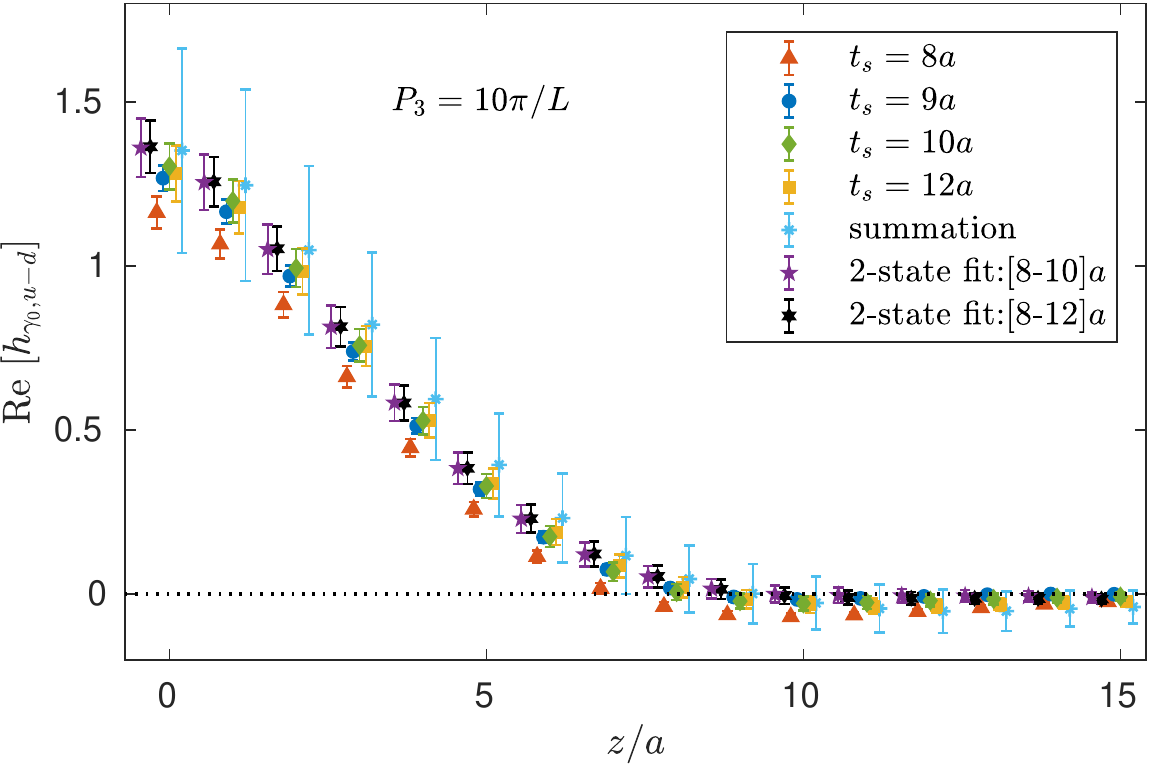}\,\,
   \includegraphics[width=0.49\textwidth]{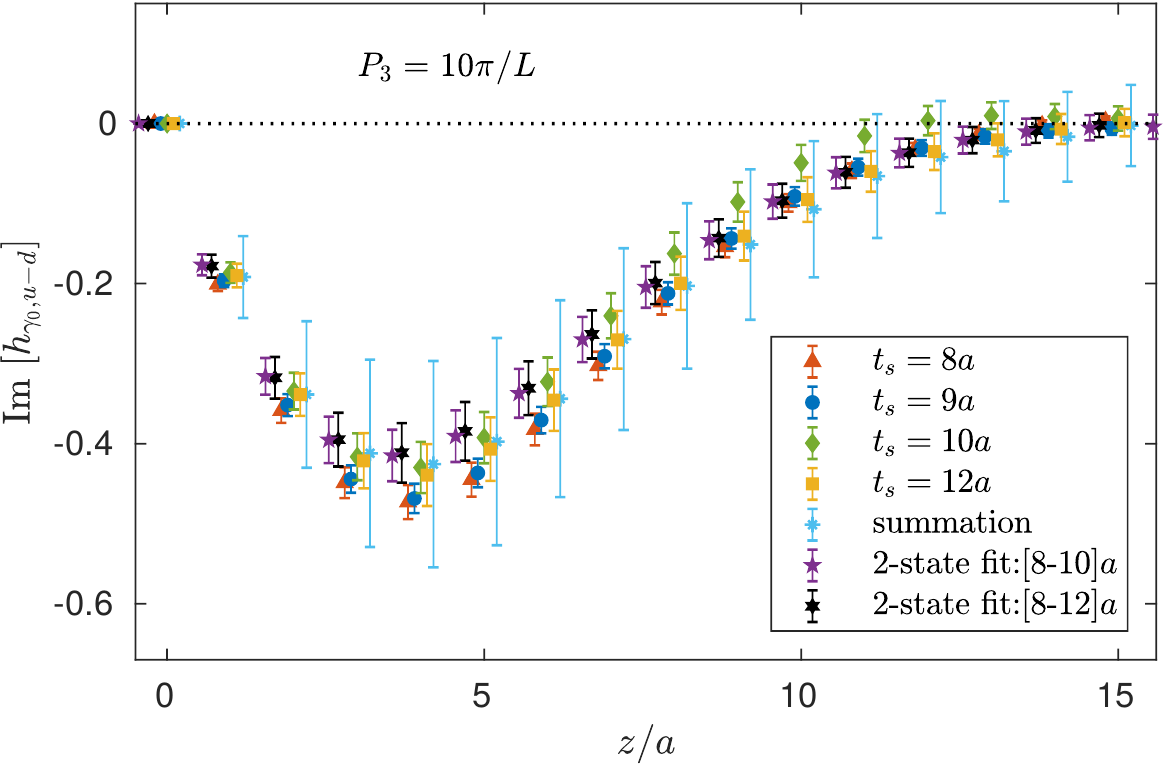}
\end{center}
\vspace*{-0.5cm} 
\caption{Real (left) and imaginary (right) part of the matrix element for the unpolarized PDF from the plateau method (points labeled with appropriate $t_s$), the summation method (using $t_s\geq9a$) and sequential 2-state fits (to $t_s{=}8a,9a,10a$ and to all $t_s$). The nucleon is boosted with $10\pi/L\simeq 1.38$ GeV.}
\label{fig:excited}
\end{figure}

\begin{figure}[h!]
\begin{center}
   \includegraphics[width=0.49\textwidth]{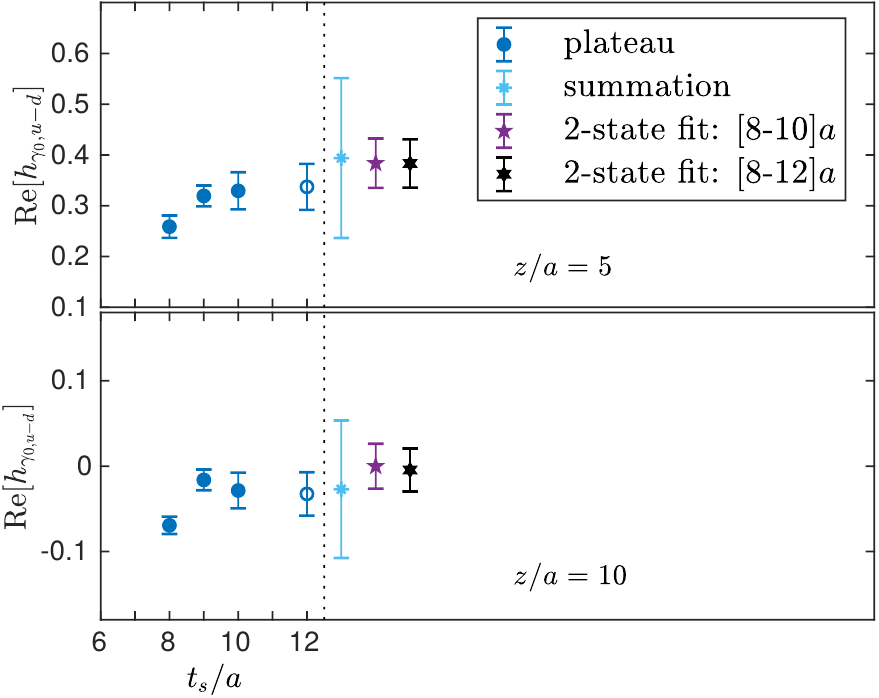}\,\,
   \includegraphics[width=0.49\textwidth]{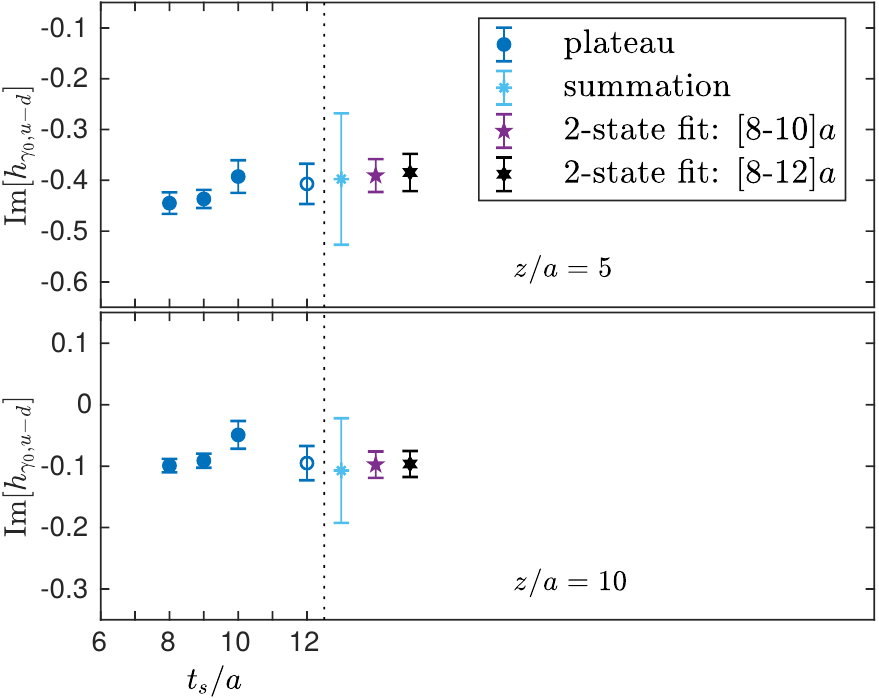}
\end{center}
\vspace*{-0.5cm} 
\caption{Real (left) and imaginary (right) part of the matrix element for the unpolarized PDF at fixed $z/a{=}5$ (upper) and $z/a{=}10$ (lower). We compare results from the plateau method, the summation method (using $t_s\geq9a$) and sequential 2-state fits (to $t_s{=}8a,9a,10a$ and to all $t_s$). The final value that we use for extracting PDFs is indicated with an open symbol. The nucleon is boosted with $10\pi/L\simeq 1.38$ GeV.}
\label{fig:g0_z5z10}
\end{figure}

In Fig.~\ref{fig:excited}, we collect the results of the analyses using the three aforementioned procedures, for all $z/a$ values, and in Fig.~\ref{fig:g0_z5z10} we present part of the same data for two selected values of the Wilson line length, $z/a{=}5$ and $z/a{=}10$, for better visibility.
Excited states effects are clearly visible for $t_s{=}8a$, especially in the real part when using the plateau method.
For $t_s{=}9a$ and $10a$, we observe small tension in the imaginary part -- the plateau values are consistently lower for $t_s{=}9a$ and consistently higher for $t_s{=}10a$, with respect to the two-state fits.
We note that, in particular, there is tension between the two-state fit using $t_s{=}8a{-}10a$ and the plateau values for $t_s{=}10a$, at $z/a{\approx}10$, which fails to satisfy our criterion of consistency between two-state fits and plateau fits.
Hence, data at source-sink separations below $12a$ are likely still contaminated by excited states, although statistical fluctuation as the source of the trend cannot be excluded.
We find that the results extracted from the plateau method using $t_s{=}10a$ and $t_s{=}12a$ are in agreement with each other and the ones from $t_s{=}12a$ are compatible with those extracted using the all-$t_s$ two-state fits for all values of $z/a$ and for both the real and the imaginary parts. 
The errors in the summation method are too large to draw any meaningful conclusions.
Given that this investigation is carried out for the largest value  of the momentum, we can take as our final value for the nucleon matrix elements of the unpolarized operator  the data at $t_s{=}12a$ for all the momentum values.
We note that increasing the nucleon boost would need a similarly thorough re-analysis on the effects of the excited states.
Likewise, increased statistical precision could reveal excited states contamination even at this nucleon momentum.
Thus, we emphasize that the attained conclusion about ground state dominance is valid only within the present statistical uncertainties of $\mathcal{O}(10)\%$.

\subsubsection{Matrix elements of the helicity and transversity PDFs}
In this subsection, we discuss the excited states effects for the two other Dirac structures used in our work -- axial and tensor, associated with the helicity and transversity PDFs, respectively.
We perform a similar analysis as for the unpolarized case at the largest momentum, $P_3{=}10\pi/L$, and using again four values of the source-sink separation, namely $t_s/a {=} 8,9,10,12$, or in physical units $t_s \simeq \{0.75, 0.84, 0.94, 1.13\}$~fm. 
The numbers of measurements are listed in Table~\ref{tab:pol_statistics}. 
The methodology of our investigation is the same as for the unpolarized case, i.e.\ for each $t_s$ value, we perform single-state fits within a plateau region, and we combine data at different $t_s$  using the summation method approach, as well as two-state sequential and simultaneous fits. 

\begin{table}[ht!]
\begin{center}
\renewcommand{\arraystretch}{1.4}
\renewcommand{\tabcolsep}{6pt}
\begin{tabular}{c | c | c | c | c}
 $P_3=10\pi/L$ &$t_s=8a$  & $t_s=9a$ &$t_s=10a$ & $t_s=12a$ \\
\hline\hline
\hsn$N_{\rm conf}$ & 36  & 50 & 88 & 811\\
\hline
\hsn$N_{\rm meas}$ & 3240  & 4500 & 7920 & 72990\\
\hline
\end{tabular}
\caption{\small{Statistics used in the study of excited states for the polarized cases ($\gamma^5\gamma^3$: helicity, $\sigma^{3j}$: transversity) at $P_3{=}10\pi/L$.}}
\label{tab:pol_statistics}
\end{center}
\end{table}

\begin{figure}[h!]
\begin{center}
   \includegraphics[width=0.49\textwidth]{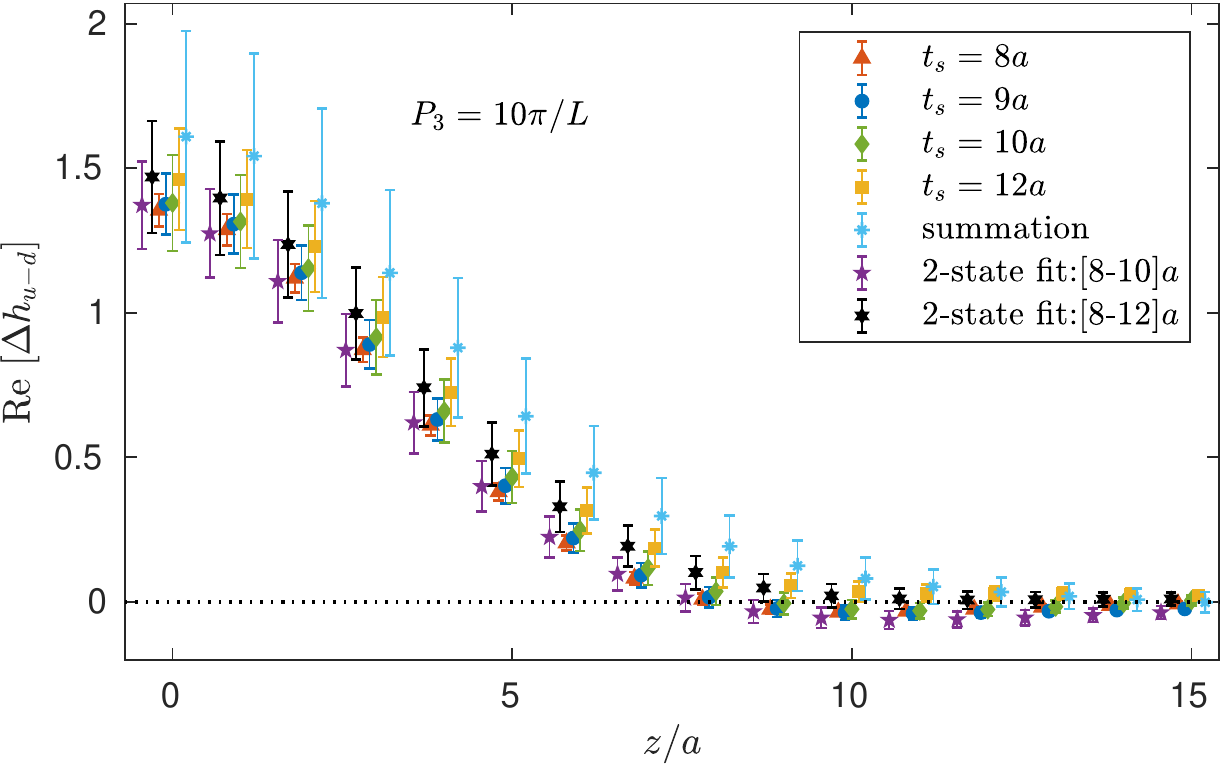}\,\,
   \includegraphics[width=0.49\textwidth]{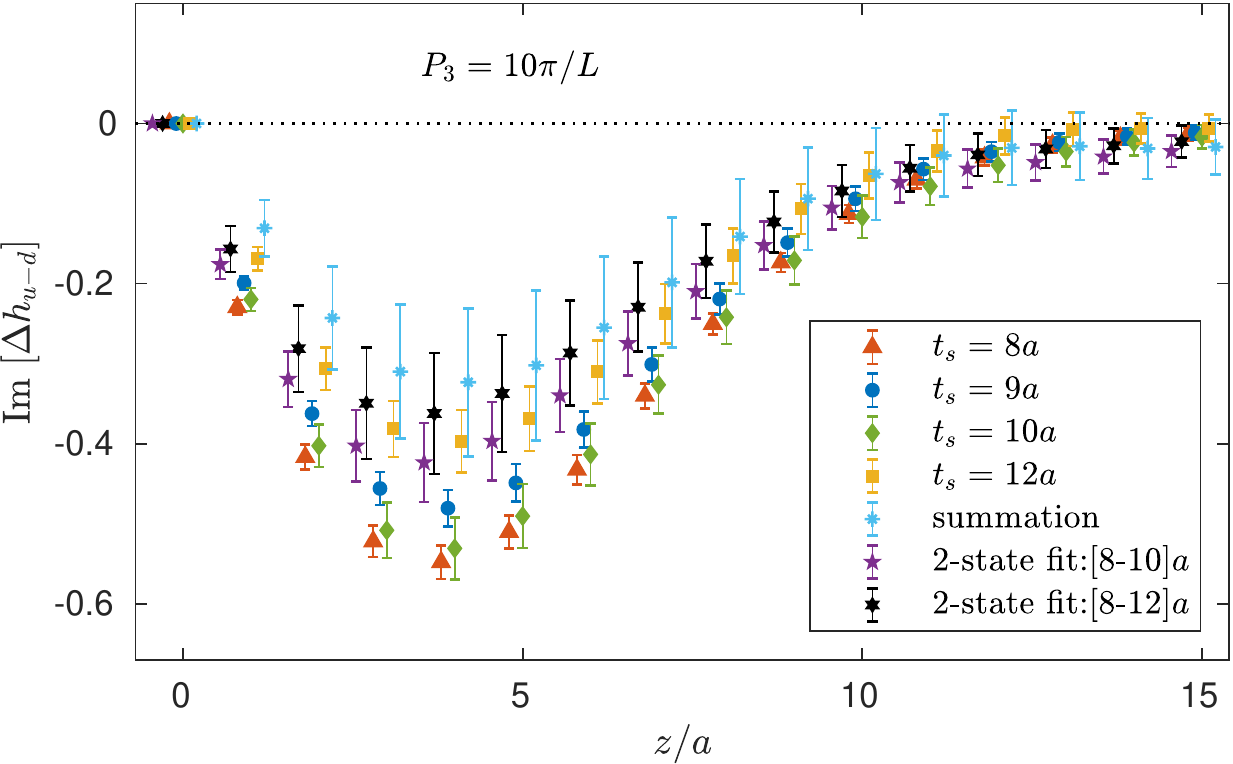}
\end{center}
\vspace*{-0.5cm} 
\caption{Real (left) and imaginary (right) part of the matrix element for the helicity PDF from the plateau method (points labeled with appropriate $t_s$), the summation method (using all $t_s$) and sequential two-state fits (to $t_s{=}8a,9a,10a$ and to all $t_s$). The nucleon is boosted with $10\pi/L\simeq 1.38$ GeV.}
\label{fig:excited2}
\end{figure}
\begin{figure}[h!]
\begin{center}
 \includegraphics[width=0.49\textwidth]{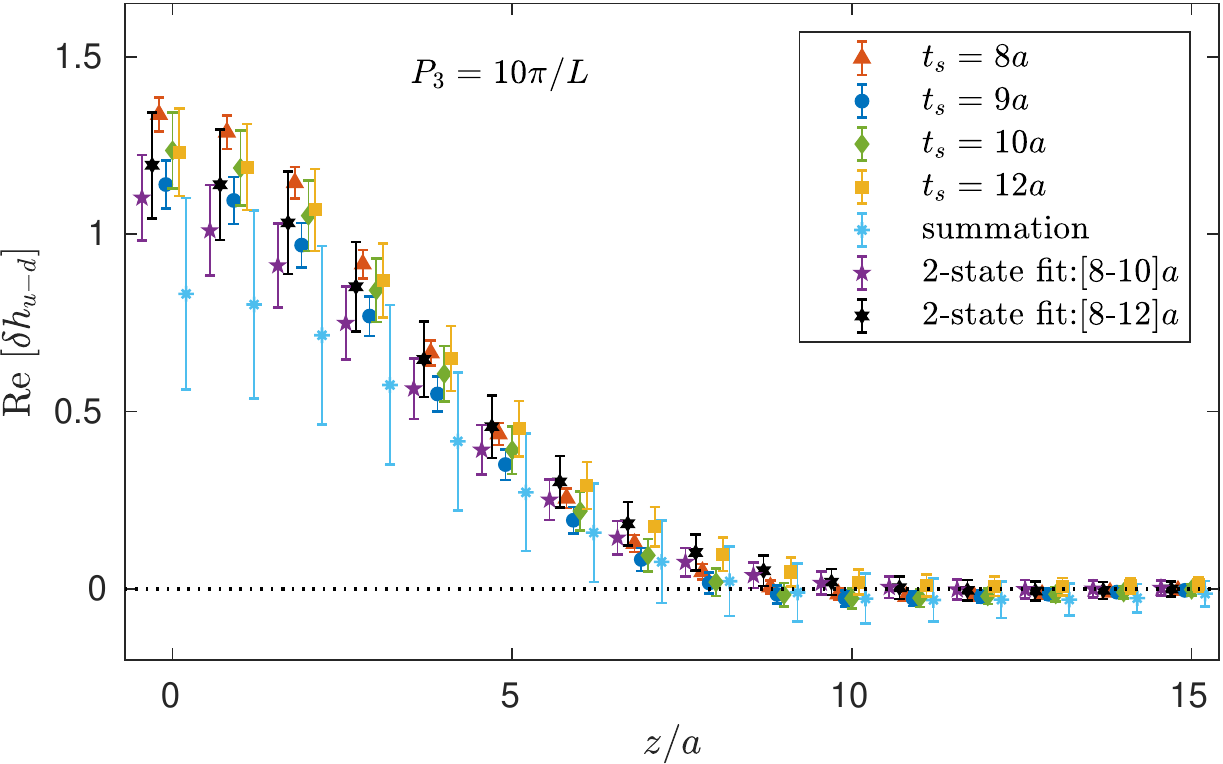}\,\,
   \includegraphics[width=0.491\textwidth]{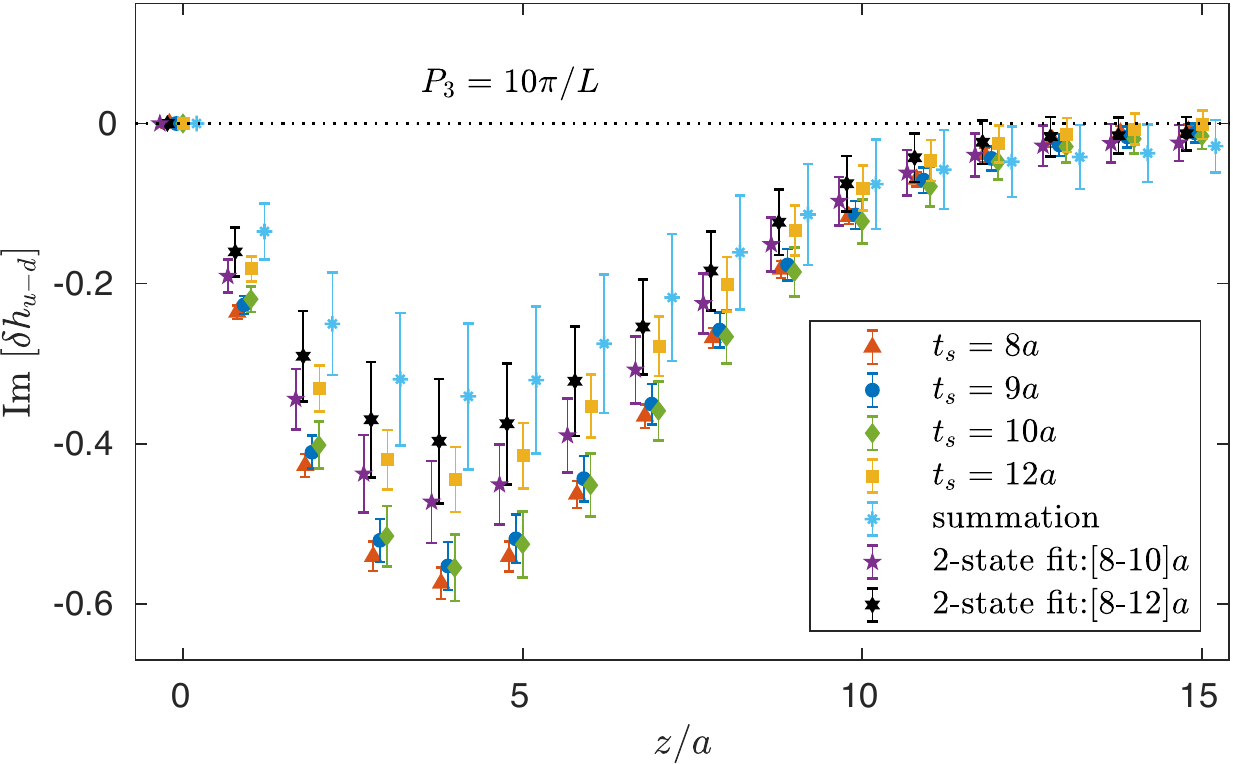}
\end{center}
\vspace*{-0.5cm} 
\caption{Real (left) and imaginary (right) part of the matrix element for the transversity PDF from the plateau method (points labeled with appropriate $t_s$), the summation method (using all $t_s$) and sequential two-state fits (to $t_s{=}8a,9a,10a$ and to all $t_s$). The nucleon is boosted with $10\pi/L\simeq 1.38$ GeV.}
\label{fig:excited3}
\end{figure}
\begin{figure}[h!]
\begin{center}
   \includegraphics[width=0.49\textwidth]{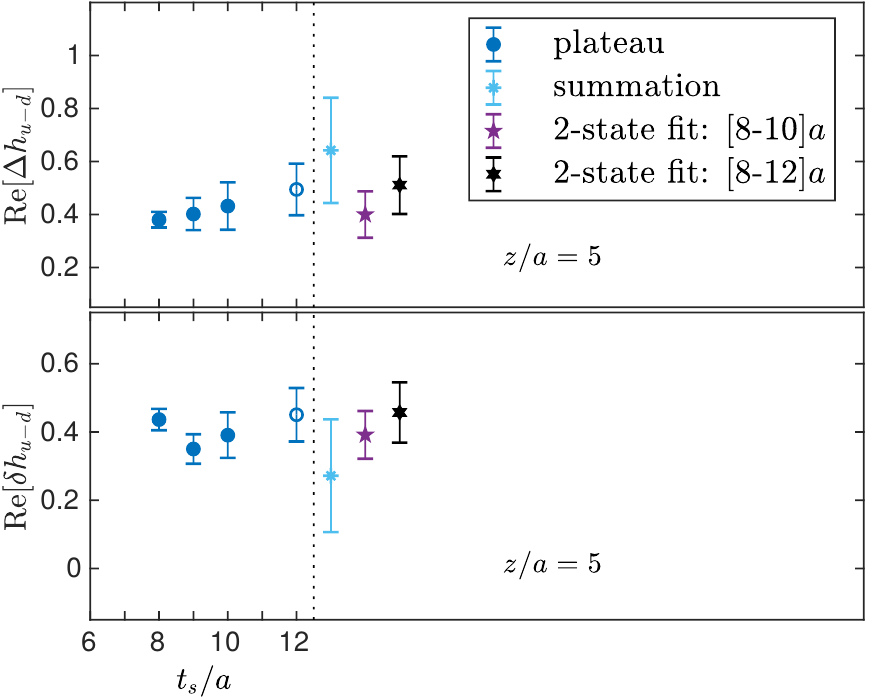}\,\,
   \includegraphics[width=0.49\textwidth]{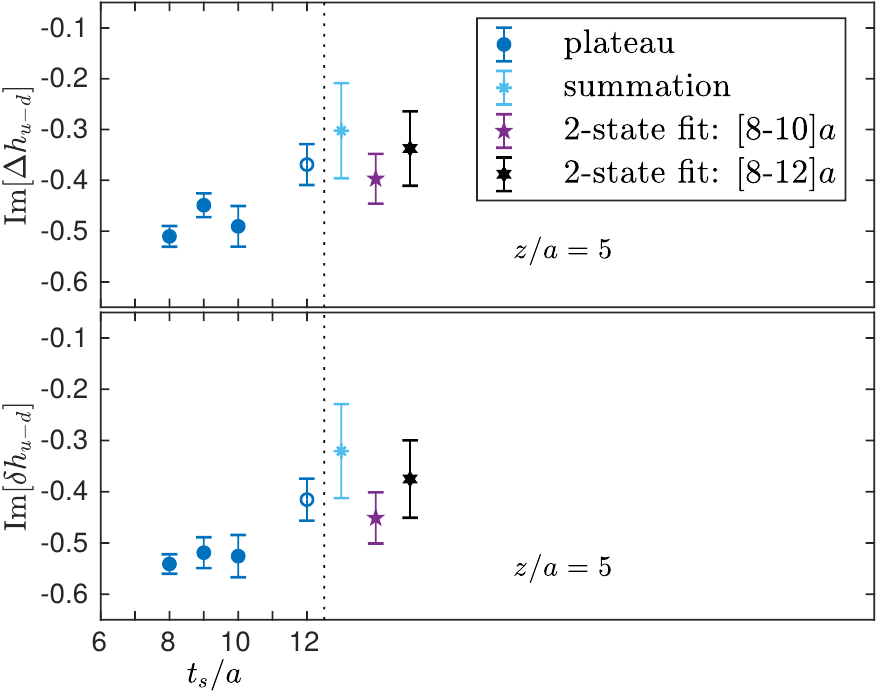}
   \includegraphics[width=0.49\textwidth]{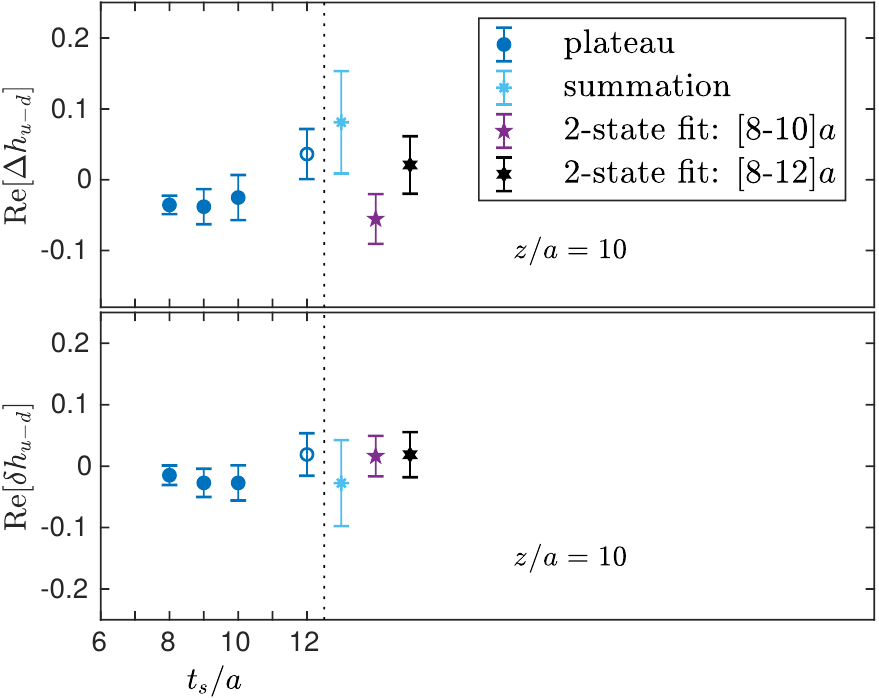}\,\,
   \includegraphics[width=0.49\textwidth]{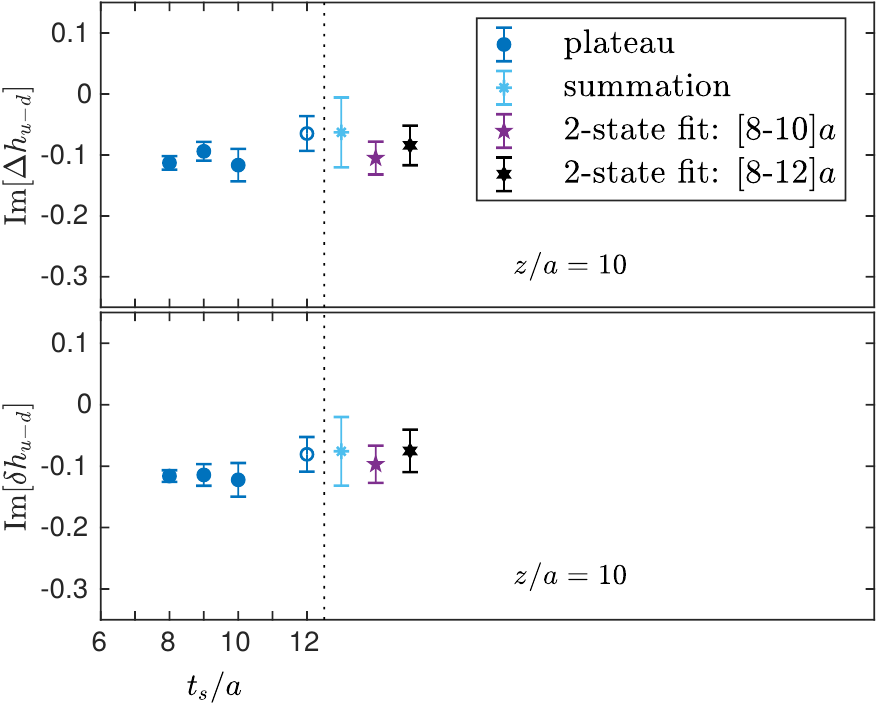}
\end{center}
\vspace*{-0.5cm} 
\caption{Real (left) and imaginary (right) part of the matrix element for the helicity and transversity PDFs at fixed $z/a{=}5$ (upper) and $z/a{=}10$ (lower). We compare results from the plateau method, the summation method (using all $t_s$) and sequential 2-state fits (to $t_s{=}8a,9a,10a$ and to all $t_s$). The final value that we use for extracting PDFs is indicated with an open symbol. The nucleon is boosted with $10\pi/L\simeq 1.38$ GeV.}
\label{fig:pol_z5z10}
\end{figure}

In Figs.~\ref{fig:excited2} and \ref{fig:excited3}, we plot the full $z$-dependence of bare matrix elements for helicity and transversity PDFs, respectively.
Fig.\ \ref{fig:pol_z5z10} displays a zoom for two selected values of $z/a$, i.e.\ $z/a{=}5$ and $z/a{=}10$.
In this case, the fits using the summation method have good quality ($\chi^2/{\rm d.o.f.}{\lesssim}1$) even when including $t_s{=}8a$ and hence, we use all source-sink separations for this approach.
For the real part, the two-state fits and the summation method yield results compatible with plateau fits for $t_s{=}12a$. 
The only observed tensions in the real part are rather small, possibly statistical fluctuations, and are visible for the plateau fits using $t_s/a{=}8,9$ versus the summation method, for intermediate and large $z/a$ in the helicity case.
However, for the imaginary part, we consistently observe a rather strong dependence of the plateau values extracted when using $t_s{=}12a$ as compared to those from $t_s/a{=}8,9,10$, with 2-$\sigma$ to 3-$\sigma$ tensions. 
The latter are also incompatible with the results extracted using two-state fits to all $t_s$ and the summation method, both of which are consistent with the plateau results for  $t_s{=}12a$.
Moreover, the two-state fits using $t_s/a{=}8,9,10$ are incompatible with plateau fits for $t_s{=}10a$ (see right panel of Fig.~\ref{fig:excited2}), which again violates our criterion for consistency, necessitating the use of $t_s{=}12a$.
This behavior reinforces the conclusion reached in the case of the bare matrix elements for the unpolarized quasi-PDF, namely that the tensions observed between results from source-sink separations below 1 fm are indeed manifestations of excited states contamination and ground state dominance is achieved, to around 10\% statistical accuracy, only at $t_s{=}12a$.
Hence, to extract PDFs, we take the plateau results for $t_s{=}12a$ as our preferred ones, since they are more precisely determined as compared to both the results extracted using the two-state and the summation approaches, but show full consistency with them. 
The more severe excited states effects observed in the cases of helicity and transversity are in accordance with observations connected to the extraction of the nucleon axial and tensor charges, where excited states contamination is more significant as compared to the case of the vector operator.
Thus, studies that aim at a higher precision in the determination of quasi-PDF at increasing values of  the nucleon boost would need large statistics to account
for the increased errors resulting from having a nucleon state with large boost, but also for investigating excited states. 

\subsubsection{Remarks and computational costs}
This concludes our investigation of excited states effects.
We emphasize that the spectrum of nucleon excitations is rich, particularly for a boosted nucleon with quarks of physical masses.
Thus, one method of extracting bare matrix elements can be misleading, as the fitted energy gap between the ground state and the explicitly modeled first excited state suggests there are tens of excited states.
In such a situation, excited states need to be suppressed by going to large enough source-sink separations and robust statements can be made only when they are based on compatible results between different methods -- in our case the plateau fits at the largest source-sink separation of around 1.1 fm, the two-state fits and the summation method.
We note that the largest $t_s{=}12a$ is crucially needed to establish this compatibility.
Having only $t_s{=}8a,9a,10a$ in the two-state fits and comparing to plateau fits at $t_s{=}10a$, one would conclude significant tensions between the former and the latter.
This is best illustrated in the imaginary part of bare matrix elements for the helicity quasi-PDF, see the right panels of Fig.~\ref{fig:excited2}.

Finally, we would like to make some remarks on the computational resources that go well beyond what one needs for typical hadron structure calculations.
The quasi-PDFs will reproduce the light-cone PDFs in the limit of large boosts. 
How large the boost should be, needs further investigation.
Within the lattice QCD formulation, as already explained, one cannot increase the momentum of the nucleon to arbitrarily large values. 
The reasons are:
\begin{enumerate}
\item 
As the momentum increases, the signal-to-noise ratio rapidly deteriorates, despite the utilization of special methods, such as smearing techniques that reduce the noise. 
We find that to increase the momentum from $6\pi/L\approx0.83$~GeV to $8\pi/L\approx1.11$~GeV, we need to increase the statistics by a factor of around 3.7 (for unpolarized PDFs) and around 6 (helicity and transversity PDFs) and to increase to $10\pi/L\approx1.38$~GeV by an additional factor of 3.7 or 6, respectively, in order to keep the statistical error approximately the same at $t_s{=}12a$.
\item A careful study of excited states must be carried out and becomes increasingly more difficult as the momentum increases, because of the denser spectrum. 
As the source-sink separation increases, the statistical errors grow exponentially, e.g.\ for $t_s{=}12a$, we needed a factor of about 10-12 (similar for all Dirac structures) more statistics for the same error as for $t_s{=}10a$ (at the largest boost). 
It is imperative to have large enough source-sink separations for at least three values of $t_s$ with comparable errors to perform a reliable analysis of excited states effects and extract the ground state matrix element.
\end{enumerate}
Therefore, in order to reach a nucleon momentum of e.g.\ 2 GeV at $t_s{=}12a$, for unpolarized PDFs, we  estimate that one would need $\mathcal{O}(100)$ million corehours (Mch) on a typical supercomputer, as compared to $\mathcal{O}(5)$ Mch at $P_3\approx1.38$~GeV studied in this work and for momentum 3 GeV, we would need $\mathcal{O}(10000)$ MCh.
For the polarized PDFs, the projected estimate for $P_3\approx3$~GeV reads $\mathcal{O}(10^6)$ Mch.
In addition, one may have to increase the source-sink separation to account for the increased excited states contamination. 
If $t_s{=}14a$ is used, then the computational resources for a boost of 3 GeV would be $\mathcal{O}(10^5)$/$\mathcal{O}(10^7)$ million corehours for the unpolarized/polarized case, which is prohibitively expensive, given the computers available presently.
These requirements may be alleviated with possible development of better algorithms, enhancing the signal-to-noise ratio at large nucleon boosts and large source-sink separations.
Another way to milden the need for huge computational resources is the derivation of two-loop matching and conversion factors, foreseen in the near future.
In this way, quasi-PDFs may be robustly connected to light-cone PDFs already at lower momenta.

We emphasize that a way to go is not to increase the nucleon boost uncontrollably, relying on precise data only at low source-sink separations.
Failing to keep the statistical errors approximately constant as one increases $t_s$ may introduce uncontrolled systematic errors, since the fits will be determined mostly by the more accurate data at small $t_s$.
As we have shown, it is essential that  all source-sink separations have approximately equal errors for a reliable extraction of the matrix elements and this is the criterion we adopt. Unlike other studies~\cite{Chen:2018xof,Lin:2018qky,Liu:2018hxv}, we do not rely solely on two-state fits of data from source-sink separations with widely varied errors, which are thus dominated by the precise data at the small values of $t_s$.
Such an approach can be uncontrolled and lead to a systematic bias in the final results.

\section{Renormalization}
\label{sec:renorm} 

Renormalization is needed in order to relate the bare lattice QCD matrix elements to physical results removing ultra-violet divergences, as well as finite dependence on the lattice action~\footnote{Elimination of residual dependence on the lattice formulation requires continuum extrapolation.}. 
In the case of quasi-PDFs, one needs to also eliminate additional divergences arising due to the utilization of operators with a finite Wilson line.
Renormalization of  Wilson loops has been addressed long time ago using dimensional regularization (DR) for smooth contours~\cite{Dotsenko:1979wR}, as well as for contours containing singular points~\cite{Brandt:1981kf}. 
Based on arguments valid to all orders in perturbation theory, it was demonstrated that smooth Wilson loops in DR are finite functions of the renormalized coupling, while the presence of cusps and self-intersections introduces logarithmically divergent multiplicative renormalization factors, referred to as $Z$-factors. 
More importantly, it was shown that other regularization schemes are expected to lead to further $Z$-factors, which are power-law divergent with respect to the dimensionful ultraviolet cutoff. 
This also appears in the lattice formulation, where a divergence arises as a function of the lattice spacing, that increases exponentially with the length of the Wilson line as  ${\sim}\,e^{{z}/{a}}$. Such a divergence must be removed  prior to the extrapolation to the continuum limit. 
We describe in this section our renormalization program that includes the removal of the exponential divergence. A recent work on the perturbative renormalization of  Wilson-line fermion operators of the type given in Eq.~(\ref{Op}) has identified the mixing pattern among non-local straight Wilson-line operators, and led to the development of an appropriate renormalization prescription for both the multiplicative renormalization and the mixing coefficient~\cite{Constantinou:2017sej}. The non-perturbative renormalization program that we developed is a generalization of the RI$^\prime$-scheme~\cite{Martinelli:1994ty}, appropriate for operators incluing a Wilson line~\cite{Alexandrou:2017huk}~\footnote{For alternative approaches, using, e.g. an auxiliary field method~\cite{Green:2017xeu}, see the recent review of Ref.~\cite{Cichy:2018mum}.}. The $Z$-factors are extracted by imposing the following conditions
\bea
\label{renorm}
\frac{Z^{\RI}_{\cal O}(z,\mu_0,m_\pi)}{Z^{\RI}_q(\mu_0,m_\pi)}\frac{1}{12} {\rm Tr} \left[{\cal V}(z,p,m_\pi) \left({\cal V}^{\rm Born}(z,p)\right)^{-1}\right] \Bigr|_{p^2{=}\mu_0^2} {=} 1\, ,\\
Z^{\RI}_q(\mu_0,m_\pi) \frac{1}{12} {\rm Tr} \left[(S(p,m_\pi))^{-1}\, S^{\rm Born}(p)\right] \Bigr|_{p^2=\mu_0^2}  = 1 \,. \hspace*{1.4cm}
\eea
We use the general notation $Z_{\cal O}$ and consider ${\cal O}{=}V_0,\,A,\,T$ corresponding  to the unpolarized, helicity and transversity operators, respectively. $Z_{\cal O}$ and $Z_q$ are the renormalization functions of the operator and the quark field, respectively. Both $Z_{\cal O}$ and $Z_q$ are scheme and scale dependent, and are expected to have some dependence on the pion mass. Also, $Z_{\cal O}$ is a function of the length of the Wilson line, $z$. ${\cal V}$ is the amputated vertex function of the operator and $S$ the fermion propagator, while ${\cal V}^{{\rm Born}}$ and $S^{{\rm Born}}$ are the corresponding tree-level values. Note that this condition is applied independently for each value of $z$. The RI$^\prime$ renormalization scale, $\mu_0$, is chosen to be democratic in the spatial directions, that is $a\mu_0{=}\frac{2\pi}{L_s}(n_t+\frac{\pi}{2},n,n,n)$, which minimizes the ratio $P4{\equiv}{\sum_i \mu_i^4}/{(\sum_i \mu_i^2)^2}$ and suppresses discretization effects~\cite{Constantinou:2010gr,Alexandrou:2015sea}. 

For the calculation of renormalization functions, we employ the momentum source method~\cite{Gockeler:1998ye,Alexandrou:2015sea} that has the advantage of yielding results of high statistical accuracy. This method requires new inversions for each momentum used, but significant reduction in the gauge noise is observed, which by far outweighs the additional computational cost. Data are produced using the three $N_f{=}2$ ensembles given in Table~\ref{Table:Z_ensembles} that have a different value of the pion mass. The twisted mass parameters of the light quarks in the sea and valence sectors have been set equal (isospin limit and unitary setup). Even though the ensemble simulated with the smallest value of the pion mass has a larger volume, the $Z$-factors are obtained at the same values of $(a\mu_0)^2$ and $P4$. In the current work, we improve our previous analysis on the $Z$-factors of one-derivative operators~\cite{Alexandrou:2017huk} by performing:
\begin{itemize}
\item[$\bullet$] a chiral extrapolation using the three ensembles at different value of the pion mass;
\item[$\bullet$] a fit on the chiral data to eliminate residual dependence on the RI$^\prime$ scale. We use several values of the RI$^\prime$ scale that cover the range $(a\,\mu_0)^2 \in [1 - 4]$. An extensive study on the choice of the renormalization scale and the corresponding systematic uncertainties can be found in Ref.~\cite{Alexandrou:2017huk}. 
\end{itemize}
\begin{table}[h]
\begin{center}
\renewcommand{\arraystretch}{1.5}
\renewcommand{\tabcolsep}{5.5pt}
\begin{tabular}{ccc}
\hline\hline 
$\beta=2.10$, & $c_{\rm SW} = 1.57751$, & $a=0.0938(3)(2)$~fm \\
\hline\hline\\[-3ex]
{$48^3\times 96$}  & {$\,\,a\mu = 0.0009$}  & $\,\,m_\pi = 130$~MeV     \\
\hline
{$24^3\times 48$}  & $\,\,a\mu = 0.003$     & $\,\,m_\pi = 235$~MeV     \\
\hline
{$24^3\times 48$}  & $\,\,a\mu = 0.006$    & $\,\,m_\pi = 340$~MeV    \\
\hline\hline
\end{tabular}
\vspace*{-0.25cm}
\begin{center}
\caption{\small{Parameters of the ensembles used to compute the Z-factors extracted in this work. For details, see Ref.~\cite{Alexandrou:2017xwd}.}}
\label{Table:Z_ensembles}
\end{center}
\end{center}
\vspace*{-0.2cm}
\end{table} 

\subsection{Pion mass dependence}

The pion mass dependence for the $Z$-factors for operators with a finite Wilson line has never been explored, and is expected that small values of $z$ will have weak dependence on $m_\pi$, as observed in the case of local operators computed within the same setup. However, it is not known how the $Z$-factors will behave when the length of the Wilson line is large. To obtain the $Z$-factors in the chiral limit, we fit the data from the three ensembles at each value of $z$. This fit is applied to the real and imaginary parts independently. The data for $Z^{\RI}_{\cal O}$ are expected to have a quadratic dependence on the pion mass (equivalently linear with respect to the twisted mass parameter) and are fitted using
\be
\label{eq:Zchiral_fit}
Z^{\RI}_{\cal O}(z,\mu_0,m_\pi) = {Z}^{\RI}_{{\cal O},0}(z,\mu_0) + m_\pi^2 \,{Z}^{\RI}_{{\cal O},1}(z,\mu_0) \,.
\ee
The chirally extrapolated value is given by the fit parameter ${Z}^{\RI}_{{\cal O},0}(z,\mu_0)$. Since the fitted data are obtained on different ensembles, we use the super-jackknife method (see, e.g., Ref.~\cite{AliKhan:2001xoi}) to correctly calculate the statistical error. This method is applicable to both correlated and uncorrelated data.
\begin{figure}[h!]
\includegraphics[scale=.89]{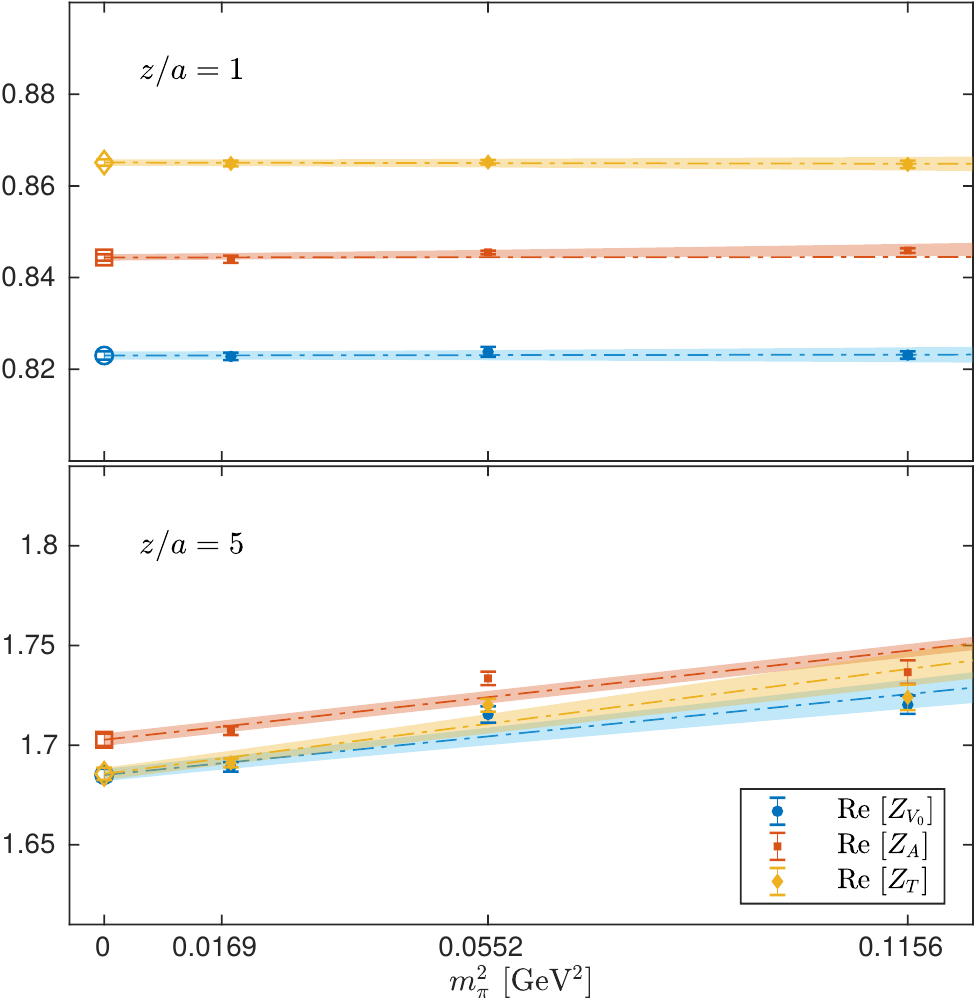}
\includegraphics[scale=.89]{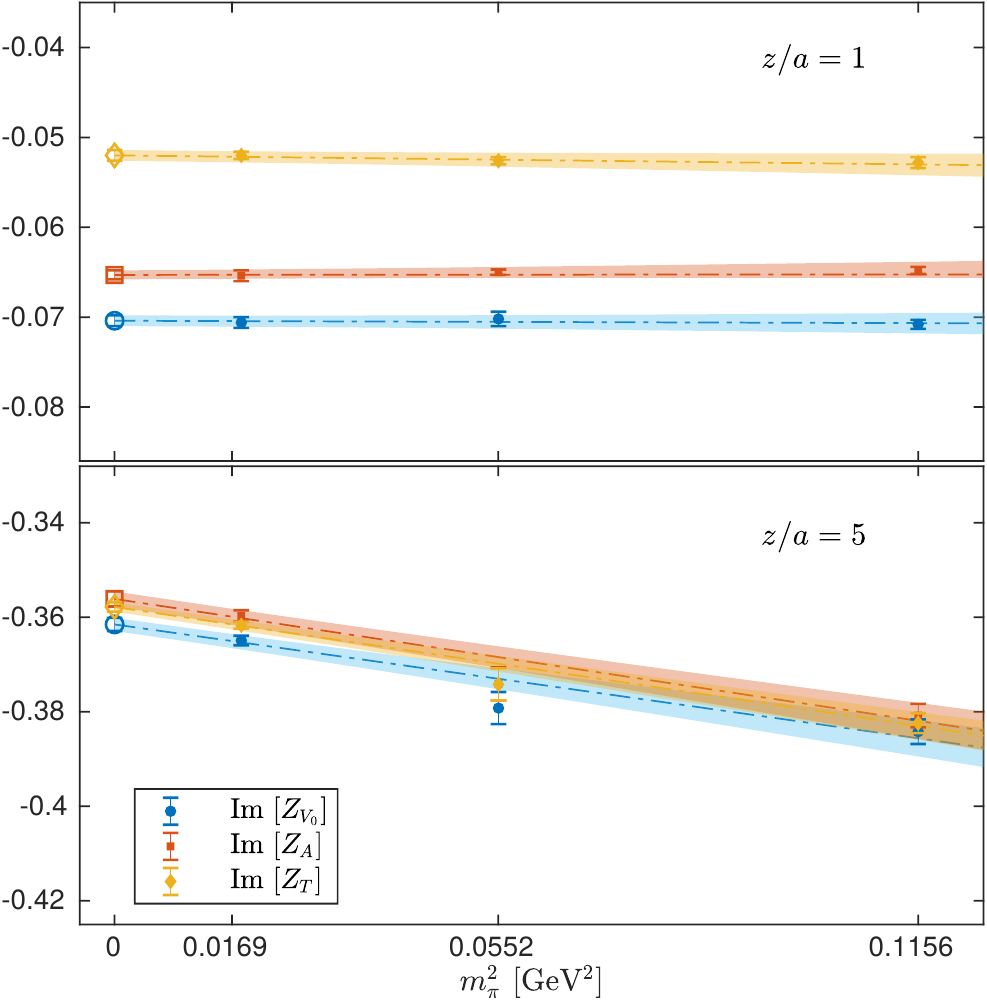}
\vspace*{-0.5cm}
\caption{Real (left) and imaginary (right) parts of the $Z$-factor $Z^{\RI}_{\cal O}(z,\mu_0,m_\pi)$  for the unpolarized (blue circles), helicity (red squares) and transversity (orange diamonds) PDFs as a function of $m_\pi^2$ of the ensemble used. The scale is $a\,\mu_0{=}2.5$. The upper and lower panels show the results for $z/a{=}1$ and $z/a{=}5$, respectively. The dashed line corresponds to the chiral fit of Eq.~(\ref{eq:Zchiral_fit}), and the open symbols are the extrapolated values, ${Z}^{\RI}_{{\cal O},0}(z,\mu_0)$.}
\label{fig:Zfactors_pion_mass_dependence}
\end{figure}

In  Fig.~\ref{fig:Zfactors_pion_mass_dependence}, we show the pion mass dependence of the real and imaginary parts of the $Z$-factors for all three operators.  For a clear presentation, we focus on $z/a{=}1,\,5$ and we plot against $m_\pi^2$ for the scale $a\,\mu_0{=}2.5$. We find that the dependence is almost constant in $m_\pi^2$ and the chiral fit yields a slope consistent with zero for both the real and imaginary parts for small values of $z$ (see, e.g., $z/a{=}1$). As $z$ increases, a non-zero slope is observed, with the dependence being  linear, as expected. The dashed line corresponds to the chiral fit of Eq.~(\ref{eq:Zchiral_fit}), and the open symbols are the extrapolated values ${Z}^{\RI}_{{\cal O},0}(z,\mu_0)$.

\subsection{Volume effects}

\begin{figure}[h!]
\includegraphics[scale=.775]{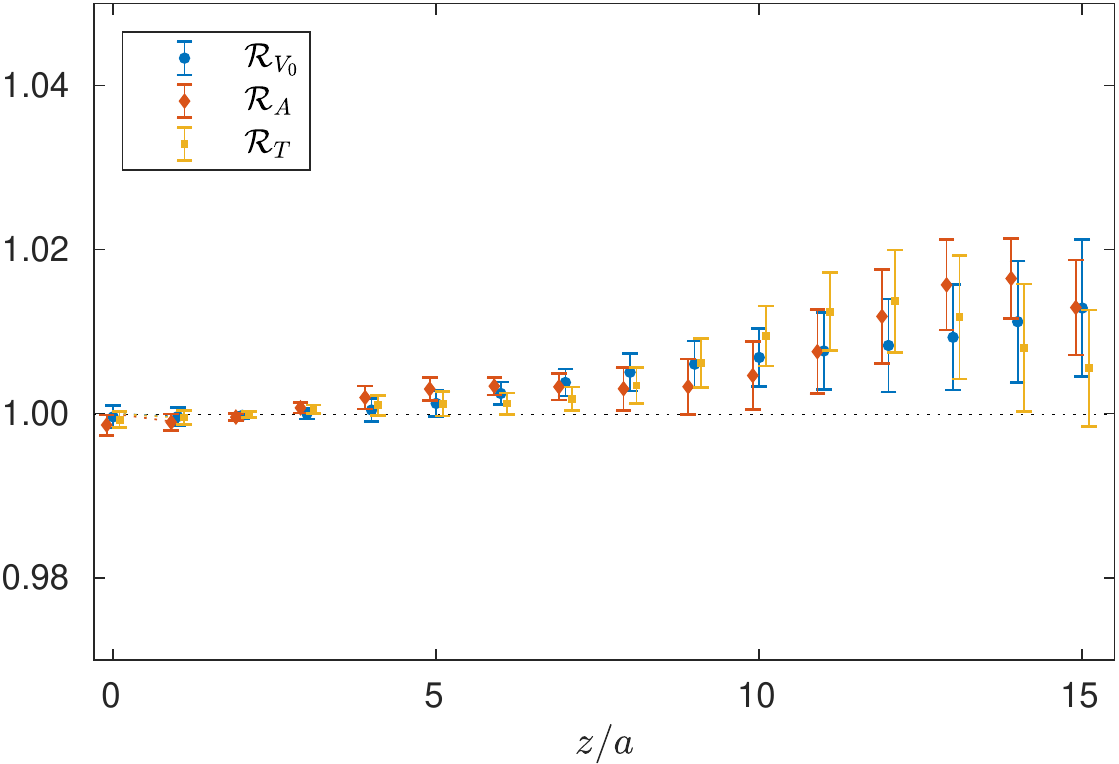}\,\,
\includegraphics[scale=.775]{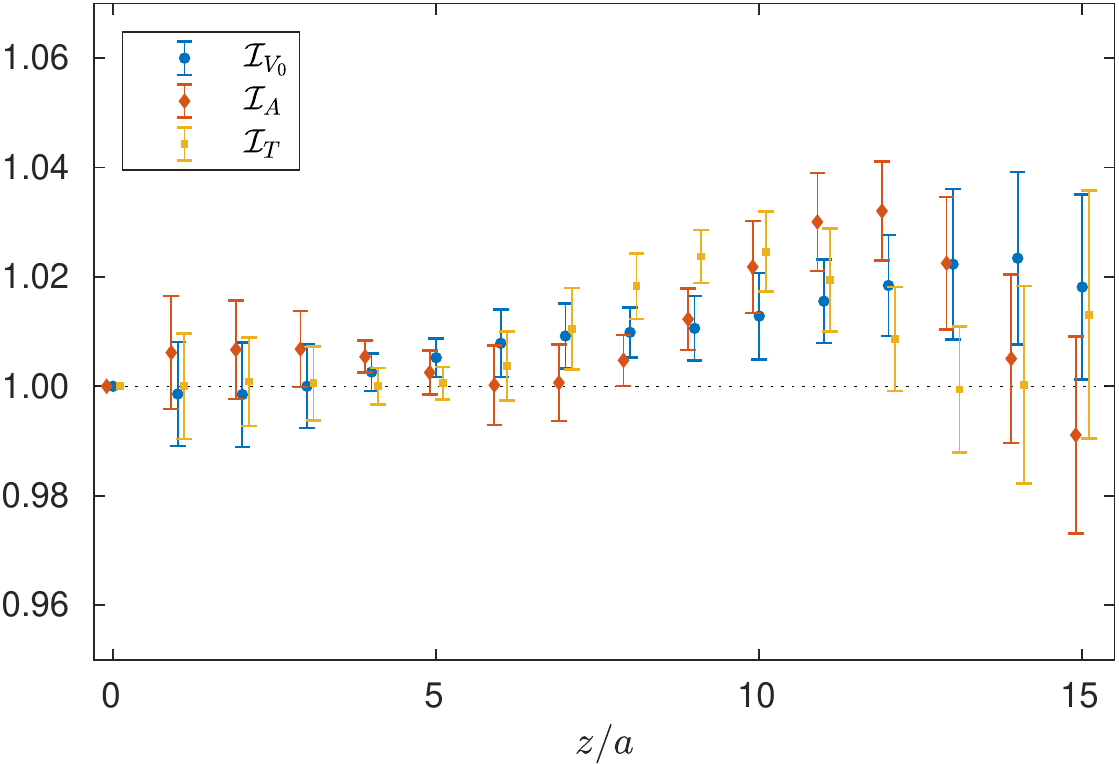}
\vspace*{-0.15cm}
\caption{${\cal R}_{\cal O}(z)$ and ${\cal I}_{\cal O}(z)$ to study volume effects between the ensembles $48^3{\times}96$ and $64^3{\times}128$, for the unpolarized (circles), helicity (diamonds) and transversity (squares) operators as a function of the length of the Wilson line, $z/a$. The RI$'$ scale is $(a\,\mu_0)^2{=}2.5$ and the pion mass is 130 MeV.}\label{fig:Zfactors_volume}
\end{figure}

Another source of systematic uncertainty entering the determination of matrix elements is due to the finite lattice extent. Finite volume effects are expected to be suppressed as $\exp(-m_\pi L)$ and based on empirical studies, $m_\pi L\geq4$ is considered sufficient in most applications. However, most lattice QCD  studies deal with matrix elements of local operators, and as discussed in Ref.~\cite{Briceno:2018lfj}, the length of the Wilson line entering the operator may enhance finite volume effects. To date, this systematic uncertainty has not been investigated due to the absence of lattice QCD computations using ensembles with different volumes, keeping the rest of the parameters constant. Such a study requires significant computational resources and is one of our future  goals. Volume effects in the renormalization functions can be easily examined due to the reduced computational resources compared to the matrix elements. Therefore, we compute the $Z$-factors using the $48^3{\times}96$ ensemble of Table~\ref{Table:Z_ensembles} and a $64^3{\times}128$ ensemble with the same pion mass. In Fig.~\ref{fig:Zfactors_volume}, we plot the ratios
\be
{\cal R}_{\cal O}(z) \equiv \frac{{\rm Re}[Z^{\RI}_{{\cal O},64}(z,\mu_0,m_\pi)]}{{\rm Re}[Z^{\RI}_{{\cal O},48}(z,\mu_0,m_\pi)]}\,,\qquad 
{\cal I}_{\cal O}(z) \equiv \frac{{\rm Im}[Z^{\RI}_{{\cal O},64}(z,\mu_0,m_\pi)]}{{\rm Im}[Z^{\RI}_{{\cal O},48} {\cal O}(z,\mu_0,m_\pi)]}\,,
\ee
as a function of the length of the Wilson line, for all three types of operators. The ratios are defined at the pion mass $m_\pi{=}$130~MeV and a renormalization scale $(a\,\mu_0)^2{=}2.5$. The additional subscript 64 and 48 indicates the lattice volume.  We find that both the real and imaginary  parts do not show a statistically significant dependence  on the volume, as the ratios take a maximum value of 1.02 and 1.03, respectively, for $z/a$ up to 15, which is well within the range of interest. In addition, we find an almost linear increase of ${\cal R}_{\cal O}(z)$, while ${\cal I}_{\cal O}(z)$  has an oscillatory dependence on the volume.

Hence, we conclude that volume effects in the $Z$-factors are small and have little effect in the renormalized matrix elements. This partly results from the fact that the bare matrix elements go to zero in the large-$z$ region. In determining the  final values for the $Z$-factors, we do not include the $64^3{\times}128$ ensemble, because the ratio $P4$ is large for most of the scales $(a\mu_0)^2 \in [1{-}4]$ compared to the smaller volumes, leading to contamination from finite-$a$ effects, as seen in Ref.~\cite{Alexandrou:2015sea}. However, for $(a\,\mu_0)^2{=}2.5$ which was used in the comparison, $P4$ is the same as the one obtained from the smaller volumes, which allows one to isolate the volume effects. Finite-$a$ effects are in fact under investigation for this class of non-local operators in lattice perturbation theory~\cite{MC_HP_artifacts}.

\subsection{Conversion to the standard and the modified $\msbar$ scheme}
\label{subsec:MSconversion}

In order to  compare renormalized lattice QCD matrix elements  with phenomenological results extracted from global analyses, the $Z$-factors must be converted to the same scheme and evolved to the same  scale as those used  in the phenomenological analyses. Traditionally, the chosen scheme is the $\overline{\rm MS}$ and the scale $\bar\mu$ is typically set to 2 GeV. The appropriate conversion factors for the non-local operators with a straight Wilson line are taken from  Ref.~\cite{Constantinou:2017sej}, where a calculation was carried out to one-loop level in perturbation theory, using dimensional regularization. Technical complications related to the non-locality of the operators under study make it very hard to extend such a calculation to higher loops, as done for local operators, usually known to  three and four loops. As a consequence, it is expected that the $Z$-factors will have a residual dependence on the initial RI$^\prime$ scale $\mu_0$. Thus, one typically computes the $Z$-factors at several values of the RI$^\prime$ scale, as set by  Eq.~(\ref{renorm}), and then uses an appropriate conversion to bring each ${Z}^{\RI}_{{\cal O},0}(z,\mu_0)$ to  ${Z}^{\MSb}_{{\cal O},0}(z,2{\rm GeV})$. The remaining dependence on $\mu_0$ can be studied with a linear fit in $(a\,\mu_0)^2$, which is the leading order of a more complicated scale dependence. The final values are obtained by taking  $(a\,\mu_0)^2\to0$ using
\be
\label{eq:Z_mu_fit}
{Z}^{\MSb}_{\cal O}(z,\bar\mu,\mu_0) = {\cal Z}^{\MSb}_{{\cal O},0}(z,\bar\mu) + (a\,\mu_0)^2 \,{\cal Z}^{\MSb}_{{\cal O},1}(z,\bar\mu) \,.
\ee
For ${Z}^{\MSb}_{\cal O}(z,\bar\mu,\mu_0)$, we use the chirally extrapolated $Z$-factors converted to $\msbar$ at the scale $\bar\mu{=}$2 GeV. For simplicity, in the notation we dropped the subscript ``0'' appearing in the fit of Eq.~(\ref{eq:Zchiral_fit}). The desired quantity is the fit parameter ${\cal Z}^{\MSb}_{{\cal O},0}(z,\bar\mu)$. 

Our renormalization program entails a new element, namely the use of a modified $\msbar$ scheme ($\mmsbar$). The development of such a scheme was motivated by the fact that the existing matching formulae do not satisfy particle number conservation (e.g.\ Ref.~\cite{Izubuchi:2018srq}). The matching using the $\mmsbar$ scheme was already presented in our recent work~\cite{Alexandrou:2018pbm,Alexandrou:2018eet}.  Here we complement the previous analyses by giving the appropriate conversion of the $Z$-factors to the $\mmsbar$ scheme, instead of $\msbar$. We find that the resulting modification is numerically very small, but moves the final values of the PDFs towards the phenomenological ones. Details on the extraction of the $\mmsbar$ are given in Sec.~\ref{sec:matching}. In a nutshell, an additional conversion factor is needed to bring ${\cal Z}^{\MSb}_{{\cal O},0}(z,\bar\mu) $ to ${\cal Z}^{\MMSb}_{{\cal O},0}(z,\bar\mu)$ via
\be
\label{eq:ZMMS}
\displaystyle {\cal Z}^{\MMSb}_{{\cal O},0}(z,\bar\mu) = {\cal Z}^{\MSb}_{{\cal O},0}(z,\bar\mu)\,  {\cal C}^{\MSb,{\rm M\overline{MS}}}\,,
\ee
which has been computed perturbatively in dimensional regularization to one-loop level and is presented in the next section. The expression for the conversion ${\cal C}_{\cal O}^{\MSb, {\rm M\overline{MS}}}$ is different for each operator under study and its general form is given by:
\begin{eqnarray}
\label{eq:CMStoMMS}
\hspace*{-0.45cm}
{\cal C}_{\cal O}^{\overline{\rm MS}, {\rm M\overline{MS}}} {=} 1+\frac{C_F g^2}{16\pi^2} \Bigg[
&{\,}&\hspace*{-0.15cm} e^{(1)}_{\cal O} +  e^{(2)}_{\cal O} \ln \left(\frac{\bar\mu ^2}{4 \mu_F^2}\right)  
+  e^{(3)}_{\cal O}   \left(\frac{i \pi  \left| \mu_F z\right| }{2 \mu_F z}-\ln (\left|
   \mu_F z\right| )-\text{Ci}(\mu_F z)-i \text{Si}(\mu_F z)+\ln
   (\mu_F z)\right)\nonumber \\
 &{+}& \hspace*{-0.15cm}  e^{(4)}_{\cal O}  \left(e^{i \mu_F z} (2 \text{Ei}(-i \mu_F
   z)+i \pi  \text{sgn}(\mu_F z)-\ln (-i \mu_F z)+\ln (i \mu_F z))\right)\Bigg]\,,
\end{eqnarray}
where $\mu_F$ is the factorization scale that is taken equal to the $\msbar$ scale, that is, $\mu_F{=}\bar\mu{=}2$ GeV. The expression also contains the special functions ${\rm Ci}$ (cosine integral), ${\rm Si}$ (sine integral) and ${\rm Ei}$ (exponential integral), as well as the sign function (${\rm sgn}$). The coefficients $e^{(i)}_{\cal O}$ depend on the operator and their numerical values are given in Table~\ref{Table:e_coeff}.

\begin{table}[h]
\begin{center}
\renewcommand{\arraystretch}{1.5}
\renewcommand{\tabcolsep}{5.5pt}
\begin{tabular}{ccccc}
\hline
${\cal O}$ & $e_{\cal O}^{(1)}$ & $e_{\cal O}^{(2)}$ & $e_{\cal O}^{(3)}$ & $e_{\cal O}^{(4)}$   \\
\hline\\[-3ex]
{$V$}  & $-$5  & $-$3 & +3 & $-$3/2   \\
\hline
{$A$}     &  $-$7  & $-$3 & +3 & $-$3/2 \\
\hline
{$T$}     & $-$4  & $-$4 & +4 & $-$4/2 \\
\hline\hline
\end{tabular}
\vspace*{-0.25cm}
\begin{center}
\caption{\small{Numerical values for the coefficients $e^{(i)}_{\cal O}$ appearing in the conversion function from $\msbar$ to $\mmsbar$ in Eq.~(\ref{eq:CMStoMMS}).}}
\label{Table:e_coeff}
\end{center}
\end{center}
\vspace*{-0.2cm}
\end{table} 

\begin{figure}[h!]
\includegraphics[scale=.885]{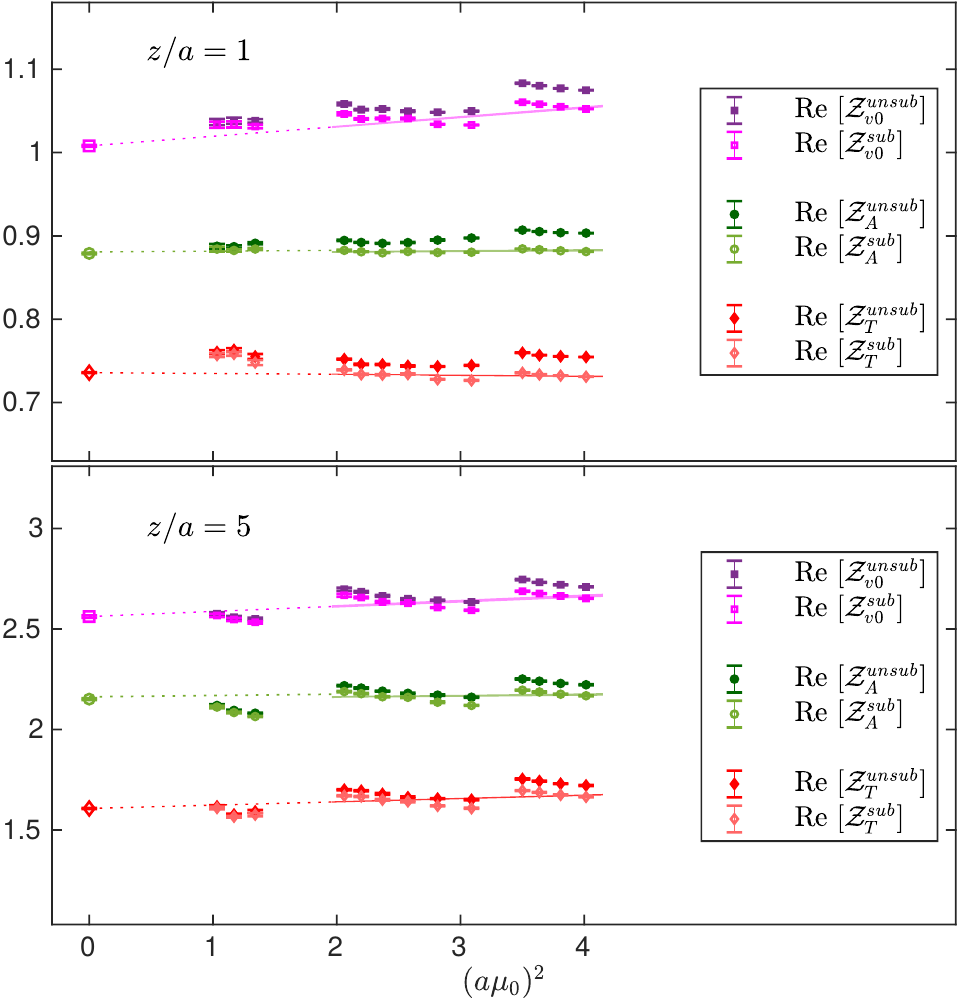}
\includegraphics[scale=.885]{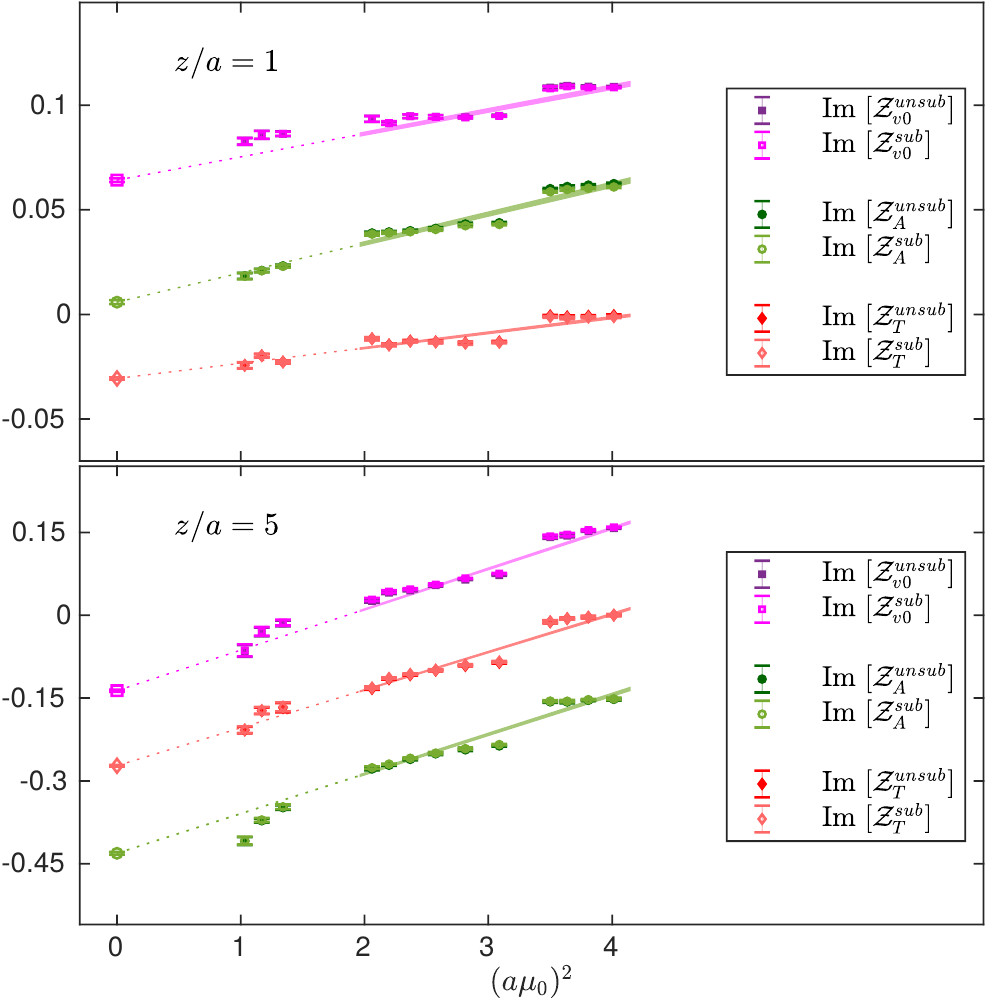}
\vspace*{-0.15cm}
\caption{Real (left) and imaginary (right) part of ${Z}^{\MMSb}_{\cal O}(z,\bar\mu,\mu_0)$ for the unpolarized (squares), helicity (circles) and transversity (diamonds) operators as a function of the initial RI$'$ scale. The upper (lower) panel corresponds to $z/a{=}1$ ($z/a{=}5$). The dashed line corresponds to the fit of Eq.~(\ref{eq:Z_mu_fit}), and the open symbols are the extrapolated values  ${\cal Z}^{\MMSb}_{{\cal O},0}$. We show results for all $Z$-factors with $Z_{V_0}$ and $Z_T$ shifted vertically by for clarity in the presentation. The filled symbols with dark color correspond to the purely non-perturbative estimates, while the corresponding lighter color open symbols have been partly improved using the perturbative results of the finite lattice spacing effects to ${\cal O}(g^2\,a^\infty)$ computed in Ref.~\cite{Alexandrou:2015sea} on the same ensembles.}
\label{fig:Zfactors_RI_scale_dependence}
\end{figure}
In Fig.~\ref{fig:Zfactors_RI_scale_dependence}, we show ${\cal Z}^{\MMSb}_{{\cal O},0}(z,\bar\mu)$ after multiplying  ${\cal Z}^{\MSb}_{{\cal O},0}(z,\bar\mu)$ with the conversion factor given in Eq.~(\ref{eq:ZMMS}). The data are shown  against $(a\,\mu_0)^2$ to demonstrate the dependence of the $Z$-factors on the initial RI$^\prime$ scale. As done for the previous figures, we choose representative values of the length of the Wilson line, namely $z/a{=}1$ and $z/a{=}5$. We find that the imaginary part  has a stronger dependence on the $\mu_0$ value. Note, however, that the imaginary part is an order of magnitude smaller than the real part  of the matrix element and thus this dependence is still suppressed in the total matrix element, especially  in the region $(a\,\mu_0)^2 \in [1-2]$, which is an indication of non-perturbative effects. Such behavior is also observed in local operators for $(a\,\mu_0)^2{<}1$. This tendency seems to affect the region $(a\,\mu_0)^2{<}2$ for non-local operators with $z/a{\ge}5$. Since we are using perturbative expressions for the conversion to the $\mmsbar$ scheme, it is important to choose a region where perturbation theory is valid, and thus, we choose the values obtained from $(a\,\mu_0)^2 \in [2-4]$ for both the real and imaginary parts. There is  a systematic uncertainty attached to the $Z$-factor due to the choice of the fit range, and here we used the ranges $(a\,\mu_0)^2 \in [1-3], [1-4], [2-3], [2-4]$ to estimate the systematic uncertainty. Even though the real and imaginary parts of the $Z$-factors do have mild dependence on the range, the final values are consistent. Therefore, we do not give any systematic uncertainty from the $(a\,\mu_0)^2{\to}0$ extrapolation.
In the same figure, we also show the improvement of the non-perturbative results when subtracting  ${\cal O}(g^2\,a^\infty)$ effects computed perturbatively in Ref.~\cite{Alexandrou:2015sea} on the same ensembles. They  are obtained by replacing Eq.~(\ref{renorm}) with
\be
\label{renorm_sub}
\frac{Z^{\RI}_{\cal O}(z,\mu_0,m_\pi)}{Z^{\RI}_q(\mu_0,m_\pi) - A^\infty_q(\mu_0)}\frac{1}{12} {\rm Tr} \left[{\cal V}(z,p,m_\pi) \left({\cal V}^{\rm Born}(z,p)\right)^{-1}\right] \Bigr|_{p^2{=}\mu_0^2} {=} 1\,,
\ee
where the denominator has been modified by subtracting the ${\cal O}(g^2\,a^\infty)$ artifacts in $Z_q$, denoted by $A^\infty_q(\mu_0)$. As expected, the differences between the subtracted results of Eq.~(\ref{renorm_sub}) and the unsubtracted results of Eq.~(\ref{renorm}) are small, and mostly affect the real part of the $Z$-factors. In addition, subtraction of the artifacts in $Z_q$ is not sufficient to eliminate the dependence on $a\,\mu_0$. For the later to be achieved, subtraction of the artifacts in ${\cal V}(z,p,m_\pi)$ is crucial, as has been discussed in Ref.~\cite{Alexandrou:2015sea} for the local operators. Such a computation for the non-local operators is not only more complicated technically, but requires the addition of stout smearing in the links of the operator resulting in lengthy expressions. This calculation is under way and will be presented in a separate publication~\cite{MC_HP_artifacts}.

\begin{figure}[h]
\centering
\includegraphics[scale=0.75]{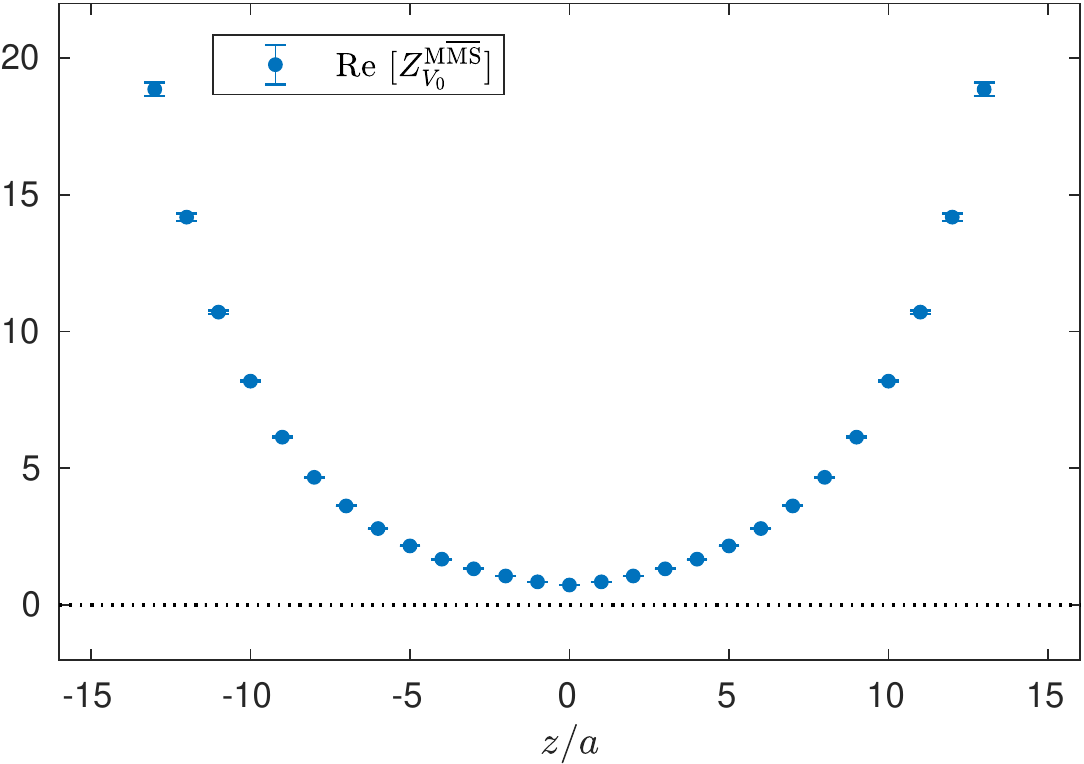} \hspace*{0.3cm} \includegraphics[scale=0.71]{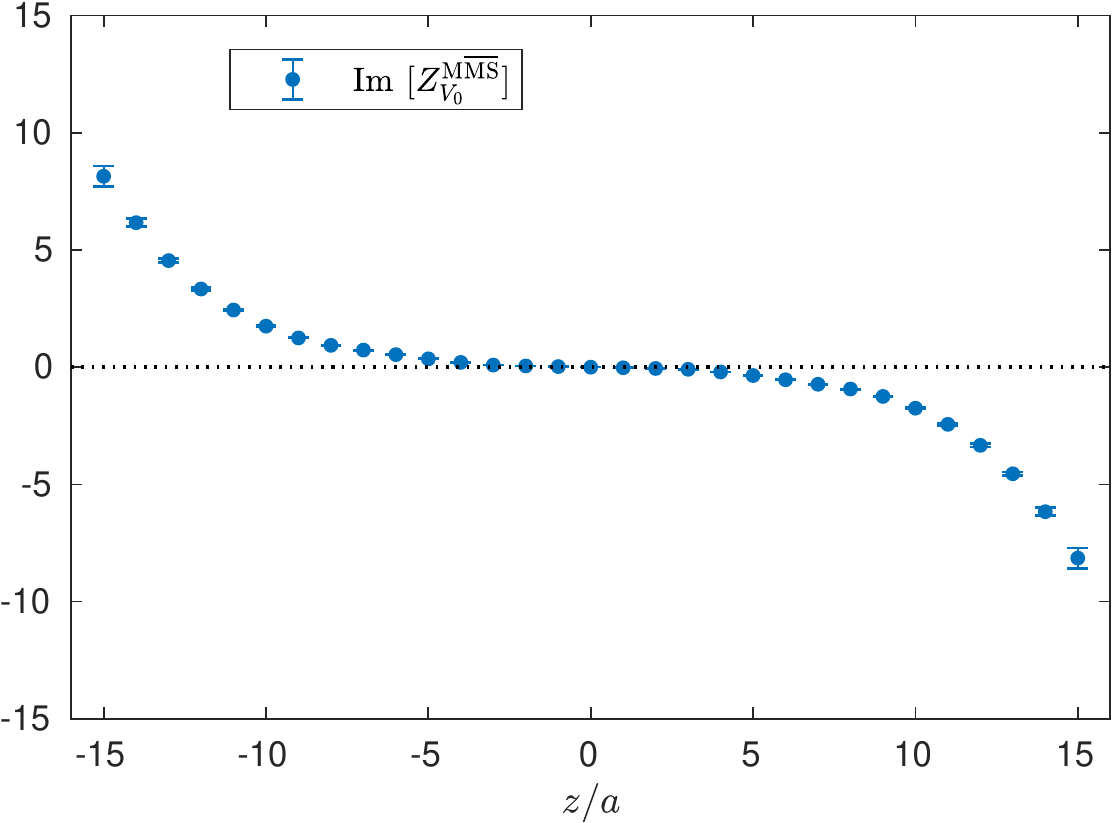}
\vspace*{-0.15cm}
\caption{\small{Final values of the $Z$-factors in the $\mmsbar$ scheme for the vector operator upon chiral and $(a\,\mu_0)^2$ extrapolation. Real and imaginary parts are shown in the left and right panels, respectively.}}
\label{fig:Z_final}
\end{figure}

The final values for the $Z$-factors are extracted by extrapolating $(a\,\mu_0)^2 {\to} 0$ linearly the $\mmsbar$ data in the region $(a\,\mu_0)^2 \in [2-4]$ (Eq.~(\ref{eq:Z_mu_fit})). They are shown in Fig.~\ref{fig:Z_final} for the vector operator. As can be seen, the  statistical errors of the $Z$-factors are much smaller as compared to those of the matrix elements, because of the momentum source method used in the calculation of the vertex functions. The real part increases significantly with $z$, which is due to the power-law divergence of the Wilson line. Note that for $z/a{>}10$, the bare matrix elements decay to zero, and thus, the renormalized matrix elements, given by the complex multiplication 
\be
h^{\MMSb}_\Gamma(P_3,z, \bar\mu) =  h^{bare}_\Gamma(P_3,z) \cdot {\cal Z}^{\MMSb}_{\Gamma,0}(z,\bar\mu)\,,
\ee
are also zero but with large statistical uncertainty originating from  the large values of the real part of the $Z$-factors. 

\begin{figure}[h]
\centering
\includegraphics[scale=0.75]{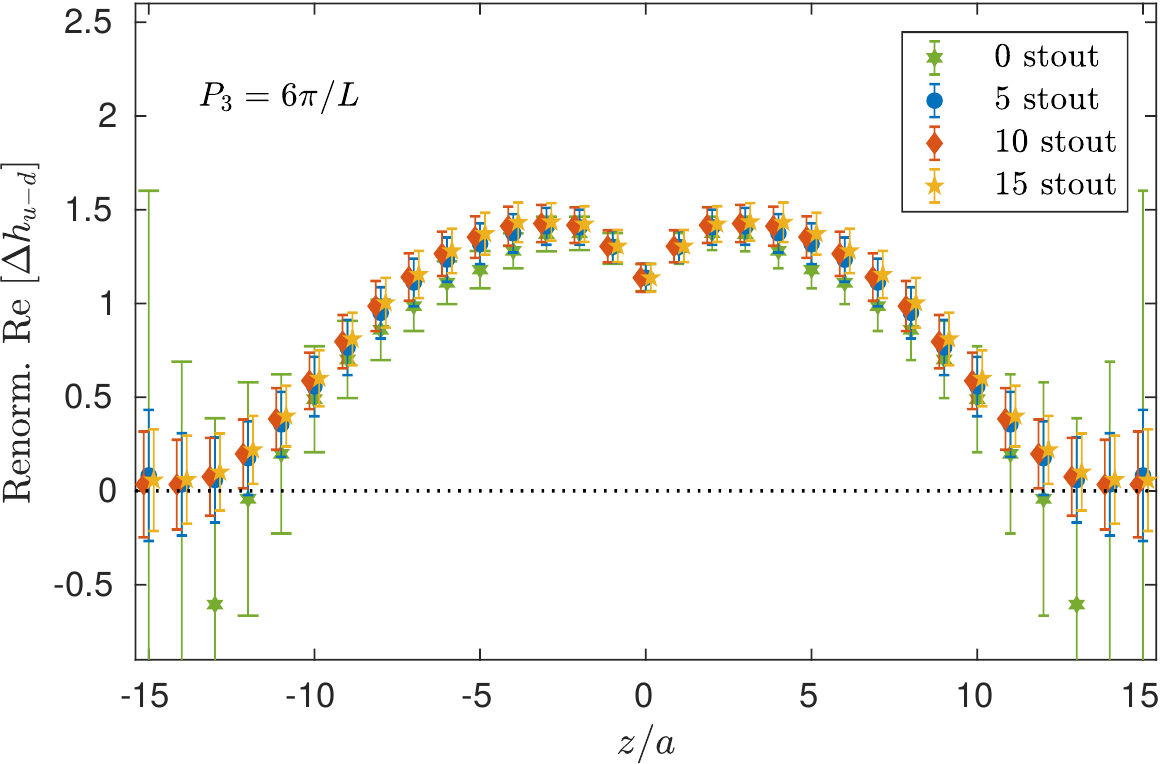} \hspace*{0.3cm}\includegraphics[scale=0.75]{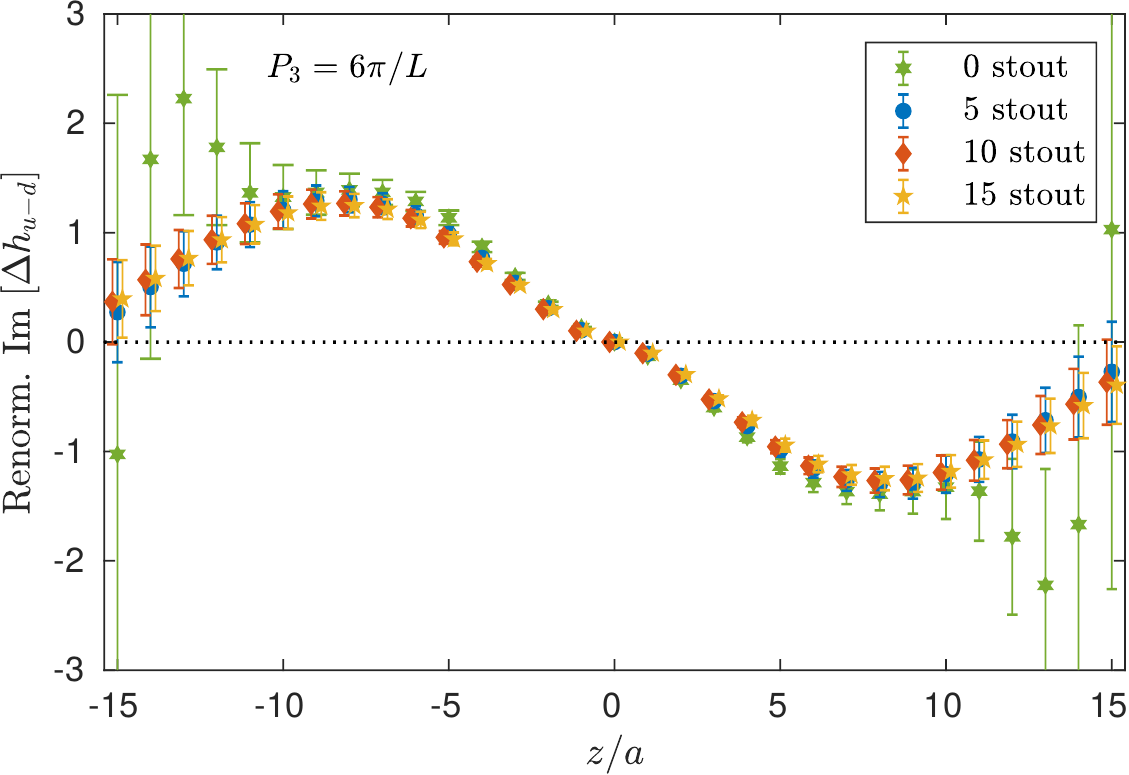}
\vspace*{-0.5cm}
\caption{\small{Real (left) and imaginary (right) part of renormalized matrix elements for the axial operator (helicity PDF) for momentum $\frac{6\pi}{L}$, as a function 
of the length of the Wilson line. Green stars/blue circles/red diamonds/orange stars correspond to 0/5/10/15 iterations of stout smearing.}}
\label{fig:ME_Renorm}
\end{figure}

Both the matrix elements  and the renormalization functions of a given operator share similar properties with respect to $z$ (symmetric real part and antisymmetric imaginary part), with the difference that the $Z$-factor decreases when a smearing technique is applied to the Wilson line, while the matrix element increases. Of course, the dependence in the smearing is non-monotonic due to the complex nature of the matrix elements and $Z$-factors, but the stout smearing dependence is expected to cancel out in the renormalized matrix elements. This is  demonstrated  in Fig.~\ref{fig:ME_Renorm}, where we compare the renormalized matrix elements for the helicity operator extracted using 0, 5, 10 and 15 stout smearing steps, in the $\msbar$ scheme at 2 GeV and for momentum $\frac{6\pi}{L}$. This confirms that the elimination of the  power divergences is correctly realized via the renormalization program, yielding compatible results for the  renormalized matrix elements for different stout iterations. We find that this hold also for the renormalized matrix elements of the unpolarized and transversity operators and therefore any stout step may be used without changing the final physical result. It is worth mentioning that the agreement is more striking upon the $(a\,\mu_0)^2 {\to} 0$ extrapolation of the $Z$-factors.

\vspace*{1cm}
 
\section{Matching to light-cone PDFs}
\label{sec:matching}

\subsection{Derivation of the matching formulae}
\label{sec:derivation}
In this section, we discuss, in detail, the matching procedure that relates the quasi-PDFs, renormalized in some scheme, to light-cone PDFs in the same or other renormalization scheme, typically chosen to be the $\MSb$ scheme.
We derive new matching formulae that relate $\MSb$-renormalized quasi-PDFs to $\MSb$-renormalized light-cone PDFs and conserve the particle number.
To satisfy this condition, we introduce a modification of the $\MSb$ scheme, i.e.\ the $\MMSb$ scheme, which was already partially discussed in the previous section, since it requires also the modification of conversion of renormalization functions.

Quasi-PDFs can be obtained as a Fourier transform (FT) of renormalized matrix elements, $h_\Gamma(P_3,z)$,
\begin{equation}
\label{eq:fourier}
\tilde{q}(x,P_3)\,=\,\frac{2P_3}{4\pi}\,\sum_{z=-z_{\rm max}}^{z_{\rm max}}\,e^{-ixP_3z}\,h_\Gamma(P_3,z)\, .
\end{equation}
To relate the quasi-PDFs $\tilde{q}(x,P_3)$ to light-cone PDFs, one relies on a perturbative matching procedure~\cite{Xiong:2013bka,Ji:2014gla,Wang:2017qyg,Stewart:2017tvs,Izubuchi:2018srq}. 
To one-loop order, and in the Feynman gauge, one needs to compute self-energy corrections, 
which include the usual self-energy plus the virtual
"sail" and "tadpole" diagrams, and the vertex corrections, with the corresponding 
the real "sail" and "tadpole" diagrams.
We use operators with four Dirac structures, namely $\gamma^0$ and $\gamma^3$ for the unpolarized distribution, $\gamma^3\gamma^5$ for the helicity, and $\gamma^3\gamma^j$, $j=1,2$  for the transversity case. 

\begin{figure}[h]
\centering
\includegraphics[scale=0.225]{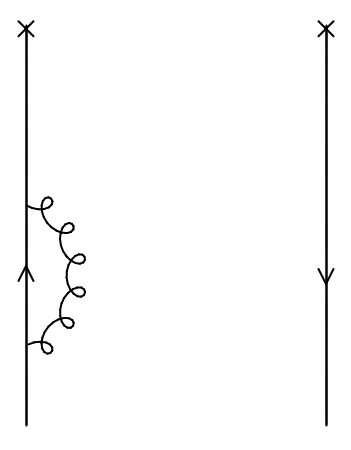} \hspace*{0.5cm}
\includegraphics[scale=0.225]{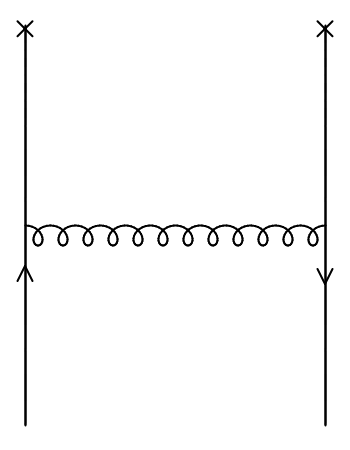} \hspace*{0.5cm}
\includegraphics[scale=0.225]{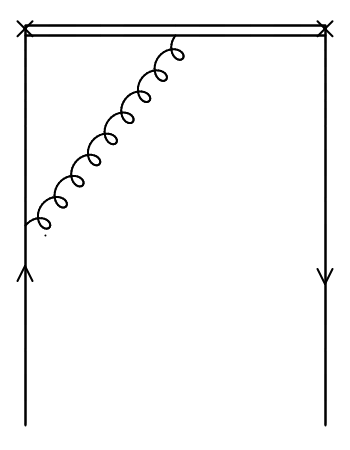} \hspace*{0.5cm}
\includegraphics[scale=0.225]{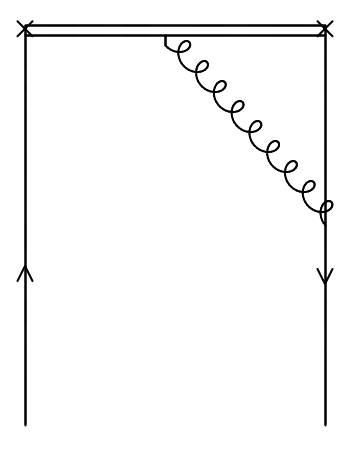} \hspace*{0.5cm}
\includegraphics[scale=0.225]{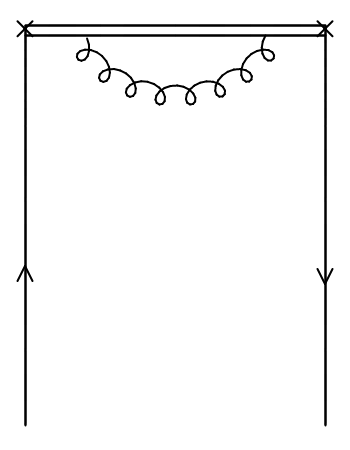} 
\vspace*{-0.25cm}
\caption{\small{One-loop diagrams entering the perturbative calculation for the matching between quasi-PDFs and light-cone PDFs.}}
\label{fig:diagrams}
\end{figure}

Since the extraction of the matching formula follows a similar process for all four operators, we use the $\gamma^0$ structure as an example and  we only present the final results for the other cases.
As already mentioned, we take the nucleon momentum in the third direction, $P{=}(P^0,0,0,P^3)$. It is assumed that, before gluon emission, the quark momentum $p$ is collinear to the nucleon momentum, i.e.\ $p{=}(p^0,0,0,p^3)$. It also obeys the Dirac equation $\slashed p u(p){=}0$, and carries a fraction of momentum $y$ of the parent hadron. 
After gluon emission, the quark has momentum $\xi p_3 {=} \xi y P_3$. 
When the bare quark distributions are dressed taking the $p_3{\rightarrow}\infty$ limit, we have the usual quark distributions in the infinite momentum frame or, equivalently, on the light cone. 
Defining $x {\equiv} \xi y$, the nonsinglet quark distribution at one-loop is given by
\begin{equation}
\label{dressed_lc_pdf}
q_{\gamma^0}(x,\Lambda) = q_{bare}(x) + \frac{\alpha_s}{2\pi}\int_x^1\frac{dy}{y}\,\left(\Pi_{\gamma^0}\left(\Lambda\right)\delta\left(1-\frac{x}{y}\right)+\Gamma_{\gamma^0}\left(\frac{x}{y},\Lambda\right)\right) q_{bare}(y)\,,
\end{equation}
where $\Pi_{\gamma^0}$ denotes the self-energy corrections, $\Gamma_{\gamma^0}$ is the vertex corrections,
and $\Lambda$ stands for the infrared (IR) and UV regulators in any given scheme.
Using DR to regulate both the IR and UV divergences, the resulting one-loop correction
for $0<\xi<1$ in the $\MSb$ scheme is
\begin{equation}
\label{renormalized_gamma_phys}
\Gamma_{\gamma^0}^{\MSb}\left(\xi,\frac{\bar\mu}{\mu_F}\right)=\frac{1+\xi^2}{1-\xi}\left(-\frac{1}{\epsilon_{IR}}+
\ln\left(\frac{\bar\mu^2}{\mu_F^2}\right)\right)\,,
\end{equation}
where the pole from the UV divergence has already been subtracted, and $\bar\mu$ is the corresponding renormalization scale in $\MSb$.
The pole from the soft IR, ${1}/{\epsilon_{IR}}$, can be absorbed in the bare distribution, at the factorization scale $\mu_F$, but it is explicitly written in Eq.~(\ref{renormalized_gamma_phys}), as it must cancel a similar pole arising in the one-loop correction to the quasi-PDF. 
The fact that the poles in $1/\epsilon_{IR}$ are the same for the light-cone PDF and the quasi-PDF, is a crucial observation that allows one to match them using perturbation theory.
There are remaining IR divergences, which are located at $\xi{=}1$ and have their origin in the emission of soft gluons. 
However, they cancel between the vertex and self-energy one-loop corrections, which are related by $\Pi_{\gamma^0}(\Lambda) {=}-\int_0^1 d\xi\, \Gamma_{\gamma^0} (\xi,\Lambda)$.

To obtain the quasi-PDF, the quark is dressed with finite $p_3$.
To simplify notation, we write the one-loop correction for positive $x$ only, that is
\begin{equation}
\label{dressed_offlc_pdf}
\tilde{q}_{\gamma^0} (x,P_3,\Lambda) = q_{bare}(x) + \frac{\alpha_s}{2\pi} \int_{0}^{1} \frac{dy}{y}\,\left(\tilde{\Pi}_{\gamma^0}\left(p_3, \Lambda\right) \delta\left(1-\frac{x}{y}\right) + \tilde{\Gamma}_{\gamma^0} \left(\frac{x}{y},p_3,\Lambda \right)\right) q_{bare}(y)\,.
\end{equation}
Because $\tilde{\Gamma}_{\gamma^0}\neq 0$ for $\xi > 1$, the lower limit of integration in Eq.~(\ref{dressed_offlc_pdf}) goes to zero, and the quasi-PDF has support for $x>1$.
Using DR to perform the integrals, the one-loop correction to the vertex is given by:
\begin{eqnarray}
\label{gamma_tilde}
&&\tilde \Gamma_{\gamma^0} \left(\xi, \frac{p_3}{\mu_F}\right) = \left\{
\begin{aligned}
&\frac{1+\xi^2}{1-\xi}\ln\left(\frac{\xi}{\xi-1}\right) + 1,~~~~&\xi>1\,,\\
&\frac{1+\xi^2}{1-\xi}\left(-\frac{1}{\epsilon_{IR}}+\ln\left(\frac{4\xi(1-\xi)(p_3)^2}{\mu_F^2}\right)\right) - \frac{2\xi}{1-\xi} + \xi,~~~~&0<\xi<1\,,\\
&\frac{1+\xi^2}{1-\xi}\ln\left(\frac{\xi-1}{\xi}\right) - 1,~~~~&\xi<0\,.
\end{aligned}\right.
\end{eqnarray}
Note that the vertex corrections have no poles when DR is used, but they are non-zero outside the physical region $0 {<} \xi {<} 1$, a behavior that is different from the one-loop correction to the light-cone PDF.
Thus, in addition to the IR divergences at $\xi{=}1$, there are UV divergences associated with the $\xi\rightarrow \pm\infty$ limits when Eq.~(\ref{gamma_tilde}) is integrated.
So far, these UV divergences have not been subtracted, and this is reflected by the $\Lambda$-dependence in the functional form of $\tilde{\Pi}_{\gamma^0}(\Lambda,p_3/\mu_F)$. 
 DR is again used to isolate the poles, in which case $d\xi \rightarrow d\xi (\xi p_3)^{d-1}(\bar\mu^2 e^{\gamma_E}/4\pi)^{1/2-d/2}$.
For the region $\xi{>}1$, the self-energy is:
\begin{equation}
\label{self_energy_xi_g_1}
\tilde{\Pi}_{\gamma^0} (\epsilon,\xi>1) = \int_1^\infty d\xi \xi^{d-1}(p_3/\bar\mu)^{d-1}(e^{\gamma_E}/4\pi)^{1/2-d/2} \left(\frac{1+\xi^2}{1-\xi}\ln\left(\frac{\xi-1}{\xi}\right)-1\right)\,,
\end{equation}
where $d{=}1{-}2\epsilon$, with $\epsilon{>}0$.
Using the plus prescription for the collinear divergence at $\xi{=}1$, Eq.~(\ref{self_energy_xi_g_1}) can be written as
\begin{equation}
\label{self_energy_divergence}
\tilde{\Pi}_{\gamma^0}(\epsilon,\xi>1) = \frac{7}{4} -
\frac{3}{4}\ln\left(\frac{p_3^2}{\bar\mu^2}\right)+\frac{3}{4}\left(\frac{1}{\epsilon} + \gamma_E - \ln\left({4\pi}\right)\right)\,.
\end{equation}
Similar computations can be done for the other regions, and the renormalized one-loop
self-energy in the $\MSb$ scheme for the quasi-PDF is given by
\begin{eqnarray}
\label{renormalized_self_energy}
\tilde \Pi_{\gamma^0}^{\MSb} \left(\xi,\frac{p_3}{\mu_F},\frac{\mu_F}{\bar\mu}\right) &=& \frac{5}{2} +
\frac{3}{2}\ln\left(\frac{\mu_F^2}{4\bar\mu^2}\right)\\
&-&\int d\xi\left\{
\begin{aligned}
&\frac{1+\xi^2}{1-\xi}\ln\left(\frac{\xi}{\xi-1}\right) + 1-\frac{3}{2\xi},~~~~&\xi>1\,, \nonumber \\
&\frac{1+\xi^2}{1-\xi}\left(-\frac{1}{\epsilon_{IR}}+\ln\left(\frac{4\xi(1-\xi)(p_3)^2}{\mu_F^2}\right)\right) - \frac{2\xi}{1-\xi} + \xi,~~~~&0<\xi<1\,, \nonumber\\
&\frac{1+\xi^2}{1-\xi}\ln\left(\frac{\xi-1}{\xi}\right) - 1- \frac{3}{2(1-\xi)},~~~~&\xi<0\,, \nonumber
\end{aligned}\right.
\end{eqnarray}
with the corresponding renormalization function, $ \tilde{Z}_\Pi^{\MSb}{=} 1
+ \frac{\alpha_s}{2\pi}C_F \frac{3}{2}\left(\frac{1}{\epsilon} + \gamma_E - \ln\left({4\pi}\right)\right)$.
Upon integration of Eq.~(\ref{gamma_tilde}) and employing the $\MSb$ scheme, the Ward identity of the resulting local current is naturally respected.

For the computation of the $x$-dependence of the distributions, however, the convolution
involving the vertex correction prevents the aforementioned cancellation to occur,
and one needs to impose a prescription to ensure conservation of the quark number for the
nonsinglet distributions. From the equations for the quark distributions, Eq.~(\ref{dressed_lc_pdf}),
and quasi-distributions, Eq.~(\ref{dressed_offlc_pdf}), and including the negative $x$ regions we
have to one-loop
\begin{equation}
\label{bare_matching_xbig1}
q_{\gamma^0}(x,\bar\mu) = \int_{-\infty}^{+\infty}\frac{dy}{|y|}\, C_{\gamma^0}^{\MSb}\left(\frac{x}{y},\frac{\bar\mu}{p_3},\frac{\bar\mu}{\mu_F}\right)\tilde{q}_{\gamma^0}(y,p_3,\bar\mu),
\end{equation}
where $C_{\gamma^0}^{\MSb}(\xi)=\delta(1-\xi) - \alpha_sC_F[(\tilde{\Pi}_{\gamma^0}^{\MSb} -
\Pi_{\gamma^0}^{\MSb})\delta(1-\xi) + \tilde{\Gamma}_{\gamma^0}\left(\xi \right) -
\Gamma_{\gamma^0}^{\MSb}\left(\xi \right)]/2\pi$. The dependence on $p_3$, $\bar\mu$, and
$\mu_F$ is implicit in $C^{\MSb}(\xi)$. $\tilde\Gamma(\xi)$ is the bare function, hence the absence of an $\MSb$
superscript. The poles of $1/\epsilon_{IR}$, on the
other hand, automatically cancel in $C_{\gamma^0}^{\MSb}$, which is explicitly written as
\begin{align}
\label{quasi_matching_MSbar}
C_{\gamma^0}^{\MSb}\left(\xi, \frac{\bar\mu}{p_3},\frac{\bar\mu}{\mu_F} \right)
= &\, \delta\left(1-\xi\right) \nonumber \\
&+{\alpha_sC_F\over 2\pi}\left\{
\begin{array}{ll}
\displaystyle \left({1+\xi^2\over 1-\xi}\ln \left({\xi\over \xi-1}\right) + 1 + {3\over 2\xi}\right)_{+(1)}- {3\over 2\xi},
&\, \xi>1
\nonumber \\[10pt]
\displaystyle \left({1+\xi^2\over 1-\xi}\left[
\ln\left({p_3^2\over \bar\mu^2}\right) + \ln\big(4\xi(1-\xi)\big)\right] - {\xi(1+\xi)\over 1-\xi}\right)_{+(1)},
&\, 0<\xi<1
\nonumber \\[10pt]
\displaystyle  \left(-{1+\xi^2\over 1-\xi}\ln \left({-\xi\over 1-\xi}\right) - 1 + {3\over 2(1-\xi)}\right)_{+(1)} - {3\over 2(1-\xi)},\quad
&\, \xi<0
\end{array}\right.\nonumber\\[5pt]
& + {\alpha_sC_F\over 2\pi}\delta(1-\xi) \left( {3\over2}\ln\left({\mu_F^2\over 4\bar\mu^2}\right) + {5\over2}\right)\,.
\end{align}
For the rest of this paper, we will employ the
usual choice of renormalization and factorization scales, $\mu_F {=} \bar\mu$.
Eq.~(\ref{quasi_matching_MSbar}) has an imbalance when integrated, because $\tilde{\Gamma}_{\gamma^0}$
outside the physical region picks up a logarithmic divergence in the momentum fraction as
$\xi\rightarrow \pm \infty$, while this divergence has already been removed
from the self-energy part in Eq.~(\ref{renormalized_self_energy}). This implies
an anomalous non-conservation of the quark number, even if the classical
current is conserved. In Ref.~\cite{Izubuchi:2018srq}, $C_{\gamma^0}^{\MSb}$ has also been computed
and a plus prescription at infinity was used to cancel the UV divergences
between the vertex and self-energy corrections. This explains the source of the
difference between our result in Eq.~(\ref{quasi_matching_MSbar}) and that in Eq.~(68)
of Ref.~\cite{Izubuchi:2018srq}. Namely, in our case, the term proportional to the Dirac $\delta$-function
depends on the ratio of the factorization and renormalization scales, while
in  the  case of Ref.~\cite{Izubuchi:2018srq}, it depends on the quark momentum $p_3$. Both prescriptions, however,
do not conserve quark number.

To treat the unbalanced divergence, we introduce a modified $\MSb$ scheme
($\MMSb$), which has already been discussed in the previous section and used in our previous work~\cite{Alexandrou:2018pbm,Alexandrou:2018eet}. 
In this scheme, an extra subtraction is made outside the physical region of the unintegrated vertex corrections, which, in
practice, renormalizes the $\xi$-dependence for $\xi>1$ and $\xi<0$ and removes the potential divergences,
\begin{equation}
\label{renormalization_MMS}
\tilde{Z}_{\Gamma_{\gamma^0}}^{\MMSb}(\xi) = 1 - \frac{\alpha_s}{2\pi}C_F\frac{3}{2}\left(-\frac{1}{\xi}\theta(\xi-1) - \frac{1}{1-\xi}\theta(-\xi)\right)
- {\alpha_sC_F\over 2\pi}\delta(1-\xi) \left( {3\over2}\ln\left({1\over 4}\right) + {5\over2}\right).
\end{equation}
One can write Eq.~(\ref{renormalization_MMS}) in $z$-space, noticing that $\xi$
outside the physical region is now renormalized, at the scale $\bar\mu$. The
inverse Fourier transform is then from $\xi$ to $z \bar\mu$, and the extra subtraction
in $z$-space is written as
\begin{eqnarray}
\label{renormalization_MMS_zspace}
Z_{{\Gamma}_{\gamma^0}}^{\MMSb}(z\bar\mu)&=& 1 - \frac{\alpha_s}{2\pi}C_F\left(\frac{3}{2}\ln\left(\frac{1}{4}\right)+\frac{5}{2}\right) \nonumber \\
&+&\frac{3}{2}\frac{\alpha_s}{2\pi}C_F\left(i\pi \frac{|z \bar\mu|}{2 z \bar\mu} - {\rm Ci}(z \bar\mu) + \ln(z \bar\mu) - \ln(|z \bar\mu|)- i {\rm Si}(z \bar\mu)\right) \nonumber \\
&-& \frac{3}{2} \frac{\alpha_s}{2\pi}C_F e^{i z \bar\mu} \left(\frac{2 {\rm Ei}(-i z \bar\mu)- \ln (-i z \bar\mu) + \ln (i z \bar\mu) + i \pi {\rm sgn}(z \bar\mu)}{2}\right).
\end{eqnarray}
For consistency, Eq.~(\ref{renormalization_MMS_zspace}) has been also applied to the renormalization functions to bring them to the $\mmsbar$ scheme, as described in Sec.~\ref{subsec:MSconversion}.
Thus, these renormalization functions are obtained as follows: the $Z$-factors calculated in the RI$'$ scheme are converted to the $\MSb$ scheme according to the perturbative formulae of Ref.~\cite{Constantinou:2017sej} and then multiplied by the factor given in Eq.~(\ref{eq:CMStoMMS}) to convert them to the $\MMSb$ scheme.
An important self-consistency check is that the expression of Eq.~(\ref{renormalization_MMS_zspace}) must cancel the $z\rightarrow 0$ divergence in $\ln \left(z^2\right)$ present in the $\MSb$ scheme \cite{Constantinou:2017sej,Ishikawa:2017faj}. Indeed, in the limit $z\rightarrow 0$,
one has:
\begin{equation}
\label{eq:c0}
Z_{\Gamma_{\gamma^0}}^{\MMSb}(z\rightarrow 0)= 1 - \frac{\alpha_s C_F}{2\pi}\left(\frac{3}{2}\ln\left(\frac{\bar\mu^2 z^2 e^{2\gamma_E}}{4}\right)+ \frac{5}{2} \right) = Z_{\Gamma_{\gamma^0}}^{\rm ratio}(z\bar\mu),
\end{equation}
i.e.\ the $Z$-factor of the vertex corrections at $z{=}0$ is the same in our scheme and in the ``ratio'' scheme introduced in Ref.~\cite{Izubuchi:2018srq} and both cancel the divergence.
The latter scheme was proposed as an alternative to our solution of the current conservation problem when using the pure $\MSb$ expression of Eq.~(68) of Ref.~\cite{Izubuchi:2018srq}.
The ``ratio'' scheme is another modification of the $\MSb$ scheme that also needs an additional conversion factor in order to bring  the renormalization functions in the $\MSb$ scheme  into the ``ratio'' scheme. 
The difference with respect to our solution is that the form of the matching kernel (or of $Z_{\Gamma_{\gamma^0}}^{\rm ratio}(z\bar\mu)$ after doing the Fourier
transform to $z$-space) implies that the $\xi$-dependence of the matching equation in the physical region is also renormalized, while our approach does not modify this region. 
Thus, the ``ratio'' scheme can induce potentially large modifications to the matched PDF, as we show below numerically.

We apply $\tilde{Z}_{\Gamma_{\gamma^0}}^{\MMSb}(\xi)$ to Eq.~(\ref{quasi_matching_MSbar})
to obtain a matching kernel that keeps the norm of the nonsinglet
distributions unchanged. We also include the results for the $\gamma^3$ and
$\gamma^3\gamma^5$ Dirac structures. Because quarks are taken to be massless,
the one-loop corrections are the same for the $\gamma^3$ and $\gamma^3\gamma^5$ cases.
Compared to the $\gamma^0$ case, a shift of $+2(1-\xi)$ inside the plus prescription
in the physical region of Eq.~(\ref{quasi_matching_MSbar}) is needed, and
also, the factor of 5/2 in the last line of Eq.~(\ref{quasi_matching_MSbar}) becomes 7/2.
The matching equation is then written as:
\begin{equation}
\label{MMS_matching_equation}
q^{\MSb}(x,\bar\mu) = \int_{-\infty}^{+\infty}\frac{dy}{|y|} C^{\MMSb}\left(\frac{x}{y},\frac{\bar\mu}{p_3}\right)\tilde{q}^{\MMSb}(y,p_3,\bar\mu),
\end{equation}
with the matching kernels, for the different Dirac structures, given by
\begin{align}
\label{quasi_matching_MMSbar}
C_{\gamma^0,\gamma^3,\gamma^3\gamma^5}^{\MMSb}\left(\xi, \frac{\bar\mu}{p_3} \right)
= &\, \delta\left(1-\xi\right) \nonumber \\
&+{\alpha_sC_F\over 2\pi}\left\{
\begin{array}{ll}
\displaystyle \left({1+\xi^2\over 1-\xi}\ln \left({\xi\over \xi-1}\right) + 1 + {3\over 2\xi}\right)_{+(1)}, &\, \xi>1,
\nonumber \\[10pt]
\displaystyle \left({1+\xi^2\over 1-\xi}\left[
\ln\left({p_3^2\over \bar\mu^2}\right) + \ln\big(4\xi(1-\xi)\big)\right] - {\xi(1+\xi)\over 1-\xi} + 2\iota (1-\xi)\right)_{+(1)},&\, 0<\xi<1,
\nonumber \\[10pt]
\displaystyle  \left(-{1+\xi^2\over 1-\xi}\ln \left({-\xi\over 1-\xi}\right) - 1 + {3\over 2(1-\xi)}\right)_{+(1)}, \quad
&\, \xi<0,
\end{array}\right.\nonumber\\.
\end{align}
where $\iota{=}0$ for $\gamma^0$, and $\iota{=}1$ for $\gamma^3$ and $\gamma^3\gamma^5$.
We note that an alternative procedure to get $\MSb$-renormalized light-cone PDFs is to directly match to them from RI-renormalized quasi-PDFs.
The relevant formulae for this were derived in Ref.~\cite{Stewart:2017tvs} for the operators with Dirac structures $\gamma^3$ and $\gamma^3\gamma^5$ and in Ref.~\cite{Liu:2018uuj} for $\gamma^0$.
Such a procedure is equivalent, to one-loop order, to the one described above, but with different higher-order effects.
This choice is investigated numerically below, by comparing the direct matching from a variant of the RI scheme to $\MSb$ with a two-step procedure that first brings the renormalization functions to the $\MMSb$ scheme and then performs the $\MMSb$ to $\MSb$ matching.

For the transversity case, the computation is similar, albeit simpler, because the usual one-loop vertex correction is zero in the Feynman gauge. 
The quark self-energy is not zero, but the unbalanced terms cancel each other in the matching equation, since they are the same for $\Pi_{\gamma^3\gamma^j}$ and $\tilde{\Pi}_{\gamma^3 \gamma^j}$. Thus, $Z_{{\Gamma}_{\gamma^3\gamma^j}}^{\MMSb}(z\bar\mu)$ is
given by the same functional form as Eq.~(\ref{renormalization_MMS_zspace}), with the replacement of the numerical factors 3/2 and 5/2 by 2. $C_{\gamma^3\gamma^j}^{\MMSb}$ is given by:
\begin{eqnarray}
\label{eq:kernel}
C_{\gamma^3\gamma^j}^{\MMSb}\left( \xi, \frac{\bar\mu}{p_3} \right)=\delta(1-\xi)+\frac{\alpha_s}{2\pi}C_F\left\{
\begin{array}{ll}
\displaystyle \left[\frac{2 \xi}{1-\xi}\ln\left(\frac{\xi}{\xi-1}\right) + \frac{2}{\xi}\right]_{+(1)},
&\, \xi>1,
\\[10pt]
\displaystyle \left[\frac{2\xi}{1-\xi}\left(
\ln\left(\frac{p_3^2}{\bar\mu^2}\right)+\ln(4\xi(1-\xi))\right) - \frac{2 \xi }{1-\xi}  \right]_{+(1)},
&\, 0<\xi<1,
\\[10pt]
\displaystyle \left[-\frac{2\xi}{1-\xi}\ln\left(\frac{\xi}{\xi-1}\right) + \frac{2}{1-\xi}\right]_{+(1)},
&\, \xi<0\,.
\end{array}\right.
\end{eqnarray}
The matching kernel for the transversity has been calculated previously,
with a hard cut-off and a quark mass to regularize the UV and the IR divergences,
respectively, but without renormalization~\cite{Xiong:2013bka}. 
More recently, following the idea of Ref.~\cite{Stewart:2017tvs} to match directly from RI-renormalized quasi-PDFs to light-cone PDFs in the $\MSb$ scheme,
the matching for the transversity  was also derived in Ref.~\cite{Liu:2018hxv}.

\subsection{Comparison of matching in the $\MSb$, $\MMSb$ and ratio schemes}
\label{sec:MMS}

Our choice for the matching to extract the light-cone PDFs from quasi-PDFs   is to use a minimal modification of the $\MSb$ scheme that ensures current conservation.
Thus, the matching formula is applied on a quasi-PDF renormalized in a modified $\MSb$ scheme, the $\MMSb$ scheme, to yield the corresponding light-cone PDF in the pure $\MSb$ scheme. 
The difference between $\MSb$ and $\MMSb$ schemes was also taken into account at the level of the renormalization functions in Sec.~\ref{sec:renorm}. 
Note that in our previous work~\cite{Alexandrou:2018pbm,Alexandrou:2018eet}, we did not modify the conversion and, thus, our quasi-PDFs were renormalized in the standard $\MSb$ scheme.
The formulae for the conversion modification, derived in the current paper, were not available at the time of our previous work and we argued that the numerical effect of the modification is subleading with respect to other uncertainties (as the modification in the $\MSb$ is minimal and only in the unphysical regions). Moreover, it disappears in the infinite momentum limit.
Having now derived the relevant formulae to convert renormalization functions from the $\MSb$ scheme to $\MMSb$, we  test  this and validate that it is indeed a small effect.

\begin{figure}[h!]
\begin{center}
\includegraphics[scale=0.815]{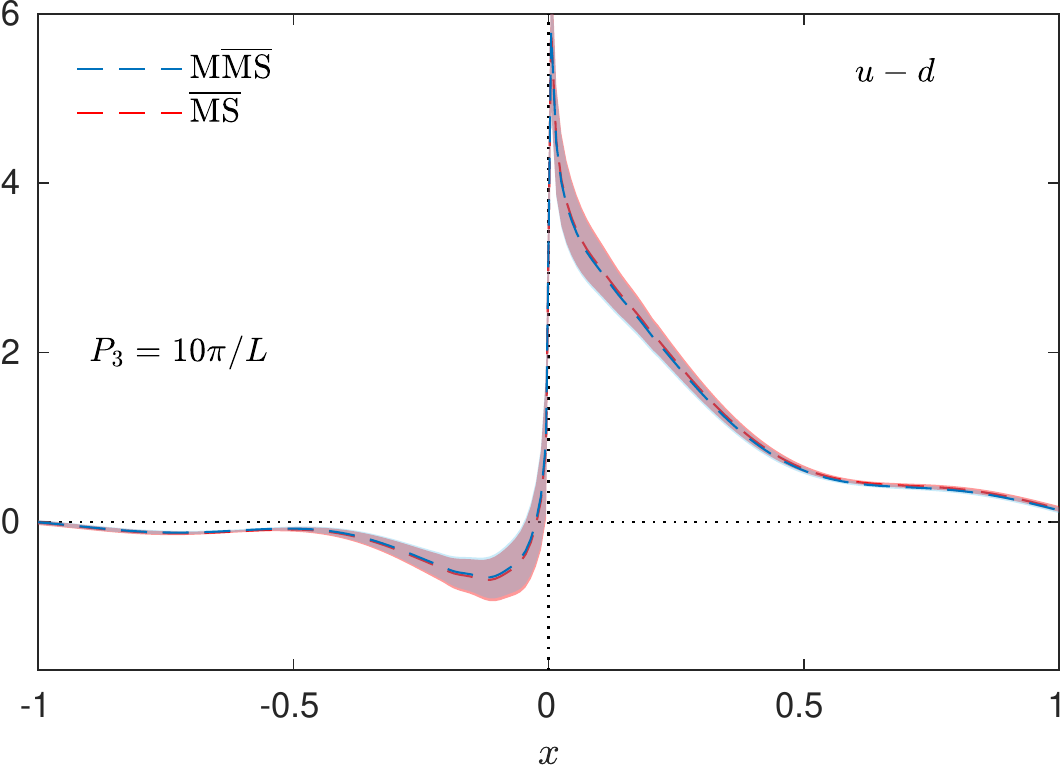}\,\,\,
\includegraphics[scale=0.815]{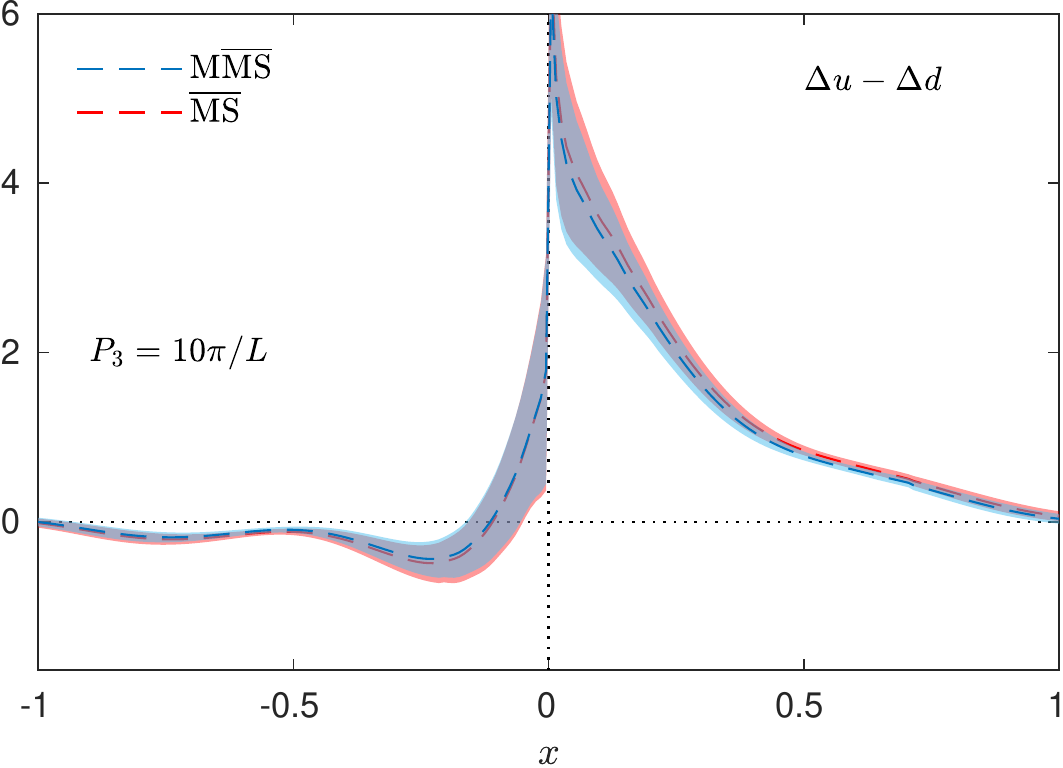}\\[2ex]
\includegraphics[scale=0.815]{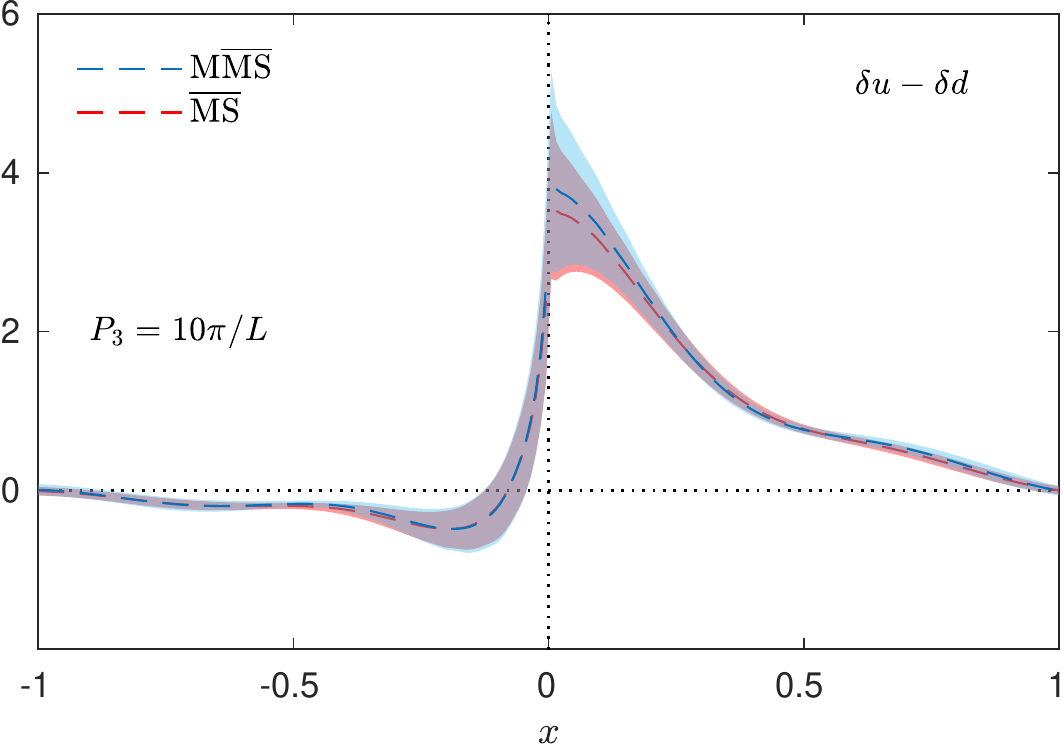}
\end{center}
\vskip -0.5cm
\caption{Comparison of matched PDFs from quasi-PDFs renormalized in the $\MSb$ (red band) scheme and in the $\MMSb$ (cyan band) scheme, for the unpolarized (top left panel), helicity (top right panel) and transversity (bottom panel) cases and for nucleon momentum $P_3{=}10\pi/L$.}
\label{fig:MMS}
\end{figure}

In Fig.\ \ref{fig:MMS}, we compare matched PDFs obtained from quasi-PDFs renormalized in the $\MSb$ scheme, where the additional conversion of Eq.~(\ref{eq:CMStoMMS}) is not included, and in the modified $\MMSb$ scheme.
As anticipated, the numerical effect is very small, particularly in the unpolarized case.
Nevertheless, it is important to take the conversion modification into account to have a self-consistent procedure.
We note that the modification brings the matched PDFs slightly towards phenomenological extractions, which is reassuring.

\begin{figure}[ht!]
\begin{center}
\includegraphics[scale=0.825]{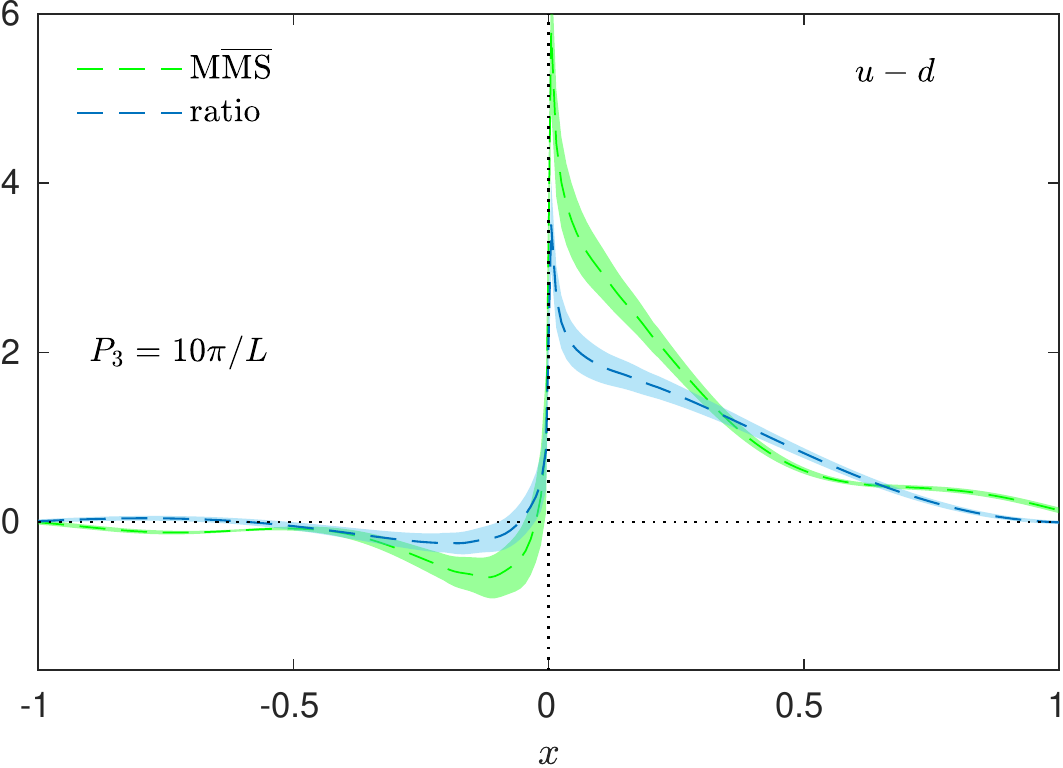}
\end{center}
\vskip -0.5cm
\caption{Comparison of matched PDFs from quasi-PDFs renormalized in the $\MMSb$ scheme (green band) and in the ``ratio'' scheme (cyan band) for the unpolarized case. Nucleon momentum is $P_3{=}10\pi/L$.}
\label{fig:ratio}
\end{figure}

Another possibility to obtain a matching formulae with current conservation is to use the procedure proposed in Ref.~\cite{Izubuchi:2018srq}, the so-called ``ratio'' scheme, already mentioned  in Sec.~\ref{sec:derivation}.
It consists in a different way of changing the $\MSb$ scheme to achieve current conservation, including a modification of the physical region.
Thus, the effect on the matched PDFs is expected to be larger numerically than when using the $\MMSb$ scheme.
We show the comparison of the matching from our previous procedure and from the ``ratio'' scheme in Fig.\ \ref{fig:ratio}.
Indeed, the ``ratio'' scheme  has a noticeable effect on the matched PDF, which is particularly large in the small-$x$ region.
Theoretically, both the ``ratio'' and $\MMSb$ schemes are valid modifications of the $\MSb$ scheme.
Obviously, they are equivalent at the one-loop level, but they have significantly different higher-order effects.
Hence, they can be both used to match quasi-PDFs to light-cone PDFs and the difference between them can serve as an estimate of truncation effects at one loop.
For the remainder of the paper, we use the $\MMSb$ scheme, as the one for which the higher-order effects are likely to be smaller, since this scheme is a milder modification of the $\MSb$ scheme, i.e.\ one that does not affect the physical region in the matching.

\subsection{Comparison of $\MSb\rightarrow\MSb$ and RI$\rightarrow\MSb$ matching}
\label{subsec:RI2MS}
As we discussed above, our renormalization and matching procedure proceeds in two steps.
First, the renormalization functions computed in the RI$'$ scheme are converted to the $\MMSb$ scheme and evolved to a chosen scale, $\bar\mu{=}2$ GeV, using one-loop perturbative formulae derived in Ref.~\cite{Constantinou:2017sej}, and including the additional conversion from the $\MSb$ scheme to the $\MMSb$ scheme, derived in this work.
The resulting $\MMSb$ $Z$-factors are used to renormalize the bare matrix elements, which are then Fourier-transformed from  $z$-space to $x$-space, yielding the renormalized quasi-PDF in the $\MMSb$ scheme.
In the second step, a matching procedure  is applied that brings the $\MMSb$-renormalized quasi-PDF to the light-cone PDF in the $\MSb$ scheme, at the same renormalization scale $\bar\mu$.
We will refer to this procedure as a two-step procedure.

In Ref.~\cite{Stewart:2017tvs}, a one-step procedure was proposed. It consists in applying the renormalization functions computed in the RI scheme to the bare matrix elements and taking the Fourier transform of RI-renormalized matrix elements to obtain a quasi-PDF renormalized in the RI scheme.
The  matching procedure then serves to simultaneously  bring the quasi-PDFs to the light-cone PDFs and to convert from the RI scheme to the $\MSb$ scheme, evolving them from the given RI scale to the reference scale of 2 GeV.
For this procedure, we use the formulae derived in Ref.~\cite{Liu:2018uuj} and consider the operator with the $\gamma_0$ Dirac structure.
We choose the variant with the $\slashed{p}$ projection that differs from the projection that we used in our variant of the RI$^\prime$ scheme. 
We refer to this choice of the projection as RI$_{\slashed{p}}$.

\begin{figure}[ht!]
\begin{center}
\begin{minipage}[t]{0.495\linewidth}
    \includegraphics[width=\textwidth]{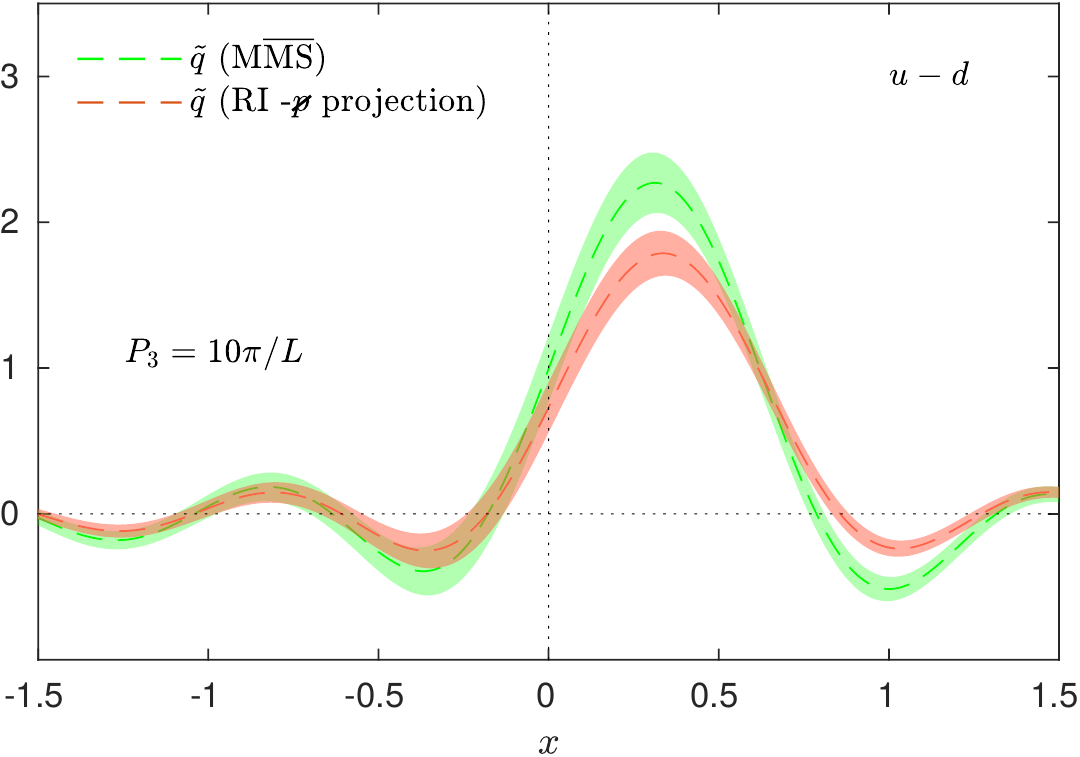}
  \end{minipage}
\begin{minipage}[t]{0.495\linewidth}
    \includegraphics[width=\textwidth]{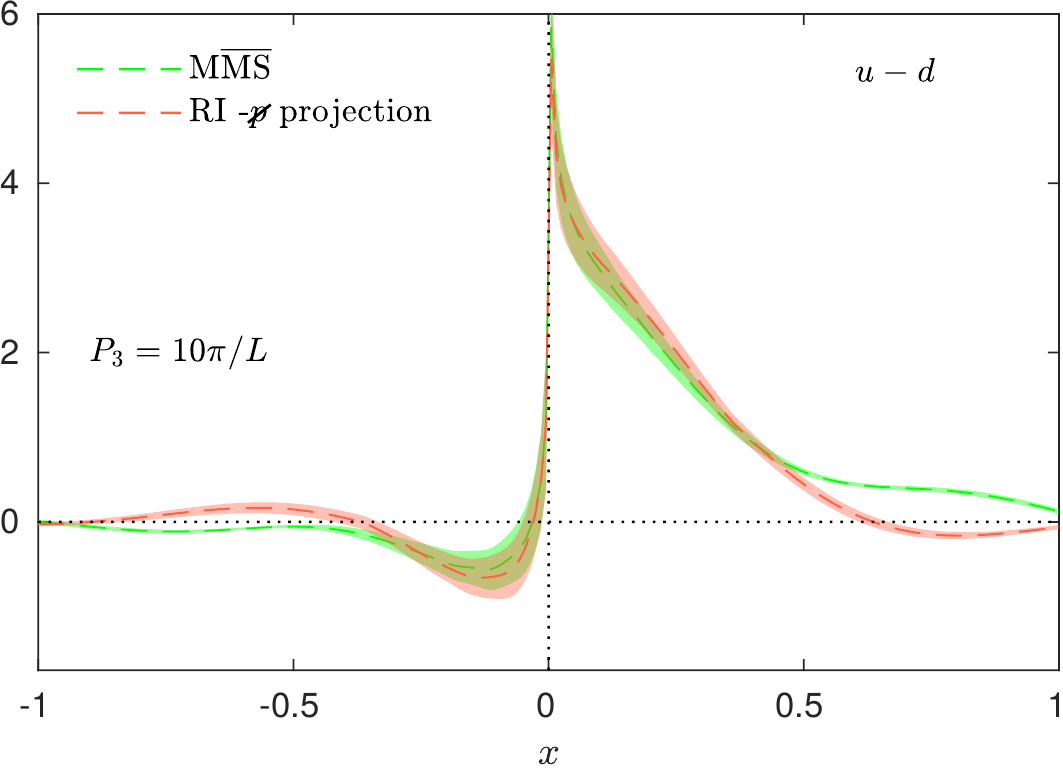}
  \end{minipage}
\end{center}
\vspace*{-0.5cm} 
\caption{Left panel: comparison of renormalized quasi-PDFs in the RI scheme with $\slashed{p}$ projection (red band) and the $\MMSb$ scheme (green band).
Right panel: matched PDFs obtained from the same bare matrix elements, but with different scheme conversion, evolution and matching procedures: either one-step (RI$\rightarrow\MSb$) or two-step (RI$'\rightarrow\MMSb\rightarrow\MSb$) (see text for details). Unpolarized case, nucleon momentum $P_3{=}10\pi/L$.}
\label{fig:RIvsMS}
\end{figure}

The comparison of the one-step and two-step procedures is presented in Fig.~\ref{fig:RIvsMS}.
In the left panel, we show our quasi-PDFs, renormalized in the RI$_{\slashed{p}}$ scheme and in the $\MMSb$ scheme.
The corresponding renormalization functions are applied to the same lattice data for bare matrix elements.
The results of application of the one-step and the two-step procedures are given in the right panel.
We observe that the results of both procedures are fully compatible for small positive and negative $x$.
For large positive $x$, the two-step procedure gives a PDF that goes to zero more slowly, while the PDF from the one-step procedure crosses zero at $x\approx0.6$ and remains negative until $x\approx1$ (reaching a minimum of around -0.16 at $x\approx0.8$).
In the large negative $x$ region, the situation is analogous -- the one-step procedure  leads to the PDF approaching zero  from below and the two-step  from above.

Both the one-step and the two-step procedures use a one-loop computation in continuum perturbation theory.
Obviously, both contain higher-order contributions that cannot be quantified unless the two-loop formulae are available.
Hence, neither of them can be considered to be the preferred one.
It might happen that one of the procedures evinces smaller higher-loop effects without any theoretical arguments to support this, but it is not possible to say which one.
The comparison between them is, thus, useful, as it reveals systematic effects due to higher-order terms, which may differ for different $x$ regions.
The present study suggests that the large-$|x|$ regions suffer more from such higher-order effects.
As Fig.~\ref{fig:RIvsMS} indicates, there may be significant two-loop effects in the conversion/evolution/matching procedures.
Thus, a two-loop computation is indeed necessary.
The two-step procedure allows for separating the conversion, the evolution and the matching and hence,  a two-loop computation even for one of these, which is simpler than the full computation of the one-step procedure, can provide some insight.

\subsection{Truncation of the Fourier transform}
\label{subsec:truncation}

\begin{figure}[h!]
\begin{center}
\includegraphics[width=0.98\textwidth]{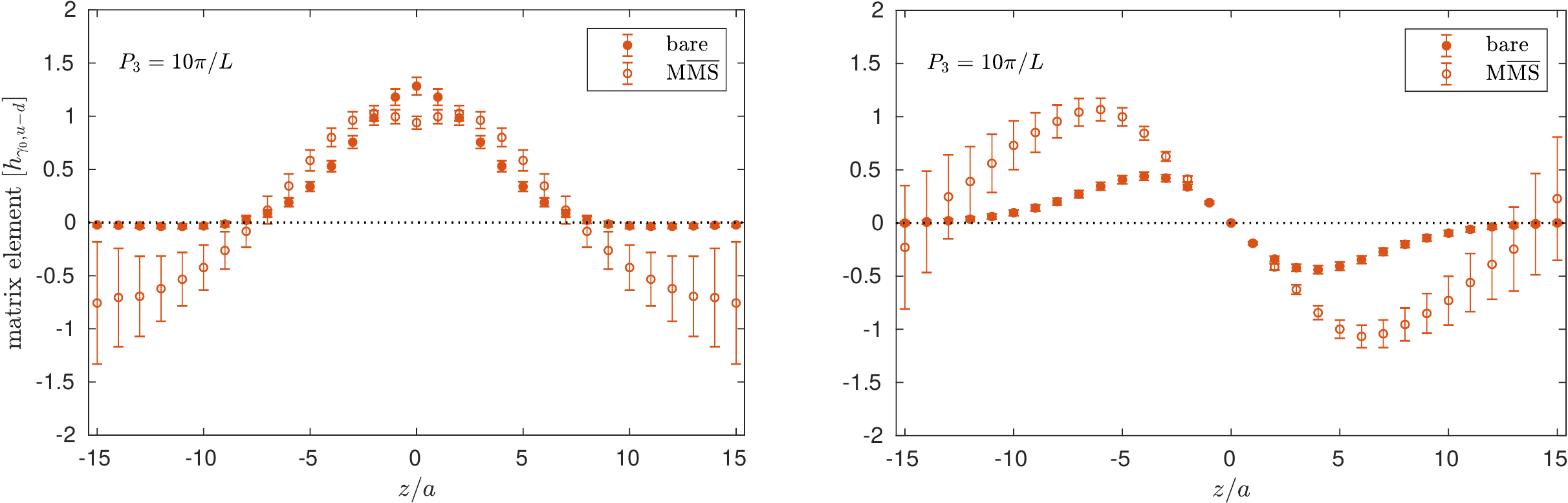}

\includegraphics[width=0.98\textwidth]{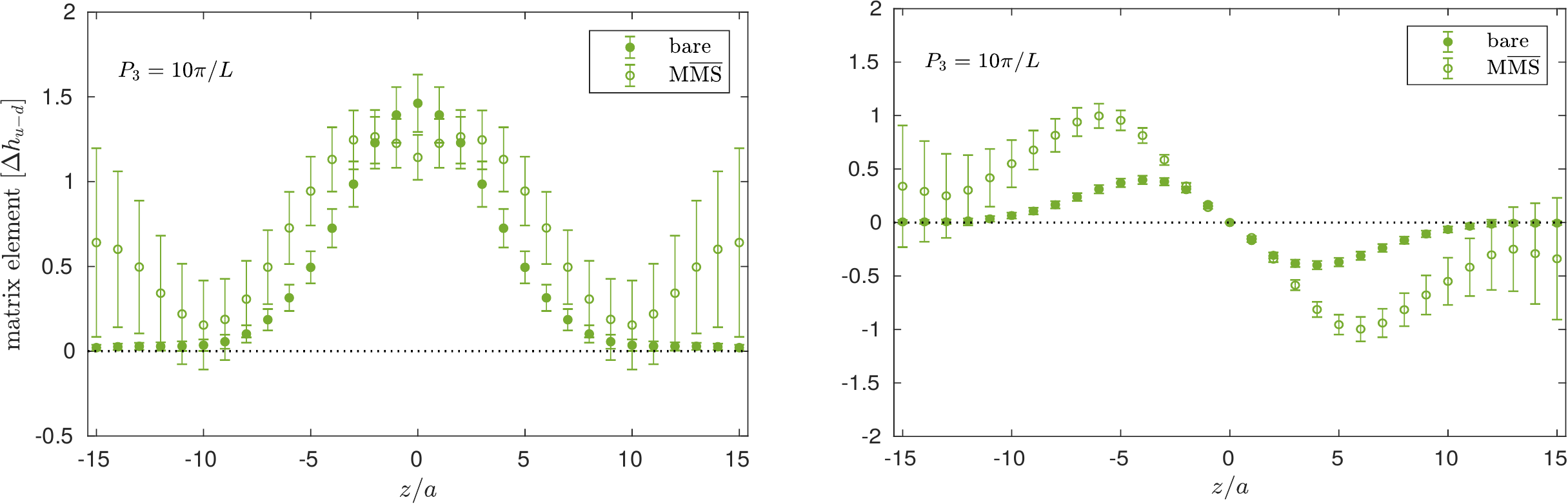}

\includegraphics[width=0.98\textwidth]{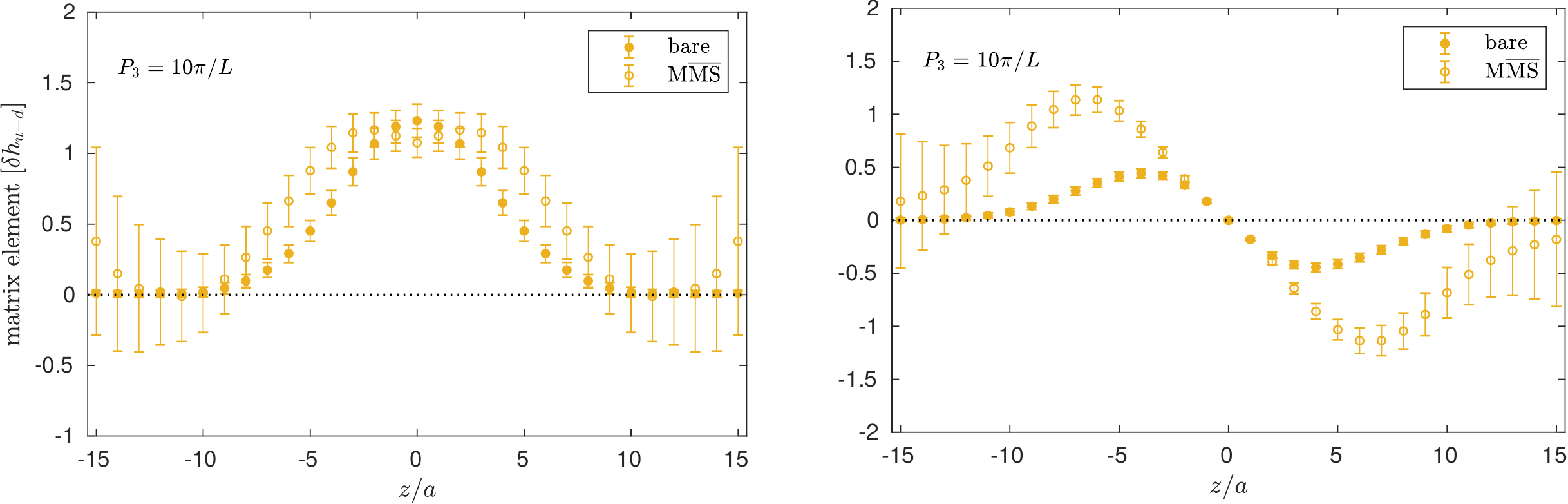}
\end{center}
\vspace*{-0.5cm}
\caption{Real (left) and imaginary (right) parts of the bare (filled symbols) and $\MMSb$-renormalized (open symbols) matrix elements for the unpolarized (upper row), helicity (middle row) and transversity PDFs (lower row); nucleon momentum $P_3{=}10\pi/L$.}
\label{fig:MEp5}
\end{figure}

We turn now to the investigation of systematic effects related to the Fourier transform (FT) that is applied on the renormalized matrix elements in $z$-space, giving the quasi-PDFs in $x$-space.
In principle, the FT integrates over all Wilson line lengths from zero to infinity, whereas lattice provides data for a finite number of discrete lengths $z/a$. 
Hence, one needs to decide about the maximum value of $z/a$ taken in the discretized FT integral. 
Ideally, it should be a value for which both the real part and the imaginary part of the matrix elements have decayed to zero.
For unrenormalized matrix elements, the choice of such a value poses no problem.
As can be seen in Fig.~\ref{fig:MEp5}, the bare matrix elements are zero for any $|z/a|\gtrsim15$.
However, renormalized matrix elements involve multiplication of the bare ones by complex $Z$-factors that mix bare real and imaginary parts, according to
\begin{equation}
\label{eq:Re}
{\rm Re}[h^{\rm ren}]={\rm Re}[Z]\,\, {\rm Re}[h^{\rm bare}]-{\rm Im}[Z]\,\, {\rm Im}[h^{\rm bare}]\,,
\end{equation}
\begin{equation}
\label{eq:Im}
{\rm Im}[h^{\rm ren}]={\rm Re}[Z]\,\, {\rm Im}[h^{\rm bare}]+{\rm Im}[Z]\,\, {\rm Re}[h^{\rm bare}].
\end{equation}

\begin{figure}[h!]
\begin{center}
\begin{minipage}[t]{0.49\linewidth}
    \includegraphics[width=\textwidth]{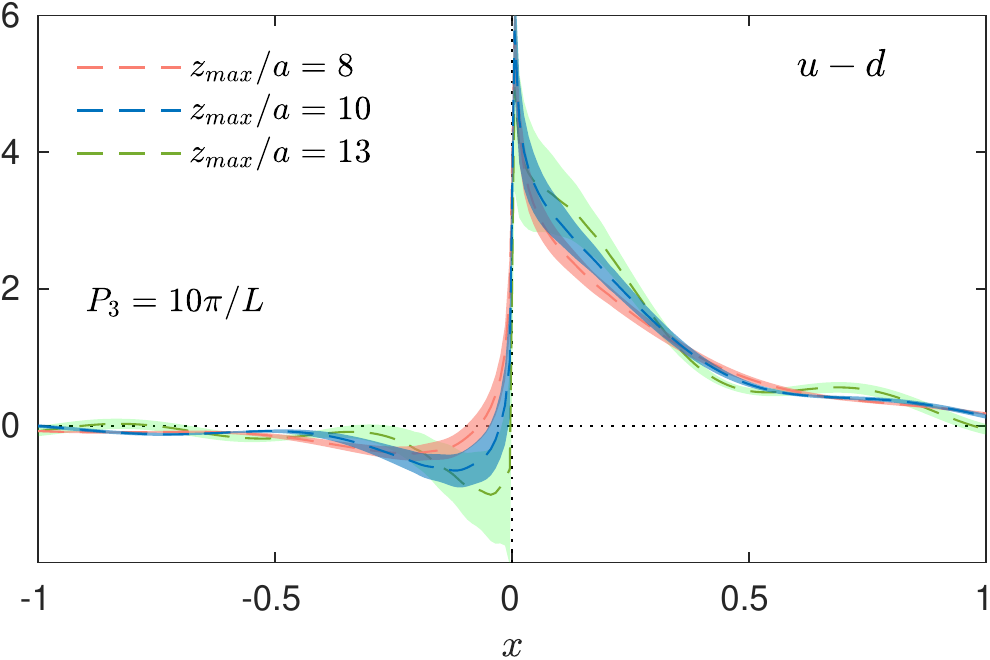}
  \end{minipage}
  \begin{minipage}[t]{0.495\linewidth}
    \includegraphics[width=\textwidth]{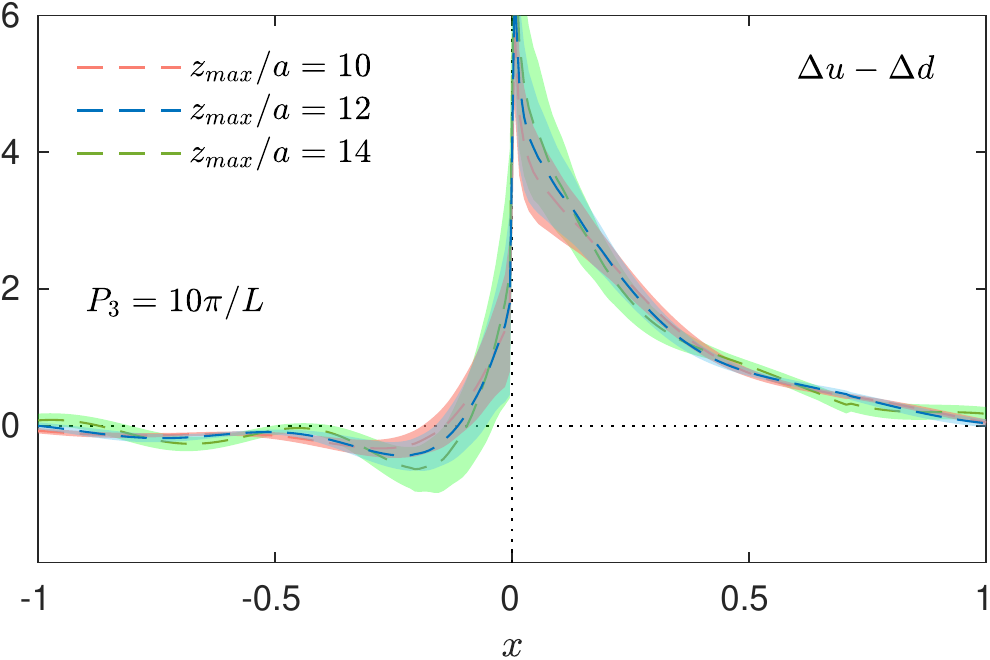}
  \end{minipage}
  \begin{minipage}[t]{0.495\linewidth}
    \includegraphics[width=\textwidth]{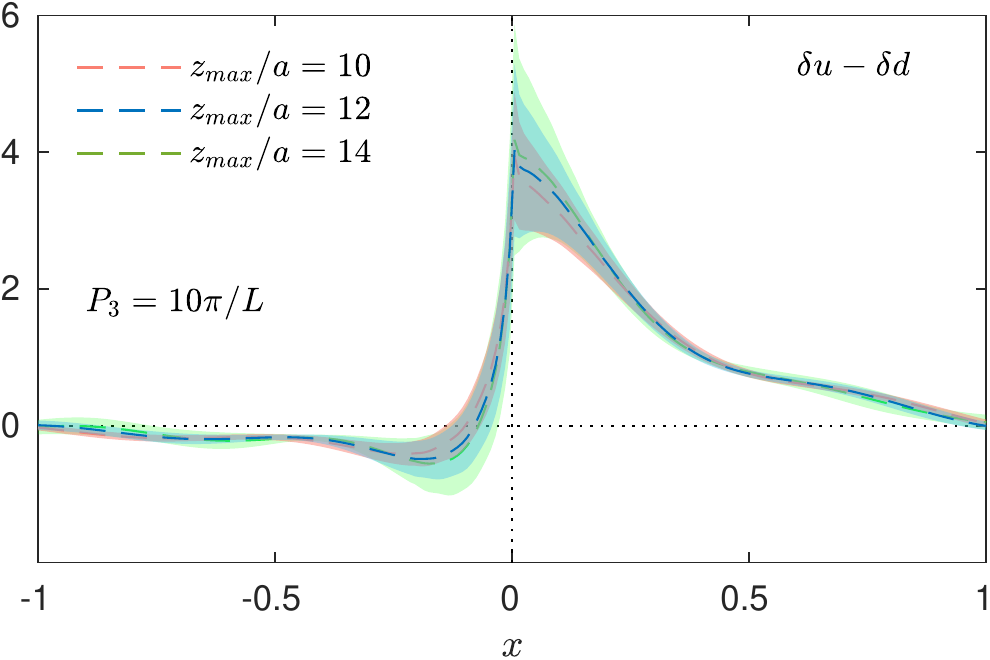}
  \end{minipage}
\end{center}
\vspace*{-0.5cm} 
\caption{Matched unpolarized PDF (upper right), helicity PDF (lower left) and transversity PDF (lower right) for different values of the cutoff $z_{\rm max}/a{=}$10 (red), 12 (blue), 14 (green)). Our final choice is always the middle value value of $z_{\rm max}/a$ from the ones shown. Nucleon momentum is $P_3{=}10\pi/L$ for all plots.}
\label{fig:zmax}
\end{figure}

As a result, the real part of renormalized matrix elements can be non-zero even if the real part of bare matrix elements has already decayed to zero. This stems from an unphysical truncation effect in the perturbative conversion between the intermediate RI$^\prime$ scheme $Z$-factors and the $\MSb$ ones.
Analogously, the imaginary part of renormalized matrix elements gets contaminated by ${\rm Im}[Z]$ that multiplies the real part of bare matrix elements and leads to non-physical contributions.
Moreover, the large values of $Z$-factors at large Wilson line lengths amplify the bare matrix elements and even if the latter are compatible with zero, this amplification introduces very large noise to the data, which propagates through the FT and matching to the final PDFs.
These effects call for a truncation of renormalized matrix elements at some justified value of $z/a$, that we call $z_{\rm max}/a$. 
In Fig.~\ref{fig:zmax}, we illustrate the $z_{\rm max}/a$-dependence of the resulting final PDFs for all operators  and for our largest momentum.
The smallest (largest) value of $z_{\rm max}/a$  is chosen according to where the real (imaginary) part of the renormalized matrix elements is compatible with zero.

For the transversity case (lower plot of Fig.~\ref{fig:zmax}), we observe very little dependence on the choice of $z_{\rm max}/a$ (apart from increased statistical noise for larger values of $z_{\rm max}/a$) and we choose $z_{\rm max}/a{=}12$ as the value for our final plots.
In this case, both the real and the imaginary parts of renormalized matrix elements are compatible with zero.
Hence, the unphysical effects of truncating the Fourier transform at finite $z_{\rm max}/a$ are minimal.
For the unpolarized case, a value of $z_{\rm max}/a$ where ${\rm Re}[h^{\rm ren}]$ and ${\rm Im}[h^{\rm ren}]$ are both zero does not exist.
Hence, we observe that the variation of $z_{\rm max}/a$ leads to a visible effect in the upper left plot of Fig.~\ref{fig:zmax}.
However, matched PDFs from all considered values of the cutoff are compatible with one another.
In the end, we choose the middle value, $z_{\rm max}/a{=}10$, as the final one.
In this way, the unphysical effect of truncation is split between the real and the imaginary parts.
For the former, the matched PDF gets a contribution from unphysical negative values of ${\rm Re}[h^{\rm ren}]$ (resulting from the imaginary part of $Z$-factors being non-zero due to the truncation of the perturbative conversion between renormalization schemes).
In turn, for the latter, part of the contribution from the imaginary part of renormalized matrix elements is missing.
In the helicity case, ${\rm Re}[h^{\rm ren}]$ decays to zero at a similar value of $z/a$ as for the other cases and, contrary to the unpolarized matrix elements, does not go below zero. 
However, $|{\rm Im}[h^{\rm ren}]|$, reaches a minimum around $z/a\approx12-13$ and then increases again, to become compatible with zero at $z/a\approx17-18$, within huge errors.
As our preferred value of the cutoff, we choose $z_{\rm max}/a{=}12$, where the real part is compatible with zero and the absolute value of the imaginary part is locally minimal.
Varying $z_{\rm max}/a$ (upper right plot of Fig.~\ref{fig:zmax}) leads to changes in the matched helicity PDF that are statistically insignificant.

\begin{figure}[h!]
\begin{center}
\begin{minipage}[t]{0.495\linewidth}
    \includegraphics[width=\textwidth]{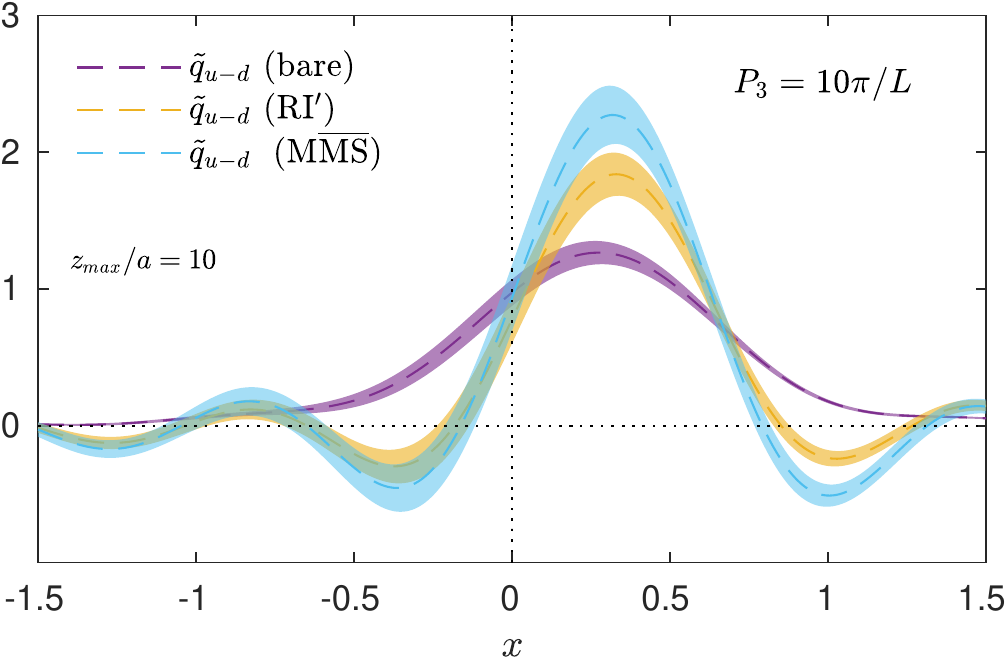}
\end{minipage}
\end{center}
\vspace*{-0.5cm}
\caption{ The unpolarized quasi-PDFs for $P_3{=}10\pi/L$. We show the  bare matrix element (purple), RI$^\prime$ (yellow) and $\MSb$ (cyan) quasi-PDFs for $z_{\rm max}{=}10a$.}
\label{fig:zmax10}
\end{figure}

Apart from the non-physical effects appearing when the renormalized matrix elements are not compatible with zero at $z_{\rm max}/a$, further non-physical effects are introduced by the FT if the decay of $h^{\rm ren}$ is not fast enough.
Namely, the periodicity of the Fourier transform leads to unphysical oscillations in the quasi-PDFs and finally in the matched PDFs.
This effect is visible in all of our final distributions as a distorted approach of PDFs to zero for $x$ between 0.5 and 1, as a mild oscillatory behavior for large negative $x$, and as an unphysical minimum in the antiquark part at $-x\approx0.1-0.2$.
Decomposing the FT into cosine and sine transforms, one can observe that the quasi-PDF is bound to be negative if matrix elements entering the transforms do not decay to zero fast enough.
The rate of decay of these matrix elements depends on the nucleon boost, as was demonstrated in Fig.~\ref{fig:mom_dep}.
For larger momenta, the decay becomes faster, i.e.\ zero is reached at a smaller value of $z/a$, both for the real and the imaginary parts.
Thus, the unphysical oscillations that we observe in the data may be attributed to insufficiently high momentum and therefore be considered higher-twist contaminations.
In Fig.~\ref{fig:zmax10}, we display the results for the unpolarized quasi-PDF for $z_{\rm max}{=}10a$, comparing bare, RI$^\prime$ and $\MMSb$ renormalized cases.
As can be seen, the oscillations do not emerge when Fourier-transforming bare matrix elements that decay to zero faster. 
The amplification of the matrix elements by the $Z$-factors is responsible for their relatively slow decay.
At this stage, one needs to remember that the $Z$-factors, as calculated now, are subject to two kinds of non-physical effects: lattice artifacts in their non-perturbative evaluation on a hypercubic lattice in the  RI$^\prime$ scheme, and truncation effects in the perturbative conversion to the $\MMSb$ scheme and evolution to the reference scale of 2 GeV.
After curing these effects, by computing higher-order conversion formula and by subtracting lattice artifacts computed in lattice perturbation theory~\cite{MC_HP_artifacts}, we expect that the oscillatory behavior will be considerably less prominent.
This is supported also by the behavior observed for the quasi-PDF renormalized using the RI$^\prime$ scheme, which has no conversion truncation effects, where we find that the oscillation is reduced with respect to the $\MMSb$-renormalized quasi-PDF, as seen in  Fig.~\ref{fig:zmax10}.
However, even in the RI$^\prime$ scheme, the bare matrix elements are still amplified by the $Z$-factors, whose real part is exponentially increasing due to the presence of the power divergence related to the Wilson line.
Thus, the oscillatory behavior in renormalized PDFs can only be avoided if the bare matrix elements decay faster, which can only happen for larger nucleon boosts.

\begin{figure}[h!]
\begin{center}
\begin{minipage}[t]{0.495\linewidth}
    \includegraphics[width=1\textwidth]{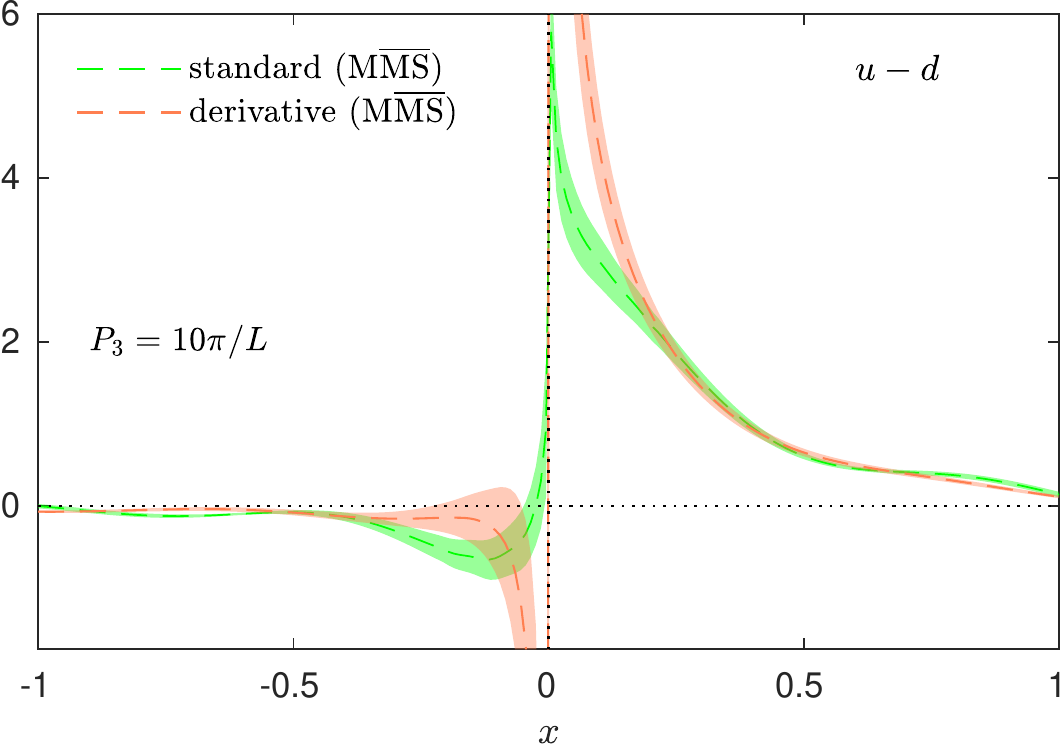}
  \end{minipage}
\begin{minipage}[t]{0.495\linewidth}
    \includegraphics[width=1\textwidth]{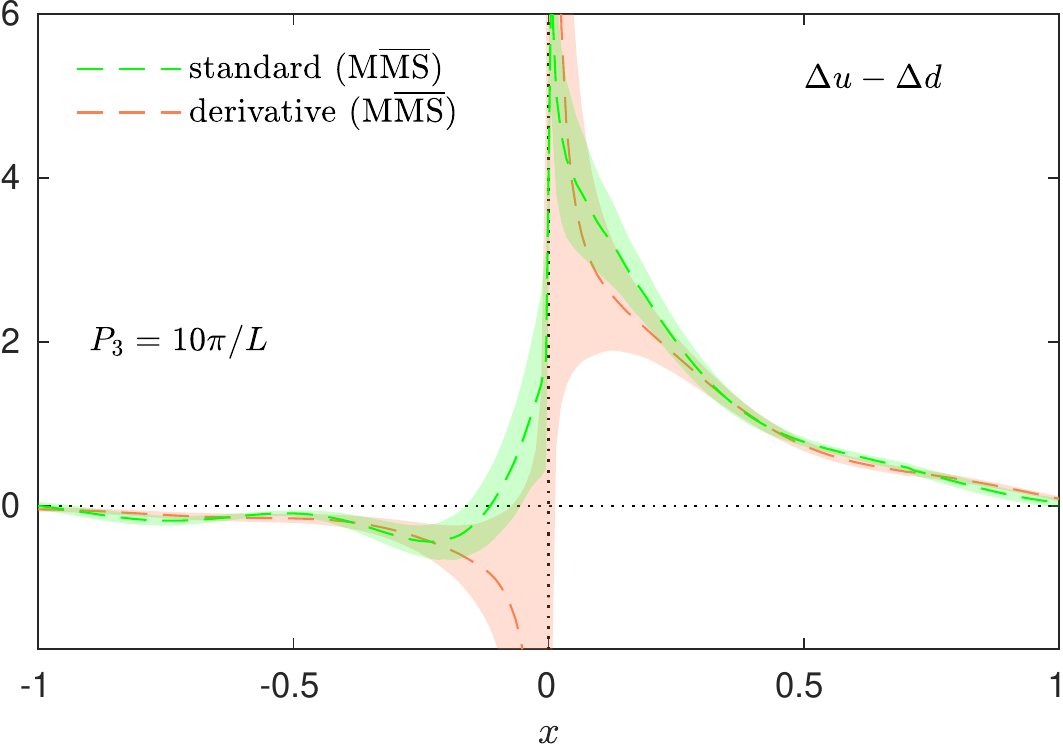}
  \end{minipage}
\begin{minipage}[t]{0.495\linewidth}
    \includegraphics[width=1\textwidth]{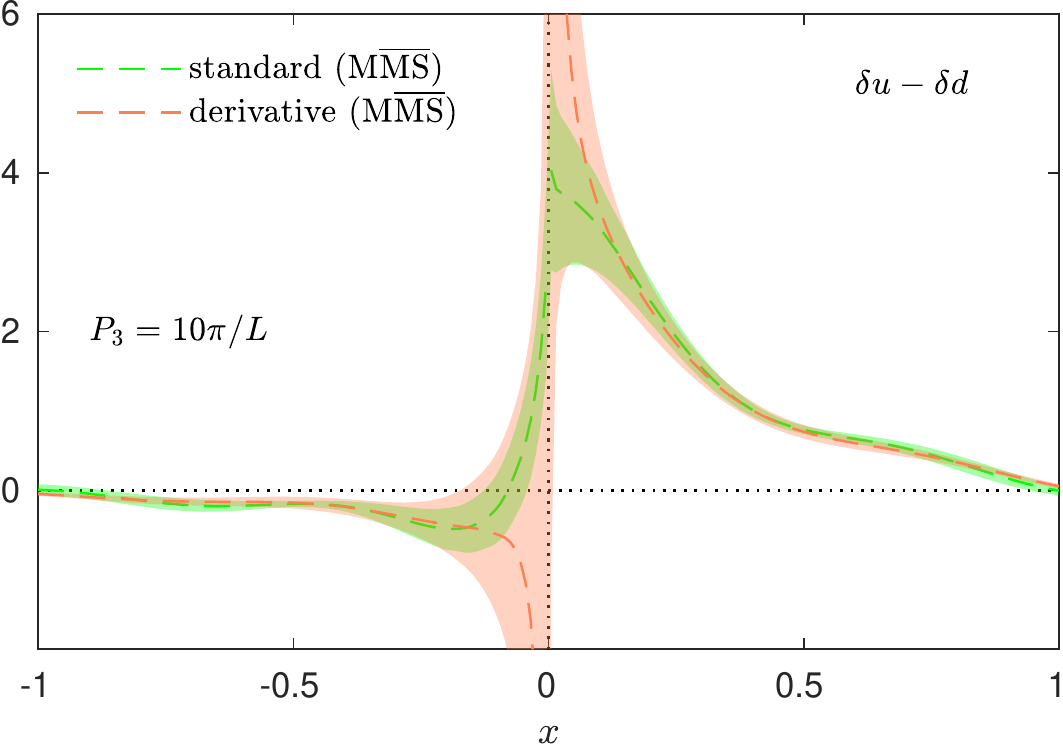}
  \end{minipage}
\end{center}
\vspace*{-0.5cm} 
\caption{Comparison of matched PDFs obtained from the same renormalized matrix elements, but with different Fourier transform definitions: the standard one (green band) or the ``derivative'' one (red band). The matching procedure is the same for both and proceeds via the $\MMSb$ scheme. We show the unpolarized (upper left), helicity (upper right) and transversity cases (lower). Nucleon momentum is $P_3{=}10\pi/L$.}
\label{fig:derivative_vs_standard}
\end{figure}

An approach that may address non-physical oscillations was proposed in Ref.~\cite{Lin:2017ani}.
The idea is to rewrite the FT that yields the quasi-PDF, using integration by parts as follows:
\begin{equation}
\label{eq:deriv}
\tilde{q}(x)=h(z)\frac{e^{ixzP_3}}{2\pi ix}\Big|_{-z_{\rm max}}^{z_{\rm max}} - \int_{-z_{\rm max}}^{z_{\rm max}} \frac{dz}{2\pi} \frac{e^{ixzP_3}}{ix} h'(z),
\end{equation}
where $h'(z)$ is the derivative of the matrix elements with respect to the Wilson line length $z$.
This step is exact, however, the proposed approximation is to neglect the surface term, even if it is non-vanishing.
The latter happens when the matrix elements have not decayed to zero at $z{=}z_{\rm max}$.
The resulting FT, given by the second term of Eq.~(\ref{eq:deriv}), thus, involves the derivatives of renormalized matrix elements, which decay to zero faster than the matrix elements themselves.
As argued above, a fast enough decay of the transformed matrix elements is sufficient to avoid oscillations.
In this so-called ``derivative'' method, the oscillations are effectively transferred to the neglected surface term.
We note, however, that this procedure is dangerous, due to the $1/x$ factor of the surface term, which leads to an uncontrolled effect for small values of $x$, where also the exponential of the surface term has a large real part.
This aspect is illustrated in Fig.~\ref{fig:derivative_vs_standard} for all three types of PDFs.
The explicit $1/x$ factor amplifies the final PDF for small $|x|$, leading to its huge values, particularly in the small positive-$x$ region. 
This is very visible in the case of the unpolarized PDF, where the results are statistically the most precise.
In the polarized cases, in the small-$|x|$ regions, one observes that the statistical errors are significantly enhanced by the ``derivative'' method.
All three final PDFs evince unphysical divergences to $-\infty$ as $x\rightarrow0$ in the antiquark part.
Furthermore, the oscillatory behavior is very mildly affected, undermining the very aim of using the ``derivative'' procedure.

The difference between our  results obtained using the standard FT and the ``derivative'' method, vanishes if $h(|z|\geq z_{\rm max}){=}0$.
Only in this  case, one can claim control over different regions of $x$.
The problem associated with renormalized matrix elements not decaying fast enough to zero or, equivalently, having a higher-twist contaminations in the PDF, cannot be artificially hidden by using the ``derivative'' method. 
This is because the problem does not disappear, since the Fourier transform that defines the quasi-PDF is still truncated and part of the truncation encoded in the surface term may be hiding physical information relevant for reconstructing the full PDF.
In other words, for a fully reliable extraction of PDFs, the surface term has to go to zero when applying actual lattice QCD data, and not by setting it to zero by hand.
Another disadvantage of using the ``derivative'' method is that it introduces additional discretization effects, which need to be controlled.
Moreover, as argued very recently in Ref.\ \cite{Karpie:2019eiq}, the ``derivative'' method does not solve the problem with the Fourier transform, shown to be ill-conditioned if truncation is necessary and the data are limited to $\mathcal{O}(10)$ values of $z/a$.
For all the above reasons, we opt not to use the ``derivative'' method  and treat the residual oscillations as an indication that the nucleon boost needs to be increased in a controlled manner to extract reliable results on the PDFs.
A possible way to improve this aspect for the currently attained nucleon momenta is advocated in Ref.\ \cite{Karpie:2019eiq}, using advanced reconstruction methods, such as Backus-Gilbert or neural networks.
It is an interesting direction for further study and we plan to explore these ideas in the future.

\section{Final results - Discussion}
\label{sec:results}

In this section, we present our final results for the collinear non-singlet quark PDFs, whose determination was guided by the extensive investigation of
 several systematic effects presented in the previous sections.
Although  it is not possible to quantify all of the systematics, this investigation clearly points to the steps that one must follow in the future  in order to arrive  at  precise extractions of PDFs from lattice QCD, with systematic uncertainties under control. We will be discussing these steps together with the presentation of the final results.

\begin{figure}[h!]
\begin{center}
\begin{minipage}[t]{0.495\linewidth}
    \includegraphics[width=1\textwidth]{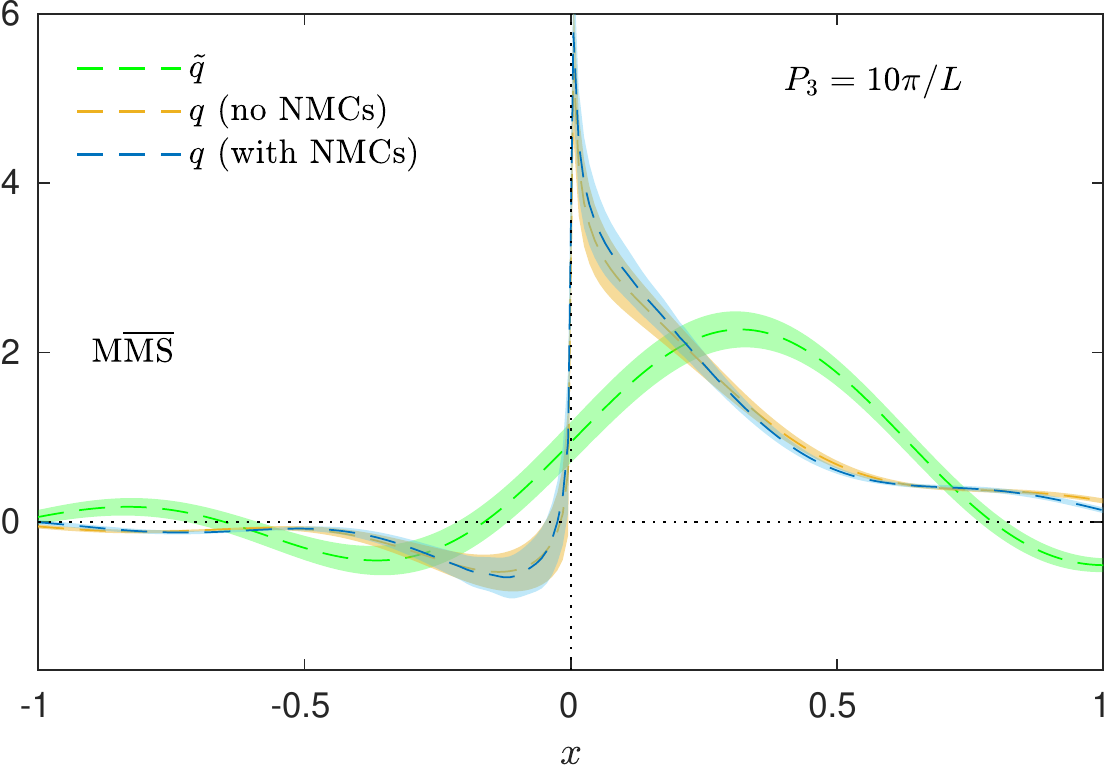}
  \end{minipage}
\begin{minipage}[t]{0.495\linewidth}
    \includegraphics[width=1\textwidth]{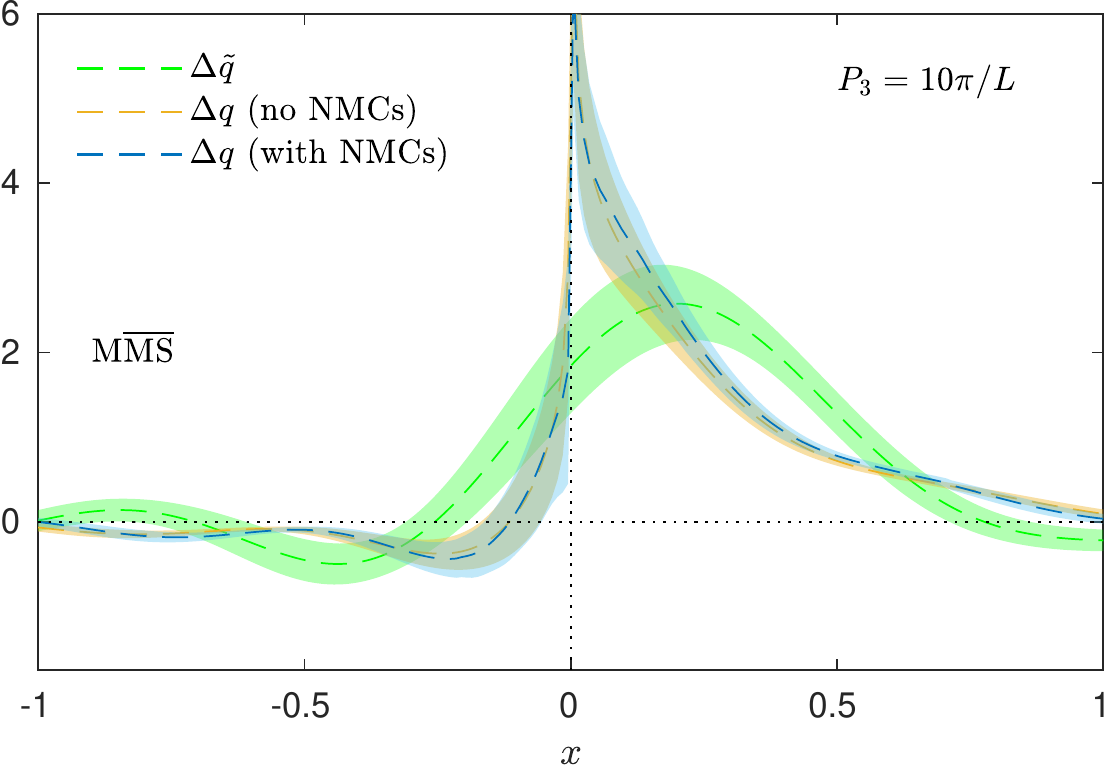}
  \end{minipage}
\begin{minipage}[t]{0.495\linewidth}
    \includegraphics[width=1\textwidth]{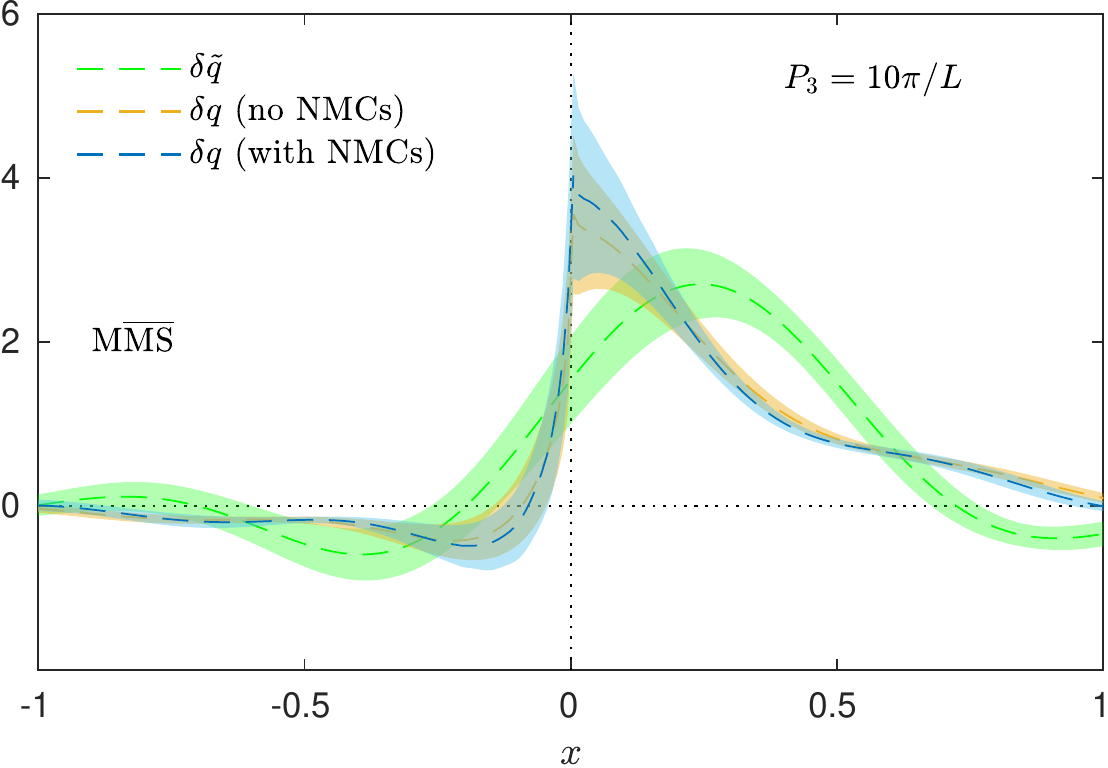}
  \end{minipage}
\end{center}
\vspace*{-0.5cm} 
\caption{Effect of matching and nucleon mass corrections in the unpolarized PDF (upper left), helicity PDF (upper right) and transversity PDF (lower). Nucleon momentum is $P_3{=}10\pi/L$ for all plots.}
\label{fig:decomposed}
\end{figure}

In Fig.\ \ref{fig:decomposed}, we show the  quasi-PDFs, the matched PDFs and the final PDFs that take into account the nucleon mass corrections (NMCs) for the largest value of the momentum.
The nucleon mass corrections  are performed according to the formulae of Ref.\ \cite{Chen:2016utp}. They are  $\mathcal{O}(m_N^2/\Lambda_{\rm QCD}^2)$ corrections to the PDF resulting from the  finite mass of the nucleon and they can be derived in a closed form to all orders.
We note that the one-loop matching that connects the  quasi-PDFs to the  light-cone PDFs is still relatively large for our largest nucleon boost.
Increasing the nucleon boost while excited states contamination is fully under control is one aspect that has to be addressed in future computations. Another is the two-loop perturbative conversion between renormalization schemes and evolution to the chosen reference scale. This will further reduce the dependence on the initial RI$'$ scale, and may be used to quantify truncation effects in the conversion factor.
NMCs, on the other hand,  are very small and negligible in most regions of $x$ and thus not expected to contribute to  large systematic effects.

\begin{figure}[h!]
\begin{center}
\begin{minipage}[t]{0.495\linewidth}
    \includegraphics[width=1\textwidth]{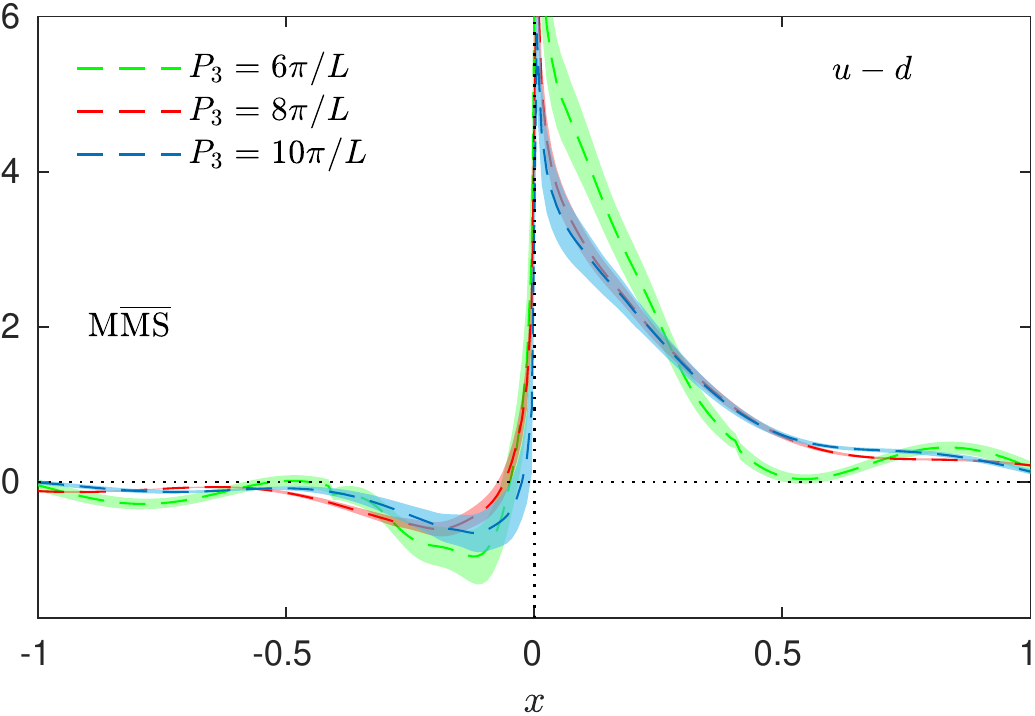}
  \end{minipage}
\begin{minipage}[t]{0.495\linewidth}
    \includegraphics[width=1\textwidth]{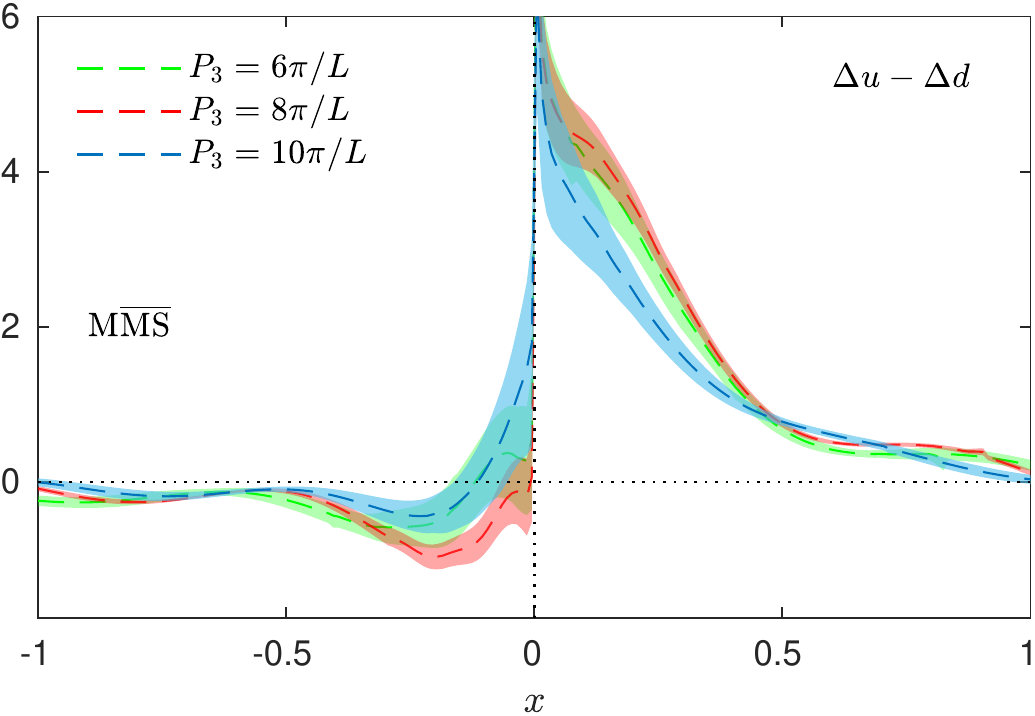}
  \end{minipage}
\begin{minipage}[t]{0.495\linewidth}
    \includegraphics[width=1\textwidth]{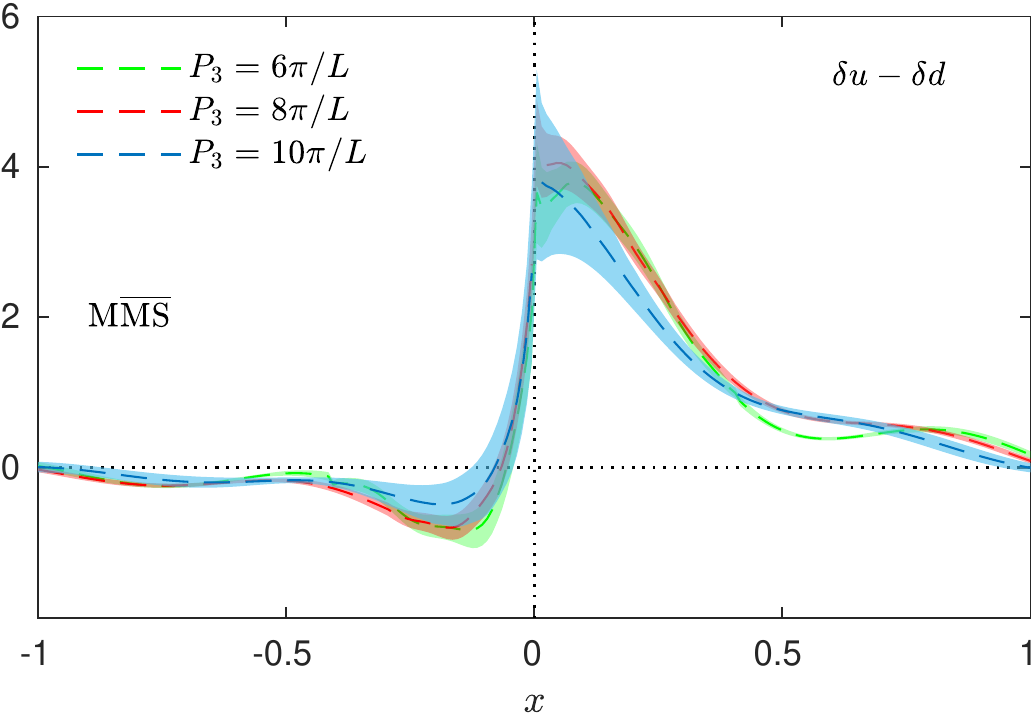}
  \end{minipage}
\end{center}
\vspace*{-0.5cm} 
\caption{Nucleon boost dependence for the unpolarized PDF (upper left), helicity PDF (upper right) and transversity PDF (lower), using $P_3=6\pi/L$ (green curve), $P_3{=}10\pi/L$ (red curve) and $P_3{=}10\pi/L$ (blue curve).}
\label{fig:PDFmom}
\end{figure}

In Fig.~\ref{fig:PDFmom} we show the  dependence of the final PDFs on the nucleon boost.
For the unpolarized PDF, we observe a strong effect when increasing the momentum from $6\pi/L$ to $8\pi/L$, which seems to converge after that yielding values
that are compatible with those obtained for nucleon boost of $10\pi/L$ over almost the whole range of  $x$.
In addition, the oscillatory behavior becomes milder as the momentum increases.
As we discussed above, this behavior results from the fact that for larger boosts, renormalized matrix elements become consistent with zero at smaller Wilson line lengths $z$, and affects the region $x{\approx}1$, as can be seen from the plots. 
For the helicity PDF, on the other hand, increasing the momentum from $6\pi/L$ to $8\pi/L$ is a minor effect, i.e.\ the final PDFs are compatible with each other for both of these momenta, over almost  the whole range of $x$.
However, increasing the momentum from $8\pi/L$ to $10\pi/L$, a significant shift is observed, especially for small values of $x$.
Thus, it is likely that higher-twist effects are more pronounced for the helicity PDF.
The  oscillatory behavior is also milder in this case as the momentum increases, and the final PDF is compatible with zero at $x{=}1$.
 For the  transversity PDF, the momentum dependence is the mildest and results using momentum $6\pi/L$ and $8\pi/L$ are compatible over a wide range of $x$-values.
Similarly to the unpolarized and helicity PDFs, the unphysical oscillations are significantly reduced and the transversity PDF vanishes at $x{=}1$ for momentum $10\pi/L$.

\begin{figure}[h!]
\begin{center}
\begin{minipage}[t]{0.495\linewidth}
    \includegraphics[width=1\textwidth]{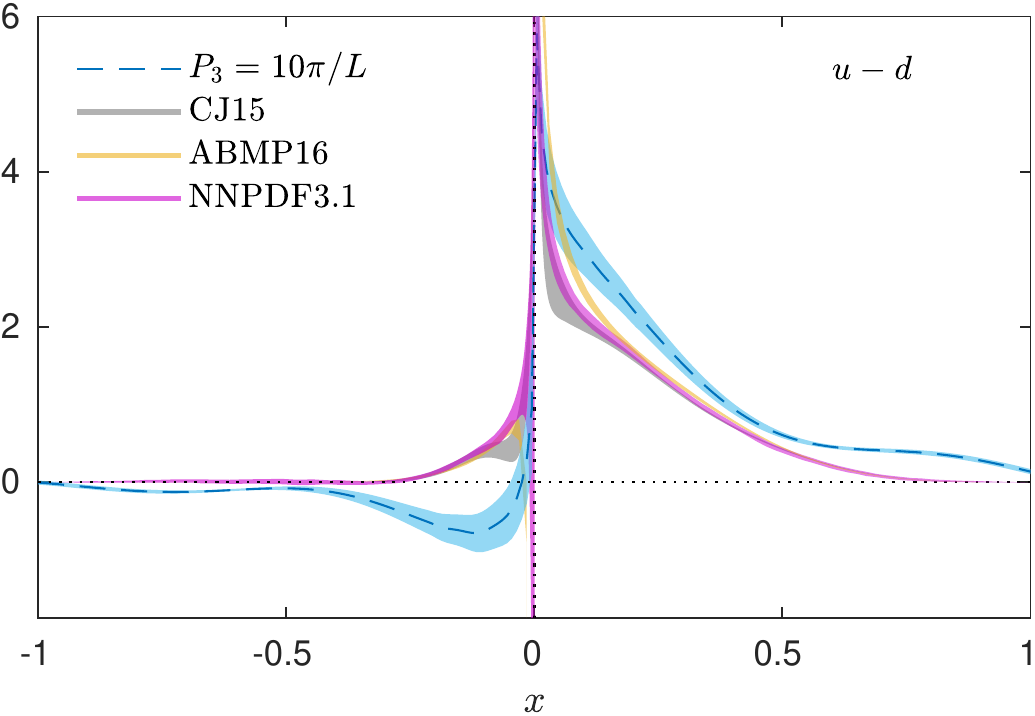}
  \end{minipage}
\begin{minipage}[t]{0.495\linewidth}
    \includegraphics[width=1\textwidth]{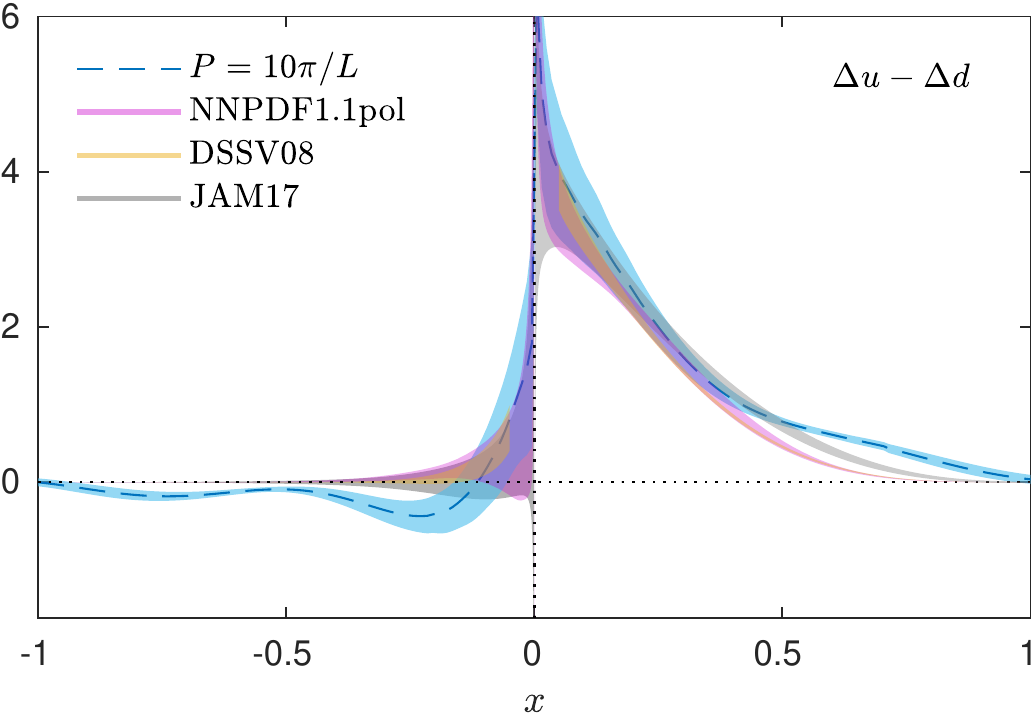}
  \end{minipage}
\begin{minipage}[t]{0.495\linewidth}
    \includegraphics[width=1\textwidth]{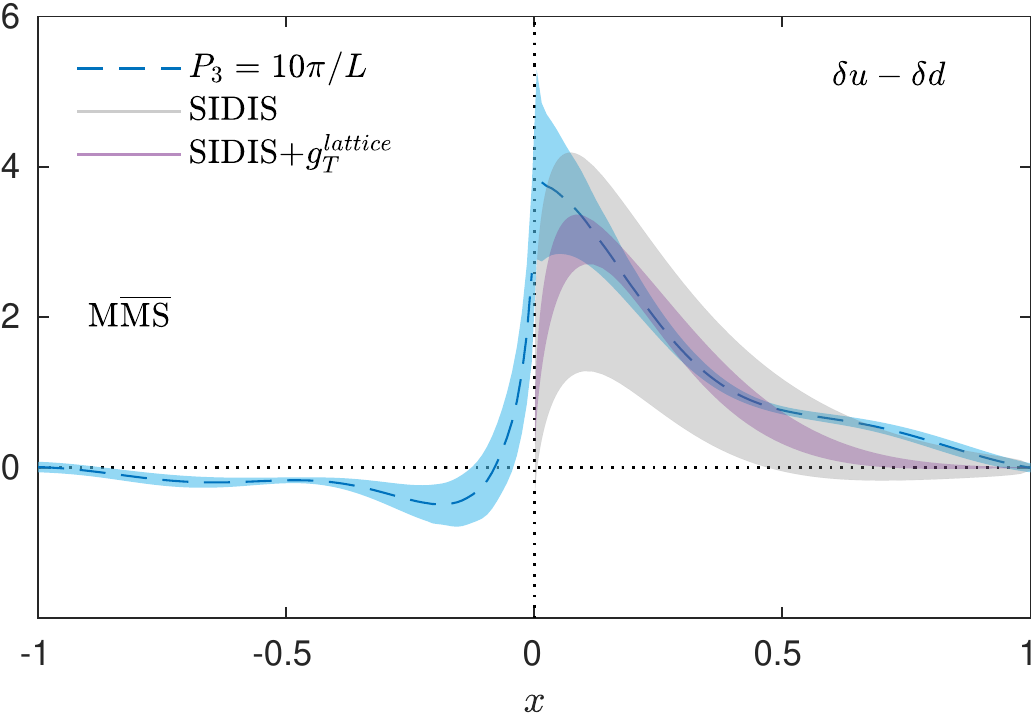}
  \end{minipage}
\end{center}
\vspace*{-0.5cm} 
\caption{Final results for the unpolarized PDF (upper left), helicity PDF (upper right) and transversity PDF (lower), using the largest momentum $P_3{=}10\pi/L$ (blue curve). The global fits of Refs.~\cite{{Alekhin:2017kpj},{Ball:2017nwa},{Accardi:2016qay}} (unpolarized) , Refs.~\cite{{deFlorian:2009vb},{Nocera:2014gqa},{Ethier:2017zbq}} (helicity) , Refs.~\cite{{Lin:2017stx}} (transversity) are shown for qualitative comparison.}
\label{fig:PDFpheno}
\end{figure}

We compare our final  PDFs at the largest momentum with the phenomenologically extracted ones in Fig.~\ref{fig:PDFpheno}. For the phenomenological
results, we use the ones by CJ15~\cite{Accardi:2016qay}, ABMP16~\cite{Alekhin:2017kpj} and  NNPDF3.1~\cite{Ball:2017nwa} for the unpolarized PDF, DSSV08~\cite{deFlorian:2009vb}, NNPDF1.1pol~\cite{Nocera:2014gqa} and JAM17~\cite{Ethier:2017zbq} for the helicity PDF and two parameterizations for the PDFs transversity, one extracted from experimental SIDIS data and one where the tensor charge computed on lattice QCD was used as input in the phenomenological analysis~\cite{Lin:2017stx}. 
We stress that the comparison with phenomenological PDFs is intended to be qualitative, since we only include statistical uncertainties.
The unpolarized PDF 
 has a similar slope to the phenomenological PDFs, but lies above them in the positive $x$-region.
Increasing the momentum should bring the PDF closer to the  phenomenological values, as demonstrated  in Fig.\ \ref{fig:PDFmom}.
Our results for the   helicity PDF are compatible with  phenomenological ones for $x\lesssim0.4-0.5$.
While this may indicate faster convergence of the quasi-PDF, it can also be the result of different systematic effects, with some of them possibly canceling out.
Likewise, our result for the transversity PDF is in agreement with both the phenomenologically extracted one, as well as the one using the lattice determination of the tensor charge as input, for $x\lesssim0.4-0.5$.
An interesting feature is that the precision of our results  is better than in both the phenomenological determinations.
Thus, lattice QCD holds the prospect of impacting significantly  our knowledge of the transversity PDF, in particular as we expect 
that in the next generation of lattice QCD computations of  PDFs  a better control of the  systematic effects will be achieved.

The errors shown in the above figures are statistical, and significant effort is needed to properly quantify systematic uncertainties, present in the various steps of the analysis. Currently, these consist the dominating source of uncertainty, and addressing them will allow one to draw final conclusions from the lattice results. Based on the current studies, it is not possible to quantify and hierarchize them. Typical systematics related to the lattice calculation include discretization effects, finite volume effects and the role of the pion mass value. The pion mass of the ensemble used here is already at its physical value and hence, there is no need for a chiral extrapolation.
The latter would be an important source of uncertainty, as the fit function is not known. We remark that, in practice, the effect of a non-physical pion mass can be large, as we demonstrated in Ref.\ \cite{Alexandrou:2018pbm}, comparing the physical pion mass PDF with one at $m_\pi\approx375$ MeV.

The parameters of the ensembles are expected to satisfy certain criteria for the range of values of the pion mass, the volume and the lattice spacing, to study uncertainties such as:

\vspace*{-0.1cm}
\noindent
$\bullet$ {\textit{Cutoff effects}}: 
A reliable control of cutoff effects requires at least three values of the lattice spacing  smaller than 0.1 fm.
Normally, cutoff effects are found to be relatively small in lattice hadron structure calculations.
In the quasi-PDF computation, the nucleon is boosted to momenta for which $P_3$ becomes significant in comparison to the inverse lattice spacing and this may lead to increased cutoff effects.
We note that  for our largest momentum, we have $aP_3{=}0.65$ which is below the lattice cutoff (unlike Refs.~\cite{Chen:2018xof,Lin:2018qky,Liu:2018hxv} where the employed nucleon momenta are significantly above the lattice cutoff), and the continuum dispersion relation is still satisfied, as shown in Fig.~\ref{fig:dispersion_relation}. Still, it is unclear to what extent the good quality of the dispersion relation translates into discretization effects of the matrix elements considered here.
 
\vspace*{-0.1cm}
\noindent
$\bullet$ {\textit{Finite volume effects}}: 
Similarly to discretization effects, finite volume effects are also usually found to be rather small in hadron structure observables.
The situation with quasi-PDFs is likely to be somewhat more complicated, since we use operators with  Wilson lines of significant length.
The volume behavior of such extended operators was considered by Brice\~no et al.\ \cite{Briceno:2018lfj} within a model using current-current correlators in a scalar theory.
Despite the fact that the model is not directly applicable to our investigation, it does provide a warning that the suppression of finite volume effects for matrix elements of spatially extended operators may change from the standard $\exp(-m_\pi L)$ to $(L^m/|L-z|^n)\exp(-m_N(L-z))$, where $m$ and $n$ are undetermined exponents, which may become dominating for large $z$. Thus, finite volume effects may turn out to be a significant source of systematics and their investigation is crucial in the future. 

\vspace*{-0.1cm}
\noindent
$\bullet$ {\textit{Systematic uncertainties in the determination of the renormalization functions}}: 
 Uncertainties also arise in the computation of renormalization functions due to the breaking of rotational invariance.
We have partly improved our work by subtracting lattice artifacts in the  renormalization factor of the quark field, computed in lattice perturbation theory (see Sec.\ \ref{sec:renorm}).
A similar subtraction for the $Z$-factors of local operators was very successful~\cite{Constantinou:2010gr,Constantinou:2014fka,Alexandrou:2015sea}, and we intend to perform such a subtraction in the future~\cite{MC_HP_artifacts}. We note that pion mass dependence and finite volume effects in the renormalization functions  studied here are found to be insignificant.

\vspace*{-0.1cm}
\noindent
$\bullet$ {\textit{Uncertainties specific to the quasi-PDF approach}}: 
The most significant uncertainty of this type is the requirement of having boosted nucleon states with values much larger than the nucleon mass.
Although we observe convergence when increasing the boost,  there are still undesirable effects from the finite value of the  momentum that
cause contamination from higher-twist effects (HTE), currently not known. 
HTE can be suppressed by going to larger boosts or alternatively explicitly computed and subtracted.
Moreover, relatively small values of the  momentum are likely to be responsible for unphysical oscillations, by making the matrix elements decay too slowly to zero, i.e.\ at rather large values of $z$.
Faster decay of these matrix elements will naturally suppress the oscillations and the only way to achieve this reliably is to increase the momentum. Other
\emph{ad hoc} prescriptions, such as the ``derivative'' method, do not actually remove the source of the oscillations. In this technique, it is unclear how to handle the small-$|x|$ region, since the ignored surface term contains an explicit $1/x$ factor. In addition, the need for discretizing the derivative of the matrix elements introduces additional cutoff effects.
Increasing the nucleon boost is difficult, as the signal is exponentially decaying at larger momenta and the latter also significantly increase excited states contributions, particularly when using simulations at the physical pion mass. As we argued, good control over excited states is necessary for extracting reliable physical results. 
Thus, a detailed study is required to  maintain ground state dominance, following the approach explained in detail in this work.
In this paper, we showed that excited states are suppressed below ${\sim}$10\% if the source-sink separation is 1.1 fm at $P_3{\approx}1.4$ GeV, as established with a detailed analysis based on three methods. We note that at the physical pion mass and at small source-sink separations, the contamination comes from tens of excited states, and thus, a one-state or a two-state fit alone are not reliable. Compatibility between one-state and two-state fits provides a strong argument that ground state dominance has been achieved.
Another possibly significant source of systematic errors are the perturbative truncation effects in the conversion to the $\MSb$ scheme, evolution to the reference scale and matching, all currently known to one-loop level. We observe that the effect of the latter is significant at our largest momentum and hence two-loop effects are likely to be substantial.
Therefore, a two-loop computation would lead to better connection to light-cone PDFs with smaller values of the nucleon momentum.

 \section{Conclusions}
\label{sec:conclusions}

This works presents a detailed investigation of the formalism employed to extract $x$-dependent PDFs within lattice QCD.
The analysis was carried out using one ensemble of $N_f{=}2$ twisted mass fermions with quark mass fixed to its physical value.
Results on the  collinear unpolarized, helicity and transversity PDFs were obtained for the isovector flavor combination. 
The quasi-PDF method used here relies on a computation of equal-time correlators involving boosted nucleons, with the momentum increased to large enough values, so that the Large Momentum Effective Theory is applicable.
In this work, we investigated three values of the  momentum, reaching a maximum of 1.38 GeV. 
All necessary components to obtain light-cone PDFs have been considered, namely, calculation of bare nucleon matrix elements, renormalization, matching to light-cone PDFs and subtraction of finite nucleon mass corrections. 
The work presented here is an extension of our earlier work~\cite{Alexandrou:2018pbm,Alexandrou:2018eet}, and includes details on technical and theoretical aspects, as well as improvements in various aspects of the calculation and examination of systematic effects.
In particular, we provide the lattice techniques used in the evaluation of the matrix elements, such as momentum smearing and methods to eliminate (within statistical accuracy) excited states contributions. 
Theoretical developments associated with the non-perturbative renormalization are presented, where a chiral extrapolation and a modification of the conversion to the $\MMSb$ scheme have been employed. 
The latter is a variant of the $\MSb$ scheme that ensures particle number conservation in the matching procedure. 
Issues related to the Fourier transform as well as different matching prescriptions were also explored.

Particular attention was given to discussing the role of systematic effects in the various steps of the analysis.
Typical systematics related to the lattice calculation that are also common in other hadron structure calculations include discretization effects and finite volume effects. These can be addressed and eliminated by simulations using additional gauge field configuration ensembles. At present, these are part of  the unquantified uncertainties, that need to be addressed in future studies. 
Another typical systematic enters lattice data if the pion mass of the ensemble is at an unphysical value. 
In this work, we use simulations at the physical point and chiral extrapolation is not needed. 
Analyses using non-physical values of the quark mass would introduce an important source of uncertainty, as the fit function to extrapolate the results to the physical point is not known.

Apart from the aforementioned systematics that are common in other hadron structure investigations, there are additional ones that are specific to the quasi-PDF approach. 
As discussed in the previous sections, increasing the nucleon boost can lead to further systematic uncertainties. 
High values of the momenta reduce the appearance of oscillations, but at the same time, increase the number of contributing excited states. 
Systematics related to the truncation of the conversion factor and the matching formula to one-loop level are non-negligible. 
In particular, having a matching formula to two-loops may lead to better convergence to light-cone PDFs at smaller nucleon momenta. 
Hence, a two-loop computation is strongly desired and would help to establish better connection to light-cone PDFs with smaller values of the nucleon momentum.

\medskip
Despite these uncertainties, this work demonstrates a tremendous  progress in the determination of PDFs from the quasi-distribution approach~\cite{Cichy:2018mum}. 
Lattice QCD results confirm the feasibility of extracting PDFs from first-principle calculations. 
The success in the quasi-PDF approach for the nucleon has also resulted in studies of other hadrons and alternative approaches, as summarized in Ref.~\cite{Cichy:2018mum}.
The theoretical and technical aspects are now well understood for the nucleon studies and addressing the systematic uncertainties is
the next step that will be enabled by using the methodology developed, in combination with the availability of increased computational resources.
In fact, our preliminary results  for an $N_f{=}2{+}1{+}1$ twisted mass ensemble with physical values of the quark masses, lattice spacing 0.082 fm and a larger volume of $64^3{\times}128$ have been shown in Ref.~\cite{Alexandrou:2018own} and demonstrate the future direction and progress of such computations. 
Furthermore, the production of another ensemble at a finer lattice spacing is already on its way.

\begin{acknowledgements}

We would like to thank all members of ETMC for their constant and pleasant collaboration. 
We also thank Y.\ Zhao for useful discussions. 
This work has received funding from the European Union's Horizon 2020 
research and innovation programme under the Marie Sk\l{}odowska-Curie grant agreement 
No 642069 (HPC-LEAP). K.C.\ was supported by National Science Centre (Poland) grant SONATA 
BIS no.\ 2016/22/E/ST2/00013. F.S.\ was funded by DFG project number 392578569.
M.C. acknowledges financial support by the U.S. Department of Energy, Office of Nuclear Physics, within
the framework of the TMD Topical Collaboration, as well as, by the National Science Foundation
under Grant No.\ PHY-1714407. 
This research used computational resources of the Oak Ridge Leadership Computing Facility (OLCF), 
which is a DOE Office of Science User Facility supported under Contract DE-AC05-00OR22725.
Additional computational resources are provided by John-von-Neumann-Institut f\"{u}r Computing (NIC) for
the JURECA supercomputer (allocation ECY00), Prometheus supercomputer 
at the Academic Computing Centre Cyfronet AGH in Cracow (grant ID \textit{quasipdfs}), Eagle 
supercomputer at the Poznan Supercomputing and Networking Center (grant no.\ 346), Okeanos 
supercomputer at the Interdisciplinary Centre for Mathematical and Computational Modelling in 
Warsaw (grant IDs gb70-17, ga71-22). 
\end{acknowledgements}

\noindent
\bibliographystyle{JHEP}
\bibliography{references}

\end{document}